\documentclass[aps, showpacs, twocolumn, notitlepage, superscriptaddress]{revtex4-1}
\usepackage{amsmath}
\usepackage{amssymb,amscd,bm,dsfont,wasysym,mathrsfs,latexsym,psfrag,accents,mathtools}
\usepackage{graphicx}
\usepackage{multirow}
\usepackage{dcolumn}
\usepackage{float}
\usepackage{color}
\usepackage{xcolor}
\usepackage{amsthm}
\usepackage{hyperref}
\usepackage{boldline}
\usepackage[normalem]{ulem}
\newcolumntype{L}[1]{>{\raggedright\arraybackslash}p{#1}}
\newcolumntype{C}[1]{>{\centering\arraybackslash}p{#1}}
\newcolumntype{R}[1]{>{\raggedleft\arraybackslash}p{#1}}

			\newcommand{\e}[1]{\begin{align}{#1}\end{align}}	
		
		\newcommand{\f}[2]{\frac{#1}{#2}}

		\newcommand{\p}[2]{\frac{\partial #1}{\partial #2}}

		\newcommand{\la}[1]{\label{#1}}

		\newcommand{\q}[1]{Eq.\ (\ref{#1})}
		\newcommand{\qq}[2]{Eqs.\ (\ref{#1}-\ref{#2})}
		\newcommand{\s}[1]{Sec.\ \ref{#1}}
		\newcommand{\fig}[1]{Fig.\ \ref{#1}}		
		\newcommand{\app}[1]{App.\ \ref{#1}}				
		\newcommand{\tab}[1]{Tab.\ \ref{#1}}

		\newcommand{\eq}{=&\;}

		\newcommand{\R}{\mathbb{R}}

\newcommand{\nabk}{\nabla_{\boldsymbol{k}}}

\newcommand{\lmt}{l^{\mt}}

\newcommand{\var}{\varepsilon}
\newcommand\as{\;\;\;\;}

\newcommand{\ba}{\boldsymbol{a}}

\newcommand{\bk}{\boldsymbol{k}}

\newcommand{\bp}{\boldsymbol{p}}

\newcommand{\br}{\boldsymbol{r}}
\newcommand{\bs}{\boldsymbol{s}}

\newcommand{\bE}{\boldsymbol{E}}

\newcommand{\bB}{\boldsymbol{B}}

\newcommand{\bK}{\boldsymbol{K}}

\newcommand{\bdelta}{\boldsymbol{\delta}}

\newcommand{\frako}{\mathfrak{o}}

\newcommand{\A}{{\cal A}}

\newcommand{\inv}{\mathfrak{i}}

\newcommand\rot{\mathfrak{c}}

\newcommand{\vectwo}[2]{\begin{pmatrix} {#1}\\{#2} \end{pmatrix}}
\newcommand{\matrixtwo}[4]{\begin{pmatrix} #1 & #2 \\ #3 & #4 \end{pmatrix}}
\newcommand{\diagmatrix}[2]{\begin{pmatrix} #1 & 0 \\ 0 & #2 \end{pmatrix}}

\newcommand{\sx}{\sigma_{\sma{1}}}

\newcommand{\sz}{\sigma_{\sma{3}}}

\newcommand{\ins}[1]{\;\;\;\;\text{#1}\;\;\;\;}

\newcommand{\cala}{{\cal A}}

\newcommand{\calh}{{\cal H}}

\newcommand{\calm}{{\cal M}}

\newcommand{\calq}{{\cal Q}}

\newcommand{\noi}[1]{\noindent (#1)}

\newcommand{\mt}{\text{-}2}

\newcommand{\braket}[2]{\big\langle #1 \big| #2 \big\rangle}

\newcommand{\braopket}[3]{\big\langle #1 \big| #2 \big| #3 \big\rangle}

\newcommand{\ket}[1]{\big|#1\big\rangle}

\newcommand{\lin}{\notag \\}

\newcommand{\ab}{\alpha\beta}

\newcommand{\bpm}{\begin{pmatrix}}
\newcommand{\epm}{\end{pmatrix}}

\newcommand{\bal}{\begin{align}}

\newcommand{\dg}[1]{#1^{\scriptstyle{\dagger}}}

\newcommand{\sma}[1]{\scriptscriptstyle{#1}}

\newcommand{\Z}{\mathbb{Z}}

\newcommand{\zpar}{Z_{\sma{\parallel}}}
\newcommand{\zper}{Z_{\sma{\perp}}}
\newcommand{\lper}{l_{\sma{\perp}}}

\newcommand{\order}{\text{O}}

\renewcommand{\H}{\mathcal{H}}
\newcommand{\barmu}{\bar{\mu}}
\newcommand{\barphi}{\bar{\varphi}}
\newcommand{\effH}{\mathfrak{H}}
\providecommand{\tabularnewline}{\\}

\begin{document}

\title{Landau quantization of nearly degenerate bands, \\
and full symmetry classification of avoided Landau-level crossings}

\author{Chong Wang}\affiliation{Institute for Advanced Study, Tsinghua University, Beijing 100084, China}
\author{Wenhui Duan}\affiliation{Institute for Advanced Study, Tsinghua University, Beijing 100084, China}\affiliation{Department of Physics and State Key Laboratory of Low-Dimensional Quantum Physics, Tsinghua University, Beijing 100084, China}\affiliation{Collaborative Innovation Center of Quantum Matter, Tsinghua University, Beijing 100084, China}
\author{Leonid Glazman}\affiliation{
Department of Physics, Yale University, New Haven, Connecticut 06520, USA}
\author{A. Alexandradinata}\affiliation{
Department of Physics, Yale University, New Haven, Connecticut 06520, USA}\affiliation{Department of Physics and Institute for Condensed Matter Theory, University of Illinois at Urbana-Champaign, Urbana, Illinois 61801, USA}

\begin{abstract}
Semiclassical quantization rules compactly describe the energy dispersion of Landau levels, and are predictive of quantum oscillations in transport and thermodynamic quantities. Such rules –- as formulated by Onsager, Lifshitz and Roth –-
apply when the spin-orbit interaction dominates over the Zeeman interaction (or vice versa), but does not generally apply when the two interactions are comparable in strength. In this work, we present a generalized quantization rule which treats the spin-orbit and Zeeman interactions on equal footing, and therefore has wider applicability to spin-orbit-coupled materials lacking a spatial inversion center, or having magnetic order. More generally, our rule describes the Landau quantization of any number of nearly degenerate energy bands -- in any symmetry class. The resultant Landau-level spectrum is generically non-equidistant but may contain spin (or pseudospin) degeneracies. To tune to such degeneracies in the absence of crystalline point-group symmetries, three real parameters are needed. We have exhaustively identified all symmetry classes of cyclotron orbits for which this number is reduced from three, thus establishing symmetry-enforced `non-crossing rules' for Landau levels.
In particular, only one parameter is needed {in the presence of} spatial rotation {or inversion}; this single parameter may be the magnitude or orientation of the field. Signatures of single-parameter tunability include (i) a smooth crossover between period-doubled and -undoubled quantum oscillations in the low-temperature Shubnikov-de Haas effect, as well as (ii) `magic-angle' magnetoresistance oscillations. We demonstrate the utility of our quantization rule, as well as the tunability of Landau-level degeneracies, for the Rashba-Dresselhaus two-dimensional electron gas -- subject to an arbitrarily oriented magnetic field.
\end{abstract}
\date{\today}

\maketitle

%n close analogy with the Wigner-von Neumann `non-crossing rule' in quantum mechanics, we find that three tunable parameters are needed to attain a Landau-level quasidegeneracy -- in the absence of crystalline point-group symmetries [cf.\ \s{sec:introducecodimension}]. 
%We have further determined symmetric variants of the ‘non-crossing rule’ for Landau levels, for any spacetime symmetry that occurs in crystalline solids

\section{Introduction}
In the semiclassical theory of a Bloch electron in a magnetic field,  quantization rules not  only serve as a compact description of the  energy dispersion of Landau levels, they also form the basis for a quantitative theory of quantum oscillations in transport\cite{SdH} and thermodynamic\cite{dHvA} quantities.  The original formulation of the rule -- by Onsager\cite{Onsager} and Lifshitz\cite{lifshitz_kosevich,lifshitz_kosevich_jetp} -- neglects the effect of spin-orbit coupling. While subsequent generalizations by Roth\cite{rotheffham,rothmag}, Mikitik\cite{Mikitik_quantizationrule}, and the present authors\cite{topoferm,100p} account for the spin-orbit effect, such quantization rules apply only when the spin-orbit splitting of energy bands is much stronger (or much weaker) than the Zeeman interaction.

\begin{figure}
\includegraphics[width=0.9\columnwidth]{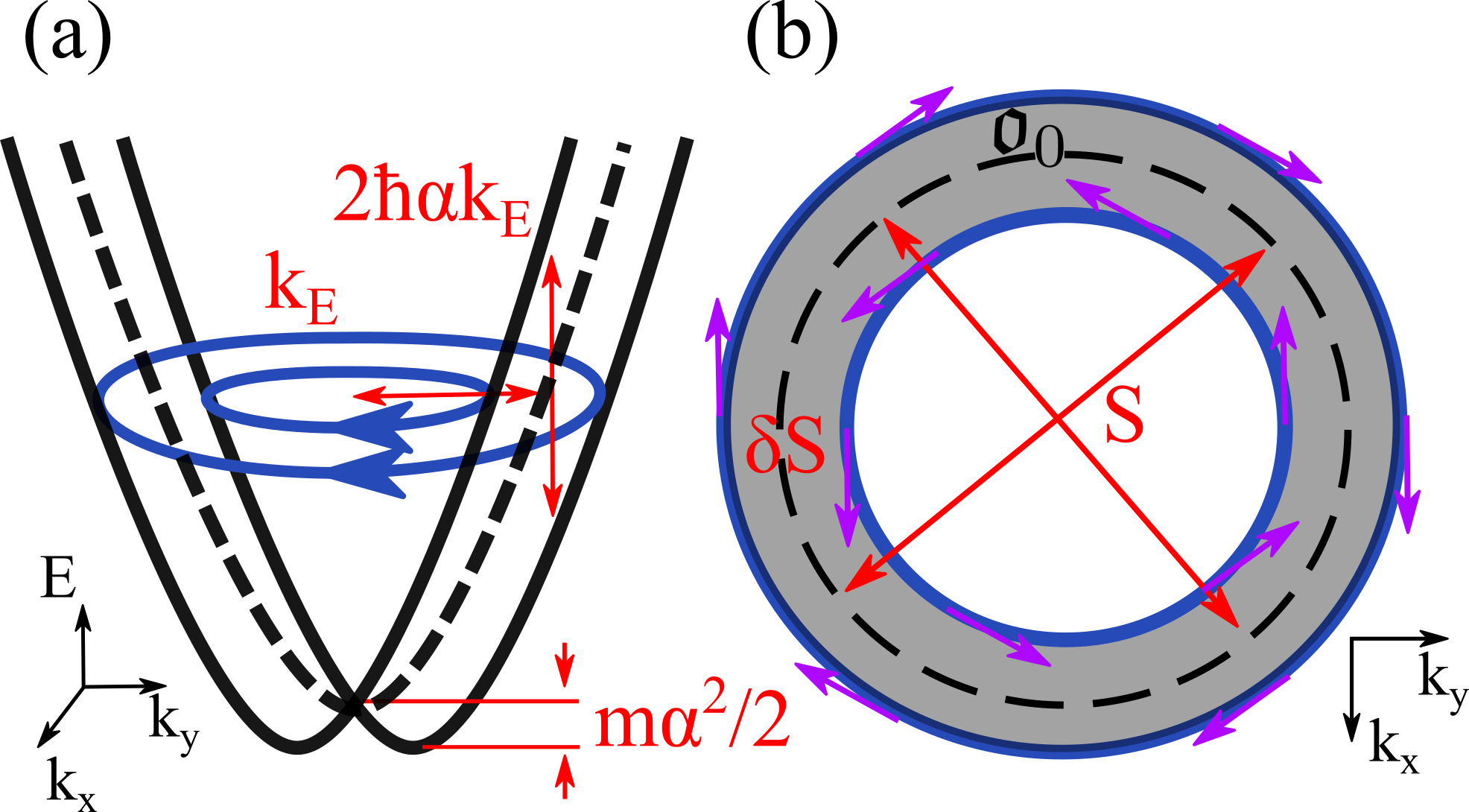}
\caption{(a) Solid (resp.\ dashed) lines illustrate the band dispersion of an electron gas with (resp.\ without) Rashba spin-orbit coupling.  The oriented circles illustrate the spin-orbit-split orbits for a magnetic field pointing to $-z$ direction. (b) The zeroth-order orbit $\frako_0$ is indicated by a dashed circle with area $S$; $\delta S$ is the differential area  between two spin-orbit-split orbits. The spin texture is indicated by purple arrows.\label{fig:orbits}}
\end{figure}

A quantization rule that treats the spin-orbit and Zeeman interactions on equal footing has yet to be formulated. Such a rule would be especially relevant to spin-orbit-coupled materials lacking a spatial inversion center, or having magnetic order. The energy bands of these materials carry a spin-half degree of freedom, and the spin degeneracy of energy bands is weakly lifted by spin-orbit coupling or magnetic ordering. A prototypical example is given by  a two-dimensional electron gas (2DEG) with Rashba spin-orbit coupling:
\begin{equation}
H_R=\frac{{\hbar^2} k^2}{2m}+\hbar\alpha  (k_{x}\sigma_{y}-k_{y}\sigma_{x}).\label{eq:Rashba-Hamiltonian}
\end{equation}
Given a magnetic field in the $-z$ direction, and focusing on energies much greater than $m\alpha^2$, the cyclotron orbits [illustrated in \fig{fig:orbits}(a)]   are split in $\bk$-space with a differential area  $\delta S$ that is much less than their average area $S$\footnote{The expressions for $\delta S$ and $S$ in \q{deltaSvsS} are asymptotically valid for $E{\gg}m\alpha^2$.}:
\e{\delta S = 4\pi m \alpha k_E/\hbar\ll S =\pi k_E^2; \;\; k_{E}:=\f{\sqrt{2m E}}{\hbar}.\label{deltaSvsS}}
$S$ is equivalently the area of the constant-energy contour in the absence of spin-orbit coupling ($\alpha{=}0$).

Our work presents a generalized rule to describe the Landau quantization of nearly-degenerate energy bands, for which  spin-orbit-split bands is but one example. By the \textit{near-degeneracy} condition, we generally mean that
the ratio of differential to average area is small: $|\delta S/S|{\ll}1$. Our proposed rule relies also on a second condition -- that $S$ is much larger than the inverse square of the magnetic length $l{=}\sqrt{\hbar/eB}$; this is the standard semiclassical condition in the Onsager-Lifshitz-Roth theory. By treating \textit{both} $|\delta S/S|$ and $1/l^2|S|$   as small parameters with finite ratio $l^2|\delta S|$, our perspective diverges from and generalizes \textit{single-parameter} semiclassical theories\cite{kohn_effham,blount_effham,rotheffham,wannier_fredkin,fischbeck_review,Mikitik_quantizationrule,topoferm,100p,gao_zero-field_2017}.

The inverse ratio $1/l^2|\delta S|$ measures the hybridization of nearly-degenerate orbits by the Zeeman interaction. In the regime $l^2|\delta S|{\gg}1$, the Onsager-Lifshitz{-Roth} quantization rules apply independently to  two concentric (but decoupled) orbits. In the opposite regime $l^2|\delta S|{\ll}1$, the spin-orbit splitting of energies is negligible, and we may then apply the known quantization rule for exactly-degenerate bands\cite{rotheffham,rothmag,topoferm,100p,Mikitik_quantizationrule}. Our unifying rule provides the smooth interpolation between these two known regimes. {Equivalently stated, we have extended the applicability of quantization rules to the intermediate regime ($l^2|\delta S|{\sim}1$) where the Zeeman interaction is comparable with the spin-orbit interaction; previously, such rules were known only when one interaction dominates over the other. This extension allows  to formulate semiclassical quantization rules for spin-orbit coupled materials in \textit{any} symmetry class -- with or without spatial inversion symmetry, with or without time-reversal symmetry. The unifying rule is presented in \s{sec:qtznrules} for nearly-degenerate bands with a spin-half (or pseudospin-half) degree of freedom. As elaborated in \s{sec:discussion}, the most general formulation of our rule is applicable  to any number of nearly-degenerate bands of any symmetry.}

In \s{sec:Rashba}, we demonstrate the utility of this rule for the Rashba{-Dresselhaus} 2DEG with an arbitrarily-oriented magnetic field. A sufficiently strong in-plane field not only moves the Dirac point but tilts the Dirac cone over -- into a type-II\cite{soluyanov_type-ii_2015, muechler_tilted_2016, bergholtz_topology_2015} Dirac point. In \s{sec:inplanezeeman}, we investigate the Landau-Zener tunneling dynamics that occur in the vicinity of a II-Dirac point; in particular, we will {demonstrate that our quantization rule can account for quantum tunneling between nearly-degenerate bands -- a phenomenon known as interband breakdown\cite{kaganov_coherent_1983,slutskin_dynamics_1968,AALG,100p}. We will also} critically evaluate a recent claim\cite{obrien_magnetic_2016} of perfect Klein tunneling at the energy of the Dirac point -- we find that such tunneling is never perfect with a proper account of the Zeeman interaction.

{\s{sec:llquasideg} applies} our quantization rule to determine how many real parameters are needed to tune to a {pseudo}spin-half degeneracy in the Landau-level spectrum. {This degeneracy may be exact if Landau-quantized wave functions transform in different unitary representations of a spatial symmetry; absent such a symmetry,} it is still possible for the repulsion between two Landau levels to be anomalously weak -- this notion of a Landau-level quasidegeneracy is formulated in \s{sec:relatedegeneracies}. In close analogy with the Wigner-von Neumann `non-crossing rule' in quantum mechanics\cite{neumann2000behaviour}, we find that three tunable parameters are needed to attain a Landau-level quasidegeneracy -- in the absence of crystalline point-group symmetries (cf.\ \s{sec:introducecodimension}). 
We have further determined symmetric variants of the Landau-level ‘non-crossing rule’, for any spacetime symmetry that occurs in crystalline solids; our results are summarized in Tab.  \ref{table:codimension-nearlydegen}-\ref{table:codimension-exactlydegen}. In particular, we highlight that in the presence of a spatial rotation or inversion symmetry,  only a single tunable parameter is needed to attain a Landau-level quasidegeneracy (cf.\ \s{sec:ctsrot}); this single parameter may be the magnitude or orientation of the field, or the Landau-level index itself.

In \s{sec:quantosc_quasideg}, we propose a signature of single-parameter tunability in the low-temperature Shubnikov-de Haas effect\cite{SdH}: a smooth crossover between period-doubled and -undoubled quantum oscillations. For a harmonic analysis of  quantum oscillations, we present in \s{sec:quantosc_equidis} a generalized Lifshitz-Kosevich formula that inputs our quantization rule and outputs harmonics. In particular, we show that destructive interference of the fundamental harmonic is always tunable by a single parameter, leading to a beating pattern in the high-temperature quantum oscillations; the beating nodes recur \textit{aperiodically} in $1/B$, due to a competition between spin-orbit splitting and the Zeeman interaction. 

We conclude in \s{sec:discussion} by summarizing  the technical results of this work, highlighting certain key equations, and presenting an outlook for future investigations.    

\section{Quantization Rule\label{sec:qtznrules}}

Let us heuristically argue for the form of such a rule; a more pedagogical and rigorous approach is elaborated in App. \ref{app:quantizationruleproof}. We consider independent electrons described (at zero field) by a translation-invariant Hamiltonian: $\hat{H}_0{+}\delta \hat{H}$. This decomposition is such that energy bands of $\hat{H}_0$ are $D$-fold degenerate at every crystal wavevector $\bk$, while $\delta \hat{H}$ is a weak perturbation that lifts this degeneracy at generic $\bk$. For conceptual simplicity, we focus on $D{=}2$ spin degeneracy, and postpone the general formulation for $D{\geq}2$ to App. \ref{app:quantizationruleproof}.  $\hat{H}_0$ is commonly exemplified by  the Schr\"odinger Hamiltonian, and $\delta \hat{H}$ by a non-centrosymmetric spin-orbit interaction; alternatively, $\hat{H}_0$ may be a centrosymmetric Pauli Hamiltonian (with spin-orbit coupling), while $\delta \hat{H}$ describes a lattice deformation that breaks the spatial-inversion symmetry.

Let us apply a magnetic field, and first consider the spin-degenerate Landau levels obtained by neglecting both $\delta \hat{H}$ and the Zeeman splitting. In the  semiclassical description,  we are interested in  field-induced dynamics within a low-energy degenerate band of $\hat{H}_0$ with  dispersion $\var(\bk)$, e.g., $\var(\bk){=}\hbar^2 k^2/2m{+}\order(k^4)$ in the Rashba model.    Landau quantization is well-known to derive from the Peierls substitution $\var(\bk){\rightarrow}\var(\bK)$\cite{peierls_substitution}, with the kinematic quasimomentum operator   $\boldsymbol{K}{:}{=}\boldsymbol{k}{+}(e/\hbar) \boldsymbol{A}(i\nabla_{\boldsymbol{k}})$ and $\boldsymbol{A}(\br)$ being the electromagnetic vector potential. Assuming that the field is aligned in the ${-}z$ direction, $[K_x,K_y]{=}i\lmt$. The semiclassical solution corresponds to  cyclotron revolutions of an electron around a constant-energy contour of $\var(\bk)$ [see Fig. \ref{fig:orbits}(b)]; this contour, equipped with an orientation given by the equation of motion ($ \hbar dk_{\alpha}/dt {=} \lmt \epsilon_{\ab}\partial \var/\partial k_{\beta} $, with $\epsilon_{xy}{=}{-}\epsilon_{yx}{=}1$), shall be referred as the zeroth-order orbit $\frako_0$.  In this work, we focus only on closed orbits which do not wrap non-contractibly around the Brillouin torus.

The quantization rule for Landau levels reflects the single-valuedness of the semiclassical (WKB) wavefunction\cite{berry_mount_review} over $\frako_0$, i.e.,  the phase accumulated by an electron (over one cyclotron period $T_c$) is an integer multiple of $2\pi$: $(l^2S(E){+}\gamma)/2\pi{\in}\Z$. The leading-order phase   is the classical action $l^2S{:}{=}{-}l^2\int_{0}^{T_c} k_x \dot{k}_y dt$; $S$ has a dual interpretation as the signed area (positive for electron like orbits) in $\bk$-space enclosed by $\frako_0$, as illustrated in Fig. \ref{fig:orbits}(b). $\gamma$ is a subleading Maslov correction\cite{keller1958} that generally equals $\pi$ times the rotation number $r$ of $\frako_0$ (i.e., the number of rotations made by the tangent vector over one cycle)\cite{100p}; for orbits that are deformable to a circle,  $r{=}{\pm} 1$. In the zeroth-order  solution, all Landau levels are spin-degenerate with a level spacing 
\e{ \var_c:=\f{2\pi}{l^2|\partial S/\partial E|}:=\f{\hbar^2}{ml^2}:=\f{h}{T_c}. \label{def:cyclotron}} 
Here, $\var_c$ and $T_c$ are respectively the cyclotron energy and period of the zeroth-order orbit, and $m$ is the effective mass. Henceforth, `degeneracy of Landau levels' should be understood as a spin degeneracy. 

To the first order, $\delta \hat{H}$ and the Zeeman interaction induce a subleading phase offset ($\lambda_a$) which generically differs for the two orthogonal spinor states (labelled $a{=}1,2$):
\begin{equation}
l^2S(E)+\lambda_a(E,B)+\gamma=2\pi n,\as  n\in \mathbb{Z},\;a=1,2. \label{eq:rule}
\end{equation}
This implies that the spin-degeneracy of Landau levels is split by approximately  $\var_c|\lambda_1{-}\lambda_2|/2\pi$. The phase offsets $\lambda_a$ can be viewed as Floquet quasienergies\cite{shirley_solution_1965} of a time-periodic, perturbative Hamiltonian ($\calh$) that governs the dynamics of the spinor wavefunction.  $\lambda_a$ is obtained from eigenvalues $\{e^{i\lambda_a}\}_{a=1}^2$ of a time-evolution propagator -- the time-ordered exponential 
\begin{equation}
\A{:}{=}\overline{\exp}\left[-\frac{i}{\hbar}\int_0^{T_c} \calh(\bk(t)) dt\right],
\label{eq:prop}
\end{equation} 
with $\bk(t){\in}\frako_0$ a periodic function determined by the semiclassical equation of motion.   $\calh$ has the two-by-two matrix form:
\begin{equation}
  \calh=\delta \var +B\bigg(M_z -g_s\mu_{B}\f{s_z}{\hbar} + e\epsilon_{\alpha\beta}\mathfrak{X}_{\beta}v_{\alpha}\bigg):=\delta \var+B\calm,\label{eq:H1}
\end{equation}
$\delta \var$ is the projection of $\delta \hat{H}$ to the degenerate band subspace of $\hat{H}_0$. $\delta \var$ is assumed to be traceless; this property is guaranteed if $\delta \hat{H}$ represents a spin-orbit coupling term that breaks the spin SU(2) symmetry of $\hat{H}_0$;\footnote{The Pauli spin-orbit coupling term ${\propto} \bs{\cdot} (\bE {\times} \bp)$ is traceless in any spin-symmetric (pseudo-)spinor basis.  Viewing the spin-orbit coupling as a small parameter in degenerate perturbation theory, the lowest-order splitting of any spin-degenerate energy level is always symmetric about the zeroth-order energy. This follows because the zeroth-order, spin-symmetric wavefunctions (associated to a spin-degenerate level) are tensor products of spin and spatial wavefunctions: $\ket{\Psi_{\pm}}{:}{=}\ket{{\pm}\hat{n}}{\otimes} \ket{\psi}$ with $\hat{n}$ a unit vector and  $\bs{\cdot} \hat{n} \ket{{\pm}\hat{n}}{=}{\pm}\ket{\pm \hat{n}}$. Hence, $\braopket{\Psi_{\pm}}{\bs{\cdot} (\bE {\times} \bp)}{\Psi_{\pm}}{=}{\pm}\hat{n}\cdot\braopket{\psi}{\bE {\times} \bp}{\psi}$.} more generally, one may always absorb the traceful part (of  the projection of $\delta \hat{H}$) into the definition of $\var(\bk)$.  In the basis of energy-split bands (which we label by ${+}$ and $-$ for bands corresponding to the outer and inner orbit respectively), $\delta \var$ is a diagonal matrix whose diagonal entries ($\delta \var_+,~\delta \var_-{=}{-}\delta \var_+$)  are related to the hybridization parameter:\footnote{The neglected term on the right-hand side of \q{spinsplitwkb} is smaller (than the kept term) by a factor $\delta v_{\pm}/v$, which is the relative change in band speed $v=|\boldsymbol{v}|=\hbar^{-1}|\partial \var/\partial \bk|$ induced by the spin-orbit coupling. This factor is assumed to be small.  For example in the Rashba model, $\delta v_{\pm}/v=m\alpha/\hbar k_E{\sim}\delta S/S{\ll}1$.}   
\e{-\int_0^{T_c} (\delta \var_+{-}\delta \var_-) \frac{dt}{\hbar} \approx l^2\delta S. \label{spinsplitwkb}}

The remaining terms in \q{eq:H1} describe a generalized Zeeman interaction $B\calm$ which is the first-order-in-$B$ correction to the Peierls-Onsager Hamiltonian $\var(\bK)$\cite{rotheffham,blount_effham,kohn_effham}. Included in this interaction is the Zeeman coupling to the spin ($g_s\mu_Bs_z/\hbar$) and orbital ($M_z$) magnetic moment\cite{thonhauser_orbital_2005} of the nearly-degenerate band; the latter is expressible in terms of matrix elements of the velocity operator $\boldsymbol{\Pi}_{mn}=i\langle\psi_m|[\hat{H}_0, \hat{\boldsymbol{r}}]|\psi_n\rangle/\hbar$ between Bloch wave functions $\psi_{n\bk}$:
\e{
[M_z]_{mn}=\f{ie\hbar}{2}\sum_l'\f{\Pi_{x,ml}\Pi_{y,ln}-\Pi_{y,ml}\Pi_{x,ln}}{\epsilon_m-\epsilon_l}. \label{eq:orbitalmoment}
}
Here, the summation in $l$ does not include the two nearly-degenerate bands.  There is finally a geometric contribution to the Zeeman interaction, which involves the two-by-two non-Abelian Berry connection\cite{berry_quantal_1984,wilczek_appearance_1984} as well as the  zeroth-order band velocity: 
\e{\mathfrak{X}_{\beta,mn}{=}i\braket{u_m}{\partial u_n/\partial k_{\beta}},  \;\;\;v_{\alpha}:= \f{1}{\hbar}\p{\var}{k_{\alpha}},\label{berryconn}}
with $\{u_n\}_{n=1}^2$ the periodic component of Bloch functions. 
%which provides a geometric characterization of band wavefunctions along $\frako_0$.\\
%All matrix elements in \qq{eq:prop}{eq:orbitalmoment} are evaluated with wavefunctions of $\hat{H}_0$. 

\qq{eq:rule}{berryconn} allow us to determine $\lambda_{1,2}$ to the lowest order  in $1/l^2|S|$ and $|\delta S/S|$, and for arbitrary finite values of $l^2|\delta S|$ and $|B\calm|/\delta \var_{\pm}$. We remark that the quantization rule is invariant under basis transformations within the nearly-degenerate band subspace (see App. \ref{app:quantizationruleproof}).  In subsequent case studies, it may be analytically convenient to adopt a $\bk$-independent basis for $\calh$, if such a basis exists; the non-abelian Berry connection -- defined with respect to this basis -- vanishes.

%In other words,  \qq{eq:rule}{berryconn} is equally valid with $\{u_n\}_{n=1}^2$ replaced by energy eigenfunctions of $\hat{H}_0$ which are not simultaneously eigenfunctions of $\delta \hat{H}$. 

%In subsequent case studies, it may be analytically convenient to adopt a $\bk$-independent basis, if such a basis exists; the non-abelian Berry connection defined with respect to this basis would vanish.

%It will be useful to separate $B$-independent and $B$-dependent terms in \q{eq:H1}; the latter we denote as $B\calm$, with $\calm$ having the dimension of a magnetic moment. Let us explain each term in the order of its appearance.\\

\subsection{Recovering the Onsager-Lifshitz-Roth quantization rules in two limits}\label{sec:recoveronsager}

It is useful to compare $\delta \var$ with $B\calm$ in the basis of energy-split bands (labelled $\pm$): in the two regimes where one dominates over the other, \qq{eq:rule}{berryconn} reduce to the known Onsager-Lifshitz-Roth rules for (i) two independent, nondegenerate orbits, and for (ii) degenerate orbits.  \\

%\footnote{Though the adiabatic limit does not exist for orbits that exactly intersect at type-II Dirac points, these orbits are nevertheless describable by our quantization rule, as exemplified in the case studies of \s{sec:inplanezeeman} and \s{sec:rotsymmbreakdown}.}}

\noi{i} If $|\var_+{-}\var_-|$ is everywhere (along $\frako_0$) much greater than $B|\calm_{+-}|$, then an electron would undergo adiabatic (band-conserving) dynamics\cite{nenciu_review} along two independent, nondegenerate orbits. Such an adiabatic regime always exists as $B{\rightarrow} 0$ for nearly-degenerate orbits that do not intersect. (Such intersection occurs, e.g., at a type-II Dirac point, which is the subject of the case studies in \s{sec:inplanezeeman} and \s{sec:rotsymmbreakdown}.)
By  neglecting  $B\calm_{+-}$ in favor of intraband elements,  $\cala$ reduces to a diagonal matrix with  diagonal entries ($e^{i\lambda_{\pm}}$) given by:
\e{\lambda_{\sma{\pm}} = -\int_0^{T_c}\calh_{\sma{\pm\pm}}\f{dt}{\hbar}=\pm \f{1}{2}l^2\delta S - B\int_0^{T_c}\calm_{\sma{\pm\pm}}\f{dt}{\hbar}.  \label{phaseindependentorbit}} 
\q{eq:rule}, combined with \q{phaseindependentorbit}, may be identified as the Onsager-Lifshitz rule\cite{Onsager,lifshitz_kosevich,lifshitz_kosevich_jetp} for two independent, nondegenerate orbits of areas $S{\pm}\delta S$, and with subleading phase corrections that were first identified by Roth\cite{rothmag}\footnote{\q{phaseindependentorbit} neglects the contribution by band $-$ to the single-band orbital moment\cite{chang_berry_1996} of band $+$. This contribution is equal to $B\delta M_z=il^{-2}\epsilon_{\ab}(\delta \var_+-\delta \var_-)\mathfrak{X}^{\alpha}_{+-}\mathfrak{X}^{\beta}_{-+}$. $\mathfrak{X}^{\alpha}_{+-}$, being an off-diagonal matrix element of the position operator, is generically of order the lattice constant $a$. We then estimate $\int_0^{T_c} B\delta M_z dt/\hbar=O(a^2 \delta S)$, which is negligible compared to $l^2\delta S/2$. Note also that the single-band orbital moment simply vanishes in some symmetry classes of orbits\cite{100p}.}. With the perturbative inclusion of $B\calm_{\sma{+-}}$, $\lambda_{\pm}$ [cf.\ \q{phaseindependentorbit}] would then be expressible as a Laurent series in $B$, with a leading term proportional to $1/B$; this will be confirmed in a subsequent case study [cf.\ \q{lambdasmallB}] and applied to understand beatings in quantum-oscillatory phenomena [cf.\ \s{sec:qo}].

\noi{ii} Let us consider the opposite regime where the Zeeman splitting due to $B\calm$ is much greater than $(\delta \var_+{-}\delta\var_-)$. \qq{eq:rule}{berryconn}, with $\calh{\approx}B\calm$, may then be identified with the quantization rule (formulated previously by us\cite{topoferm,100p}) for degenerate orbits of any symmetry.\footnote{An equivalent but less general formulation was proposed earlier in Ref. \onlinecite{rothmag} and Ref. \onlinecite{Mikitik_quantizationrule} for centrosymmetric solids with time-reversal symmetry (at zero field). For comparison, it should be emphasized that \qq{eq:rule}{berryconn} applies to solids of any symmetry, including magnetically-ordered solids.}

The quantization rule, as summarized by \qq{eq:rule}{berryconn}, is the technical accomplishment of this work; the remainder of this work is focused on unpacking its physical implications. We will first demonstrate the utility of our rule in case studies where the adiabatic approximation is invalid, and $\delta \epsilon$ competes with $B\calm$. There are two conceptually different ways in which the adiabatic approximation may fail: (i) if the spin-split orbits are  symmetric under continuous rotations (as they are for the Rashba 2DEG), then adiabaticity may be lost over the entire orbit [cf.\ \s{sec:Rashba}]. (ii) For asymmetric spin-split orbits, adiabaticity may be lost in the vicinity of a single isolated point, where orbits intersect at a type-II Dirac point [cf.\ \s{sec:inplanezeeman}]. 
In \s{sec:llquasideg}, we investigate how many real parameters are required to tune $\lambda_1{=}\lambda_2$ (mod $2\pi$), which would imply a (near) degeneracy of Landau levels.  

\section{Case study: Rashba 2DEG with arbitrarily-oriented fields}

\subsection{Rashba 2DEG with an out-of-plane field}\label{sec:Rashba}

In this section, we apply our rule to the Landau levels of the Rashba model [cf.\ \q{eq:Rashba-Hamiltonian}], for which analytic, closed-form solutions exist for verification\cite{bychkov_oscillatory_1984}.

For simplicity, we assume that the Pauli matrices in \q{eq:Rashba-Hamiltonian} correspond to spin: $s_i{=}\hbar \sigma_i/2$. Let us evaluate the effective Hamiltonian $\calh(\bk)$ along the cyclotron trajectory $\frako_0$ with radius $k$. In the energy basis of the Rashba Hamiltonian $H_R$ [cf.\ \q{eq:Rashba-Hamiltonian}], 
\e{\calh(\bk)=-\hbar\alpha k\tau_z{+}\f{g_{s\perp}}{2}\mu_{\sma{B}}B\tau_x {-} \f{eB\hbar}{2m} (\tau_x-I) \epsilon_{\alpha\beta} k_{\alpha}\nabla_{\beta}\theta,\label{rashbaeffham}}
where $\tau_i$ are a new set of Pauli matrices with $\tau_z{=}{+} 1$ and ${-}1$ corresponding to the outer and inner orbits, respectively. $e^{i\theta_{\bk}}{=}(ik_x{-}k_y)/k$ is the angle of the in-plane spin-splitting field $(-\hbar\alpha k_y,\hbar\alpha k_x,0)$ of $H_R$ [cf.\ \q{eq:Rashba-Hamiltonian}], which winds once over $\frako_0$; the spin of our basis vectors likewise winds, in parallel (or anti-parallel) alignment with this field (see Fig. (\ref{fig:orbits})(b)). Terms proportional to the rate of spin rotation ($\nabk \theta$) originate from the non-Abelian Berry connection. The first term of  $\calh$ corresponds to the spin-orbit-induced splitting of the energy-degenerate bands;
the remaining terms may be interpreted as $B$-induced deformations of the bands. The orbital moment term $B M_z$, which generally couples the two-band subspace (of interest) to all other bands, is absorbed into the Zeeman spin interaction with a phenomenological g-factor $g_{s\perp}$\footnote{For simplicity, we neglect here off-diagonal elements of $M_z$ in the basis of $S_z=\pm 1/2$, with $\vec{z}$ the direction of the field.}. It is well-known that the Zeeman term tends to polarize spin in the direction of $\bB$; remarkably, the non-Abelian Berry term  also has a polarizing effect, but with opposite sign. These two effects completely cancel for free electrons (with $g_{s\perp}=2$ and $m$ equal to the free-electron mass $m_0$), but not generically in band systems. 
%is absent in any simplified two-band model.

Due to the continuous rotational symmetry of $H_R$, $\nabk \theta$ has magnitude $1/k$ and lies tangential to $\frako_0$; consequently, $\calh$ is constant along $\frako_0$, and the propagator simplifies to $\cala{=}e^{{-}i\calh T_c/\hbar}$ without time-ordering, and the phase corrections to the quantization rule are simply the eigenvalues of $\calh$ multiplied by $T_c$:
\begin{equation}
\lambda_{\pm}=\pm\f1{2}\sqrt{({l^2\delta S})^2+\pi^2(g_{s\perp}m/m_0-2)^2}+\pi. \label{eq:Rashba-lambda}
\end{equation}
This is a function of the differential area $\delta S$ [cf.\ \q{deltaSvsS}] and the effective mass $m$. The standalone $\pi$ term is the single-band Berry phase associated to the winding of $\theta_{\bk}$ by ${\pm}2\pi$. The two quantities under the square root originate from the spin-orbit energy splitting $\delta \var$ and the inter-band matrix elements of $B\calm$ (in the basis of zero-field bands). 

In the weak-hybridization regime: $l^2\delta S \gg \pi|g_{s\perp} m/m_0{-}2|$, we may expand \q{eq:Rashba-lambda} as a Laurent series in $B$:
\e{\lambda_{\pm}= \pm \f{l^2\delta S}{2} +\pi \pm \f{\pi^2}{4}\f{(g_{s\perp}m/m_0-2)^2}{l^2\delta S} + O(B^2). \label{lambdasmallB}}
\q{eq:rule}, combined with the first two terms on the right-hand-side of \q{lambdasmallB}, may alternatively be derived from  the standard Onsager-Lifshitz rules for two independent orbits, with a common $\pi$ Berry-phase correction for each orbit\footnote{The intra-band Zeeman and orbital-moment couplings vanish due spacetime-inversion symmetry\cite{topoferm}.}: $l^2(S{\pm}\delta S)/2\pi {\in}\Z$. Higher-order terms in \q{lambdasmallB} reflect the field-induced hybridization of the spin-orbit-split orbits. Eq.\ (\ref{lambdasmallB}) has been partially obtained by Mineev for the Rashba model\cite{mineev_haas--van_2005}; he missed the second term in $(g_{s\perp}m/m_0-2)$ [cf.\ \q{lambdasmallB}], due to his neglect of the interband Berry connection.

In the strong-hybridization regime where $l^2\delta S {\ll}\pi(g_{s\perp}m/m_0{-}2)$, $\lambda_{\pm}{\approx}{\pm}\pi g_{s\perp}m/2m_0$ mod $2\pi$, which corresponds to the Zeeman splitting of Landau levels by $g_{s}\mu_BB$, in the \textit{absence} of the spin-orbit interaction. 

Our quantization rule [\q{eq:rule} with \q{eq:Rashba-lambda}] may be compared with the  exact Landau-level spectrum of the Peierls-substituted Hamiltonian $H_R(\bK)$ (plus the Zeeman coupling to spin).  The exact levels\cite{bychkov_oscillatory_1984} are given implicitly by
\e{
&n_{\pm}=\frac{l^2S}{2\pi}+\frac{l^2\delta S}{16\pi}\f{\delta S}{S}\lin 
&\pm\f1{2}\sqrt{\bigg(\frac{l^2\delta S}{8\pi}\f{\delta S}{S}\bigg)^2+(l^2\delta S)^2+\pi^2\bigg(\f{g_{s\perp}m}{m_0}-2\bigg)^2},\label{eq:Rashba-exact}}
with the functional forms of $\delta S(E)$ and $S(E)$ provided in \q{deltaSvsS};  $n_{+}{\ge} 1$ and $n_-{\ge} 0$ here are integer-valued labels for two distinct sets of levels. In the near-degenerate ($\delta S/S\ll 1$) and semiclassical ($1/l^2S \ll 1$) regime, with finite $l^2\delta S$, the two terms in \q{eq:Rashba-exact} involving $\delta S/S$  may be neglected, and \q{eq:Rashba-exact} reduces to our quantization rule with $\lambda_{\pm}$ given by \q{eq:Rashba-lambda}.

With the further inclusion of the Dresselhaus spin-orbit interaction in \q{eq:Rashba-Hamiltonian}, no closed-form analytic solution (of the Landau levels) is known. We may anyway verify the validity of our approximation scheme  -- by comparison with the Landau levels obtained from numerical exact diagonalization; this  is carried out in  \app{app:approximatevsexact}.

\subsection{Case study of breakdown: Rashba 2DEG with tilted field}\label{sec:inplanezeeman}

Our next case study involves orbits that (nearly) touch at a type-II Dirac point\cite{soluyanov_type-ii_2015,muechler_tilted_2016} -- an asymmetric variant of the conventional Dirac point (as will be introduced below). Quantum tunneling localized to a type-II Dirac point  (in short, a II-Dirac point) is known to be mathematically equivalent to  Landau-Zener breakdown, with the electric force replaced by the Lorentz force\cite{AALG,obrien_magnetic_2016,kane_blount}. Here, we will demonstrate how Landau-Zener dynamics emerge from our quantization rule [\qq{eq:rule}{eq:H1}], and explore (for the first time) how these dynamics are modified by the Zeeman effect. In particular, we will show that the perfect Klein tunneling predicted by O'Brien \textit{et al}\cite{obrien_magnetic_2016} is never perfect with a proper account of the Zeeman effect.

\begin{figure}
\includegraphics[width=1.0\columnwidth]{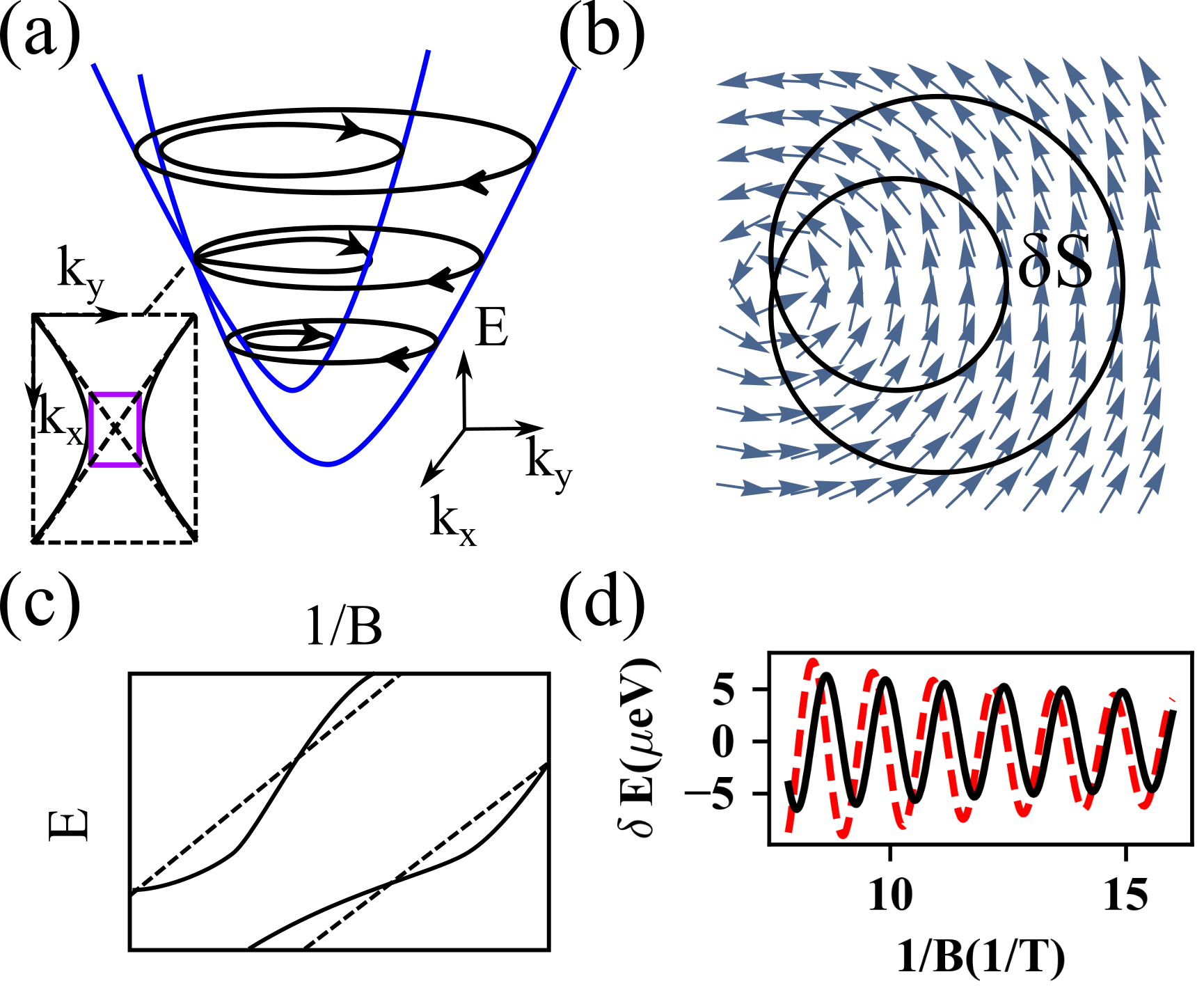}
\caption{(a) illustrates the cyclotron orbits of the Rashba 2DEG in a tilted magnetic field. The inset illustrates an enlarged view of near II-Dirac point. (b) illustrates the cyclotron orbit at the energy of the II-Dirac point; a Klein-tunneling electron will follow a trajectory that resembles the edge of a M\"obius strip. The orbit is drawn over the  spin-splitting field ($d_x,d_y$) of the degeneracy-lifting Hamiltonian $\delta\var$ [cf.\ \q{eq:RZ-Hamiltonian}]. The wavefunction along the inner (resp.\ outer) component of the M\"obius-strip orbit has its pseudospin aligned parallel (resp.\ anti-parallel) to the spin-splitting field. (c) schematically illustrates  Landau levels near the II-Dirac point (in the weak-hybridization regime). The dashed line represents the Onsager-Lifshitz-Roth quantization of the M\"obius-strip orbit [cf.\ Eq.\ (\ref{trefoilrule})], and the solid line incorporates the first-order correction [cf.\ Eq.\ (\ref{RZfirstorderE})] due to tunneling. (d) The black solid line is the plot of $\delta E_n$ [cf. Eq. (\ref{RZfirstorderE})], which is the energetic correction due to deviations from perfect Klein tunneling; for comparison, the red dashed line shows the numerically-exact deviations obtained from exact diagonalization. This comparison is made for the Landau levels closest to 0.45eV and with even Landau-level index $n$; for the parameters we have chosen ($m{=}0.076m_0$, $\alpha{=} 1.5\times10^{5}$cm/s, $Z_\parallel {=}$1meV and $g_{s\perp}=2$),  the II-Dirac point lies at 0.5eV. 
%Out-of-plane Zeeman coupling is negligibly weak.
\label{fig:RZ}}
\end{figure}

We begin again with the Rashba model [cf.\ \q{eq:Rashba-Hamiltonian}] of a rotationally-symmetric Dirac cone. To convert this Dirac point to its asymmetric variant, 
let us introduce a Zeeman coupling to an in-plane magnetic field $B_\parallel$ (parallel to the $+y$ direction):
\e{ 
H_{RZ}=\f{{\hbar^2}k^{2}}{2m}+\delta \var; \;\;\;\delta \var=d_x\sigma_x+d_y\sigma_y,\lin
(d_x,d_y):=(-\hbar\alpha k_y,\hbar\alpha k_x+\zpar),\;\;\;\zpar:= g_{s\parallel} \mu_B B_{\parallel}/2.
\label{eq:RZ-Hamiltonian}}
While this field breaks both time-reversal ($T$) and two-fold rotational ($C_2$) symmetries, the composition $C_2T$ is preserved as $\sigma_x H^*_{RZ}(\bk)\sigma_x{=}H_{RZ}(\bk),$ which constrains spin to lie in-plane at each $\bk$. In this symmetry class, Dirac nodes are irremovable but movable   in two-dimensional $\bk$-space. The in-plane Zeeman field has the dual effect of  moving the Dirac node  to the position:
\begin{eqnarray}
\bar{k}_x & = & -\frac{\zpar}{\hbar\alpha},\;\bar{k}_y=0,\;\epsilon_{0}=\frac{\zpar^{2}}{2m\alpha^2},\label{whereisdiracpoint}
\end{eqnarray}
and tilting the Dirac cone. To observe this tilt, consider the linearized Hamiltonian in the vicinity of this node: 
\e{
H & =\hbar\alpha\left[\left(\sigma_y-\frac{\zpar}{m\alpha^2}\right)\delta k_{x}-\sigma_{x}\delta k_{y}\right]+\epsilon_{0}.}
When $|\zpar|>|m\alpha^2|$, the Dirac cone tilts over in the $x$ direction, leading to a Lifshitz transition  of the constant-energy band contours -- from circular to hyperbolic (in the linearized approximation), as illustrated in Fig. \ref{fig:RZ}(a). An over-tilted Dirac cone shall be referred to as type-II. Our case study of a II-Dirac point  connects  two electron-like  pockets [cf.\ \fig{fig:RZ}(a,b)], and  differs from previous case studies\cite{obrien_magnetic_2016,AALG} that connect an electron- and hole-like pocket.

%red dashed line shows the energetic difference between Landau levels determined by numerical diagonalization and Landau levels obtained by the M\"obius quantization rule  [cf.\ Eq.\ (\ref{trefoilrule})]; this red dashed line may be compared to the perturbative formula $\delta E_n$ [cf. Eq. (\ref{RZfirstorderE})], which is plotted as a black solid line.

In addition to the in-plane field, let us also apply an out-of-plane field $B_{\perp}$, with associated magnetic length $\lper{=}\sqrt{\hbar/eB_{\perp}}$, cyclotron energy $\var^{\sma{\perp}}_c{:}{=}\hbar^2/m\lper^2{:}{=}\hbar \omega_c^{\sma{\perp}}{:}{=}h/T_c^{\sma{\perp}}$ and Zeeman coupling to the spin magnetic moment:  $\zper{:}{=}{g_{s\perp}}\mu_B B_{\perp}/2$. We shall assume  $1/\lper^2S{\ll}1$, where $S$ is the area of the zeroth-order orbit (with effective mass $m$ and radius $k_E{=}\sqrt{2mE}/\hbar$). For energies much greater than $m\alpha^2$ and $\zpar$, the nearly-degeneracy condition ($\delta S/S{\ll}1$) is met, where $\delta S$ is the differential area induced by both spin-orbit and in-plane Zeeman couplings [see \fig{fig:RZ}(b)].  The smallness of $1/\lper^2S$ and $\delta S/S$ (with finite ratio $\lper^2\delta S$) justifies our application of the quantization rule [\qq{eq:rule}{eq:H1}],  with the effective Hamiltonian: 
\begin{equation}
\calh(t)=\hbar\alpha(k_{x}\sigma_{y}-k_{y}\sigma_{x})+Z_\parallel\sigma_{y}-Z_\perp\sigma_{z}.\label{calHRZ}
\end{equation}
Here, $t$ is a time variable which parametrizes the zeroth-order orbit:
$\bk(t){=}({-}k_E \cos \omega_c^{\sma{\perp}} t, k_E \sin \omega_c^{\sma{\perp}} t )$. In the momentum-independent spin basis of \q{calHRZ}, the non-Abelian Berry connection $\mathfrak{X}$ [in \q{eq:H1}]  vanishes, hence the generalized Zeeman interaction $B_{\sma{\perp}}\calm$ simply reduces to the Zeeman coupling to the spin  moment: $Z_\perp\sigma_{z}$.

As alluded to in \s{sec:qtznrules}, the adiabatic limit of two independent orbits does not exist if the two spin-orbit-split orbits exactly intersect at a II-Dirac point. To investigate this failure of adiabaticity, let us study the Landau levels at energies close to the II-Dirac point ($\epsilon_0$). We focus first on the regime where, on average (over $\frako_0$), the energy splitting $|\delta \var_+{-}\delta \var_-|$ [cf.\ \q{eq:RZ-Hamiltonian}] is much greater than $B_\perp|\calm_{\sma{+-}}|$. Equivalently, we are in the regime where $\hbar\alpha k_{\sma{E}}{\gg}\var_c^\perp$.\footnote{The orbit-averaged energy splitting is of order $\alpha k_{\sma{E}}$ for energies near the Dirac point. $B_\perp|\calm_{\sma{+-}}|$ is simply the Berry contribution to the Zeeman interaction, and has the form of the right-most term in \q{rashbaeffham}; this term is of order $\var_c$.} This implies that the split orbits do not appreciably hybridize -- except in the vicinity of the II-Dirac point where orbits (nearly) touch.  Near this touching point (where $t{=}0$), the dynamics of a spinor electron is described by linearizing $\H$ [cf.\ \q{calHRZ}] with respect to $t$:
\e{
\calh = -\hbar\alpha k_{E}\omega^{\sma{\perp}}_c t\,\sigma_x+(Z_\parallel-\alpha k_{E})\sigma_y-Z_\perp\sigma_z,\label{effhamIIDirac}
}
The eigenbasis of $\sigma_x$ corresponds to two diabatic levels with characteristic slope $v_d{:}{=}\hbar\alpha k_E\omega^{\sma{\perp}}_c$; near $t{=}0$, the diabatic levels anticross  due to the hybridization terms proportional to $\sigma_{y,z}$. Generally, the probability of tunneling across the hybridization-induced barrier is  $\rho^2{=}\exp(-2\pi\barmu)$, with $\barmu{=}{E_g}^2/{2v_d \hbar}$ and $E_g$ the energy gap at the center of the anticrossing\cite{wittig_landauzener_2005,lifshitz_e.m._quantum_1991}; in this context,
\e{\barmu=\frac{Z_\perp^2+(Z_\parallel-\alpha k_{E})^2}{2\hbar\alpha k_{E}\var^{\sma{\perp}}_c}, \label{mu1}}
with $2\hbar\alpha k_E$ the spin-orbit splitting at energy $E$ [cf.\ \fig{fig:orbits}(a)]. In the absence of the out-of-plane Zeeman coupling ($Z_\perp{=}0$), $\bar{\mu}$ would vanish at the energy of the II-Dirac point -- leading to unit-probability Klein tunneling, as was first proposed by Ref. \onlinecite{obrien_magnetic_2016} in a different model of the II-Dirac point. However, with a proper account of $Z_\perp$, we find instead that $\bar{\mu}$ has a nonvanishing minimum: $\bar{\mu}_{\sma{\text{min}}}{:}{=} (g_{s\perp} m/m_0)(Z_\perp/4\hbar\alpha k_{E})$, which is a product of the dimensionless effective mass and the ratio of the out-of-plane Zeeman splitting over the spin-orbit splitting. This minimum may be significant for semiconductor heterostructures (e.g.,  $g_{s\perp}m/m_0{\sim} 1$ in InAs heterostructures\cite{pakmehr_g-factor_2015}) and  for heavy-fermion systems.

To determine the Landau levels, one needs not just the probability for tunneling but also the phase of the probability amplitude. The full information is encoded in a scattering matrix\cite{AALG,kaganov_coherent_1983}
\e{\mathbb{S}=\matrixtwo{\tau e^{i\barphi}}{-\rho}{\rho}{\tau e^{-i\barphi}},\;\rho=e^{-\pi\barmu},\;\tau=\sqrt{1-\rho^{2}}\label{scattmat}}
which is a connection formula that matches the two-component semiclassical (WKB) wavefunction across the Dirac point. This formula  has been expressed in the basis of Zeeman-modified energy bands [i.e., eigenvectors of Eq.\ (\ref{calHRZ})]; diagonal (resp.\ off-diagonal) elements of $\mathbb{S}$ are amplitudes for intraband transmission (resp.\ interband tunneling). The phase of the intraband amplitude is $\barphi{:}{=}\barmu{-}\barmu\ln\barmu{+}\text{arg}[\Gamma(i\barmu)]{+}\pi/4$ with $\Gamma$ the gamma function.

\begin{figure}
\includegraphics[width=1.0\columnwidth]{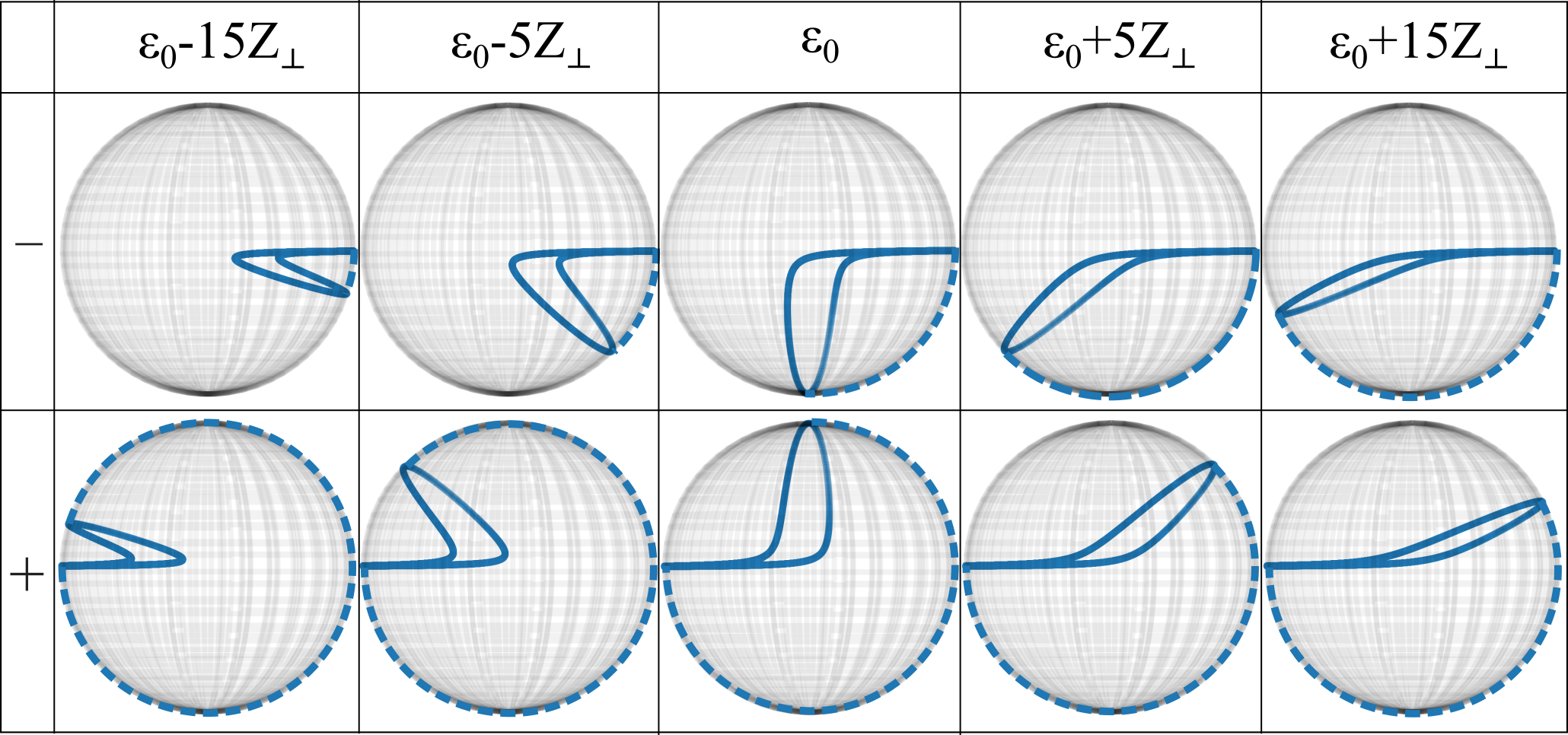}
\caption{Solid-angle representation of  the single-band Berry phase $\phi_{\pm}^B$ for inner (subscript $-$, upper panel) and outer ($+$, lower panel) orbit, at various energies close to the Dirac-point energy $\epsilon_0$ [cf.\ Eq. (\ref{whereisdiracpoint})]. The decomposition of the effective Hamiltonian [cf.\ Eq. (\ref{calHRZ})] into  Pauli matrices ($\calh{=}\sum_j\calh_j\sigma_j$) defines a three-vector $\boldsymbol{\calh}{=}({-}\alpha k_y$,$\alpha k_x{+}Z_\parallel,{-}Z_\perp$). $\phi_{\pm}^B$ equals half the solid angle subtended by   ${\mp} \boldsymbol{\calh}(\bk)$ [solid blue line] as $\bk$ is varied along the outer/inner orbit\cite{berry_quantal_1984}; the orientation of the solid angle is indicated by blue dotted lines. The out-of-plane Zeeman interaction lifts $\boldsymbol{\calh}$ out of the $x-y$ plane, but this effect is significant only over a fraction $(\sim Z_{\perp}/\hbar \alpha k_E)$  of the cyclotron period. % ${\sim}[(\bar{\mu}_{\sma{\text{min}}}/g_{s\perp})(m_0/m){\ll}1]$
\label{fig:blochsphere}}
\end{figure}

Away from the II-Dirac point, an electron undergoes adiabatic (band-conserving) dynamics. The time-ordered, coherent process of Landau-Zener tunneling/transmission and adiabatic dynamics is described by the propagator:
\e{{\A}= \matrixtwo{\tau e^{i\bar{\varphi}}}{-\rho}{\rho}{\tau e^{-i\bar{\varphi}}} \diagmatrix{e^{i\tilde{\lambda}_-}}{e^{i\tilde{\lambda}_+}}. \label{propinplanezeeman}}
The adiabatic phase $\tilde{\lambda}_{\pm}$ is the sum of a dynamical phase ${\pm} l^2 \delta S/2$ and  the single-band Berry phase $\phi^B_\pm$, for the outer ($+$) and inner ($-$) orbit respectively. At energies far above the Dirac point ($|E{-}\epsilon_0|{\gg}|Z_{\perp}|$) where the Zeeman interaction is negligible, both orbits encircle the Dirac point and hence $\phi_{\pm}^B{\approx}{\pm}\pi$; far below the Dirac point, neither orbit encircles the Dirac point, leading to $\phi_{\pm}^B{\approx}0$. At intermediate energies, $\phi_{+}^B$ (resp.\ $\phi_{-}^B$) is a continuous and decreasing (resp.\ increasing) function of energy, as supported by the Bloch-sphere argument in Fig. \ref{fig:blochsphere}.

Our quantization rule [\qq{eq:rule}{eq:H1}], with $\A$ having the form of \q{propinplanezeeman}, may be equivalently expressed as
\e{
\text{cos}\left[\frac{\Omega_{-}+\Omega_{+}}{2}\right]=\tau\,\text{cos}\left[\frac{\Omega_{-}-\Omega_{+}}{2}+\bar{\varphi}\right], \lin
\ins{with}\;\Omega_{\pm}=l^2 (S\pm \delta S/2)+\phi_{\pm}^B +\gamma_{\pm}, \label{coscos}
}
$\Omega_{\pm}$ are phases acquired by an electron in adiabatically traversing the orbits labelled $\pm$, respectively; the quantum Maslov correction $\gamma_{\pm}{=}\pi$ for circular orbits.  A quantization rule analogous to \q{coscos} was previously derived by us for a II-Dirac point that intermediates an electron- and hole-like orbit\cite{AALG}. 

In the limit $\bar{\mu}{\rightarrow} \infty$, Landau-Zener tunneling is negligible ($\tau{\rightarrow}1,\bar{\varphi}{\rightarrow}0$), and \q{coscos} simplifies to the Onsager-Lifshitz-Roth rules ($\Omega_{\pm}/2\pi{\in}\Z$) for  two independent orbits.

For any nonzero field, $0{<}\tau{<}1$ implies that the two incommensurate harmonics $(\Omega_+{\pm}\Omega_-)$ in \q{coscos} compete to produce a quasirandom Landau-level spectrum\cite{kaganov_coherent_1983}. That is to say, the spectrum is disordered on the scale of nearest-neighbor level spacings but has longer-ranged correlations. To appreciate this, let us analyze the Landau levels (near the Dirac point) perturbatively in the small parameter $\tau$. To zeroth order, \q{coscos} reduces to
\e{2l^2S(E_n^0)+\pi = 2\pi n \rightarrow \{E_n^0(B)\}_{n\in\Z}.\label{trefoilrule}}
To derive the above equation, we have utilized $\phi^B_-{+}\phi^B_+{=}0$, which is valid for any two-band model in the nearly-degenerate regime\footnote{For two band models, Berry phases of the two bands sum up to 0 on arbitrary path}; in \fig{fig:blochsphere} this is reflected in the sum of solid angles (at fixed energy) equalling $4\pi$. \q{trefoilrule} may be interpreted as a quantization rule for a trajectory on a M\"obius strip  with net area $2S{=}4\pi m E{/\hbar^2}$; the corresponding Landau levels (sketched by solid lines in the upper panel of Fig. \ref{fig:RZ}(c)) are periodic in $E{\rightarrow}E{+}\pi/l^2(\partial S/\partial E)$ and in $l^2{\rightarrow}l^2{+}\pi/S$. Both periods are approximately half that of either disconnected  orbit, in the assumed nearly-degenerate regime $\delta S{\ll}S$. The $\pi$ correction in \q{trefoilrule} is the Berry phase associated to a $2\pi$ rotation of the wavefunction spin over a single M\"obius orbit, as illustrated in Fig. \ref{fig:RZ}(b). The Maslov phase, which is generally equal to $\pi$ times the rotation number of an orbit, is trivially $2\pi$ for the M\"obius orbit.

To recapitulate, $E_n^0(B)$ [cf.\ \q{trefoilrule}] exhibits a periodicity on short scales associated to the large area of the M\"obius orbit. This short-scale periodicity is lost due to  imperfect Klein tunneling, yet a longer-range correlation (associated to the smaller differential area $\delta S$ between spin-split orbits) persists, as shown by the first-order-in-$\tau$ correction:\footnote{The general form of this perturbative calculation can be found in Sec. IX-E of Ref. \onlinecite{100p}.} 
\e{&\delta E_n(B) = \f{(-1)^n}{2\pi} \tau \var_c^\perp \cos\bigg(\f{l^2\delta S}{{2}}+\f{\phi_+^B-\phi_-^B}{2}-\bar{\varphi}\bigg) \lin
    &\approx \f{(-1)^{n+1}m\alpha^2(E-\epsilon_0)(\var_c^{\perp})^{1/2}}{2\sqrt{\pi}\hbar^2
    Z_{\parallel}^{3/2}}\cos\bigg(\f{l^2\delta S}{{2}}-\bar{\varphi}\bigg),\label{RZfirstorderE}}
with the right-hand side (of each line) evaluated at $E{=}E_n^0(B)$. The second line is an approximation for $Z_{\perp}{\rightarrow}0$, and has greater utility where $g_{s\perp}m{\ll}m_0$. For two adjacent Landau levels, we have plotted $E_n^{(0)}+\delta E_n$ as dashed lines in Fig. \ref{fig:RZ}(c). The validity of Eq.\ (\ref{RZfirstorderE}) is  further supported by a comparison with numerically-exact Landau levels, as illustrated in Fig.\ \ref{fig:RZ}(d).

\section{Tuning the (pseudo)spin-half degeneracy of Landau Levels}\label{sec:llquasideg}

\begin{figure}
\includegraphics[width=1.0\columnwidth]{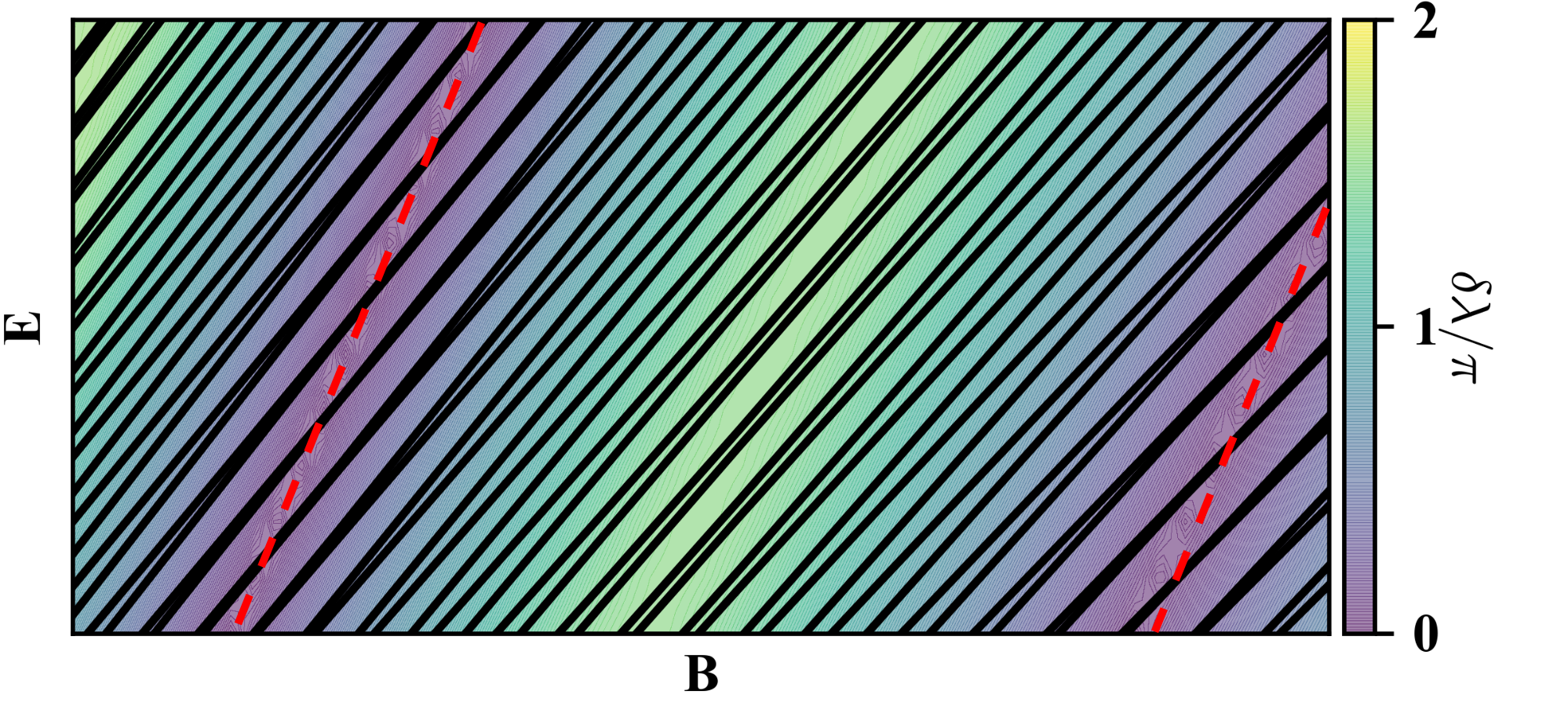}
\caption{A schematic diagram of Landau level dispersion (solid lines) with respect to $B$ for nearly-degenerate orbits. $\delta\lambda$ is plotted in the colored background. Landau level quasidegeneracies occur at $\delta\lambda\equiv 0\equiv 2\pi$ ({indicated by red dashed lines}).\label{fig:LL}}
\end{figure}

{The Landau levels derived from \q{eq:rule} are generically non-equidistant, and the spin-half (or pseudospin-half) degeneracy between Landau levels is generically lifted due to the subleading corrections ($\lambda_{1,2}$). As argued in \s{sec:qtznrules},} the level splitting is approximately $\var_c\delta \lambda/2\pi$ with $\delta \lambda=|\lambda_1-\lambda_2|$. Conversely, Landau levels are approximately spin-degenerate if $\lambda_1{=}\lambda_2$ (mod $2\pi$) -- this degeneracy condition for the eigenvalues of $\A$ typically requires fine-tuning of some Hamiltonian parameters.  

In \s{sec:relatedegeneracies}, we will relate the eigenvalue-degeneracies of  $\A$ to the (near) degeneracies of Landau levels. We then ask how many tunable real parameters are needed to attain an eigenvalue-degeneracy of $\A$. Our answer is  $3$ in the absence of crystallographic point-group symmetry, as discussed in \s{sec:introducecodimension}; the answer may reduce to $0,1$ or $2$ depending on the symmetry class of the orbit, as will be explained in \s{sec:tenfold}. Before this general symmetry classification,  we will first develop physical intuition by studying two rotationally-symmetric models [cf.\ \s{sec:singleparameterrashba} and \ref{sec:rotsymmbreakdown}] where only one tunable parameter is needed -- this can be the magnitude of the $B$ field itself.

\subsection{Relating the eigenvalue-degeneracy of $\A$ to the quasidegeneracy of Landau levels}\label{sec:relatedegeneracies}

\fig{fig:LL} illustrates a typical Landau-level dispersion in the near-degeneracy regime ($\delta S{\ll} S)$ of the Rashba 2DEG (in an out-of-plane field). Each solution of the quantization rule [cf.\ \q{eq:rule}], with fixed $a\in \{1,2\}$ and $n \in \Z$, defines a curve in $(E,B)$-space that we indicate by  a solid black line. The condition of eigenvalue-degeneracy for $\A$: \e{\lambda_1(\bar{E},\bar{B})=\lambda_2(\bar{E},\bar{B})+2m\pi, \as m\in \Z \la{eigenvaluedegeneracyA}} 
defines a distinct set of curves [red dashed lines in \fig{fig:LL}] -- for the symmetry class of the Rashba 2DEG. (For other symmetry classes, the eigenvalue-degeneracy for $\A$ may instead define a set of points, or may generically have no solution in $(E,B)$ space.)   The two sets of curves (solid and dashed) intersect at isolated points where Landau levels are spin degenerate -- to the accuracy of our quantization rule. In principle, such an intersection may correspond to an exact  degeneracy, if the associated Landau levels belong to different unitary representations of a spatial symmetry -- this subject will be partially addressed in subsequent subsections, but a systematic group-theoretic analysis is involved and will be left to future investigations.

Moving away from these intersection points, the repulsion between paired Landau levels  remains anomalously weak (in a manner that will be quantified in \qq{llquasideg}{llquasidegB} below) in the immediate vicinity of the red dashed lines, as illustrated by the purple-shaded regions in \fig{fig:LL}.  From a pragmatic standpoint, a typical experiment may not be able to distinguish an exact degeneracy from an anomalously weak degeneracy (in short, a \textit{quasidegeneracy}), due to the smearing effects of temperature and/or disorder\cite{shoenberg_magnetic_2009}. The existence of  quasidegeneracy lines (i.e., lines of solutions to \q{eigenvaluedegeneracyA}) imposes a distinct signature on the low-temperature Shubnikov-de Haas effect, as will be elaborated in \s{sec:quantosc_quasideg}.  We are thus motivated to (i) precisely quantify a quasidegeneracy, and (ii) identify the symmetry classes for which quasidegeneracies occur along lines rather than points, or rather than being generically nonexistent. The first task is carried out in the rest of this subsection, and the second will be the subject of subsequent subsections in \s{sec:llquasideg}. 

Suppose the propagator $\A$ were degenerate at $(\bar{E},\bar{B})$ [cf.\ \q{eigenvaluedegeneracyA}]. Generically, $(\bar{E},\bar{B})$ is not \emph{simultaneously} a solution of the quantization rule in \q{eq:rule}.  
That is to say, $\bar{E}$ is not generically the energy of a Landau level at field $\bar{B}$. To obtain a Landau level, we may tune either $E$ or $B$   away from $(\bar{E},\bar{B})$, and  utilize that  $\lambda_{\pm}$ are slowly-varying functions of $(E,B)$ compared to $l^2S(E)$:
\e{l^2\left|\p{S}{E}\right| \gg \left|\p{\lambda_a}{E}\right|,\, \left|S\right| \gg \left|\p{\lambda_a}{l^2}\right|;} 
these inequalities are guaranteed by the nearly-degeneracy condition ($\delta S{\ll}S$) and the smoothness of $S(E)$, as shown in \app{sec:proofLLquasideg}.  For fixed $\bar{B}$, we are therefore guaranteed to find paired Landau levels  in close proximity to $\bar{E}$, with each pair having an energetic splitting of order:
\e{\bigg|\f{E_{1}-E_2}{\var_c}\bigg|_{B=\bar{B}} \sim \order\left( \f{\partial \lambda_a/\partial E}{\bar{l}^2\partial  S/\partial E}  \right)\ll 1,\label{llquasideg}}
with $\bar{l}$ the magnetic length associated to $\bar{B}$; this is illustrated schematically in Fig.\ \ref{fig:LL}. At fixed $B$, the number of Landau levels sandwiched between two adjacent curves of \q{eigenvaluedegeneracyA} is of order $\bar{l}^2(\partial S/\partial E)/(\partial \lambda_a/\partial E).$

Alternatively, we may fix $\bar{E}$ and investigate how Landau levels disperse as a function of $B$; the analog of \q{llquasideg}  is the existence of paired Landau levels in close proximity to $\bar{l}^2$, with a splitting of order:
\e{\bigg|\f{l^2_{+}-l^2_-}{T_{{l^2}}}\bigg|_{E=\bar{E}} \sim O\left( \f{1}{S}\p{\lambda_a}{l^2} \right)\ll 1,\;\; T_{{l^2}}:=\f{2\pi}{S(\bar{E})}.\label{llquasidegB} }
$\partial \lambda_a/\partial l^2$ is estimated in \app{sec:proofLLquasideg}, and
$T_{l^2}$ above is the period (in $l^2$) of quantum oscillations -- in the absence of spin-orbit coupling. At fixed $E$, the number of Landau levels sandwiched between two adjacent curves of \q{eigenvaluedegeneracyA} is of order $S/(\partial \lambda_a/\partial l^2).$

To recapitulate, for every exact degeneracy of $\cala$, there exist, in close proximity, paired Landau levels which are  nearly degenerate in the sense of \qq{llquasideg}{llquasidegB}. Each such pair of Landau levels will be described as \textit{quasidegenerate}. 

%\red{It is possible that a subset of these quasidegeneracies are in fact exact degeneracies, owing to Landau levels carrying different quantum numbers -- a subject we elaborate in \s{sec:discussion}.}

\subsection{Codimension analysis without symmetry, and with band-conservation symmetry}\label{sec:introducecodimension}

In the absence of symmetry, eigenvalue-degeneracies of the unitary propagator $\A$ [cf. \q{eq:prop}] are attainable by tuning three real parameters. Analogously, it is well known (as the non-crossing rule in quantum theory\cite{neumann2000behaviour}) that eigenvalue-degeneracies of complex, finite-dimensional matrix Hamiltonians are also attainable by tuning three real parameters.
 To prove the non-crossing rule for two-by-two unitaries, we begin by decomposing any such matrix as $\A{=}e^{i\phi}{\cal \bar{A}}$, with  ${\cal \bar{A}}{\in}\text{SU}(2)$. Then the necessary and sufficient condition for an eigenvalue-degeneracy of $\A$ is that ${\cal \bar{A}}{=}{\pm}I$. In the canonical parametrization of $\text{SU}(2)$ over the three-sphere $S^3$: 
\e{{\cal \bar{A}}{=}\matrixtwo{r_1{+}ir_2}{r_3{+}ir_4}{{-}r_3{+}ir_4}{r_1{-}ir_2},\;\; \sum_{j=1}^4r_j^2{=}1,\;\; r_j\in \mathbb{R},  \label{s3}}
${\cal \bar{A}}{=}{\pm}I$ lie on distinct points of $S^3$, and are attainable by tuning three real parameters: $r_2{=}r_3{=}r_4{=}0$. The conclusion that three parameters need be tuned remains valid for any  non-singular parametrization of SU$(2)$.
 
Symmetry may reduce the number of tunable parameters needed to attain an eigenvalue degeneracy for $\A$. A case in point is band-conservation symmetry in the adiabatic regime (cf.\ \s{sec:qtznrules}), which restricts $\cala$ to have diagonal form  ($r_3{=}r_4{=}0$) in the basis of energy bands [cf.\ \q{phaseindependentorbit}]. The phase of $r_1{\pm}ir_2$ equals ${\pm} l^2\delta S/2$ plus field-independent corrections. This  phase -- a single parameter --  can be  tuned to attain $\cala{=}{\pm}I$. As a function of $l^2$, we deduce that Landau-level quasidegeneracies occur with a periodicity of $2\pi/\delta S$. This periodicity  has a dual interpretation -- as a periodic increment of a single flux quantum for the differential magnetic flux (between the two split orbits).

To generalize the discussion beyond band-conservation symmetry, it is instructive to adopt a geometric perspective on degeneracies. A generic SU(2) matrix that is constrained by symmetry may no longer be uniformly distributed over $S^3$ [cf.\ \q{s3}], e.g., ${\cal \bar{A}}$ with band-conservation symmetry is uniformly distributed over a phase in $S^1$. In such cases, we are  motivated to adopt \textit{symmetric coordinates} [$\{r_1,\ldots,r_{d}\}{:}{=}\br$] that independently parametrize the symmetry-constrained SU(2) matrix; the precise nature of these constraints will be developed in subsequent subsections.
In our example, we may choose the symmetric coordinate as the phase of the diagonal element of ${\cal \bar{A}}$. Such coordinates are generally not relatable -- by a non-singular coordinate transformation -- to the canonical coordinates [\q{s3}] of asymmetric special unitaries. ${\cal \bar{A}}(\br){=}{\pm}I$ then defines a set of submanifolds in $\br$-space which we refer to as \textit{degeneracy manifolds}. The \textit{codimension} of the degeneracy manifold  is defined as $d$ minus the dimension of said manifold; alternatively stated, the codimension is the number of independently-tunable parameters (in $\br$) needed to attain a degeneracy. Here and henceforth, we adopt the shorthand that  \textit{parameters} (or coordinates) will always refer to real parameters (coordinates), and \textit{codimension} always refers to a codimension of a degeneracy manifold. In our example, the degeneracy manifolds  are two points on $S^1$, and each has unit codimension. 
 
One of our main results is that  rotational symmetry (with rotational axis parallel to the field) results in degeneracy manifolds with unit codimension. This will be demonstrated by two  case studies: the Rashba 2DEG in \s{sec:singleparameterrashba}, and a model with II-Dirac points in \ref{sec:rotsymmbreakdown}. 

\subsection{Codimension reduction for the Rashba-Dresselhaus 2DEG}\label{sec:singleparameterrashba}

\subsubsection{Rashba 2DEG with continuous rotational symmetry}\label{sec:ctsrot}

To understand why rotational symmetry reduces codimension in the Landau-level problem, we return to the Rashba model [cf. \q{eq:Rashba-Hamiltonian}]. This model has a continuous rotational symmetry (denoted $\mathfrak{c}_{\infty}$), which manifests as the constancy of $\calh(\bk)$ over the circular zeroth-order orbit [cf.\ \q{rashbaeffham}]. Consequently, the propagator [cf.\ \q{eq:prop}] simplifies to the exponential form:
\e{\A{=}-\exp(-2\pi i\sum_{j}r_j\tau_j),\label{expform}}
where
\e{\br{=}\bigg(\f{g_{s\perp}m}{4m_0}{-}\f1{2},0,-\f{\hbar\alpha k_{\sma{E}}}{\var_c}\bigg) \la{rvectorrashba}} 
and $\{\tau_j\}$ are a set of Pauli matrices. Any $\text{SU}(2)$ matrix can be expressed in the exponential form [cf.\ \q{expform}] for some (possibly non-unique) $\br$, with $r_2$ not necessarily zero. We will demonstrate codimension reduction for the general form of $\A$ (having any $\br$), with the Rashba model as a particular realization.

All eigenvalue degeneracies of $\A$ occur on spheres in $\br$-space with radii $|\br|{=}\sqrt{r_1^2{+}r_2^2{+}r_3^2}$ equal to an integer ($j$)  multiple of $1/2$;  ${j}$ is odd (resp. even) for  $\A{=}I$ (resp.\ $\A{=}{-}I$).  This condition $|\br|{=}j/2$ (for any nonzero $j$) may be attained by tuning a single parameter. In other words, all nonzero-radius spheres are unit-codimensional surfaces (known as hypersurfaces), whose stability rely on $\mathfrak{c}_{\infty}$ symmetry. These hypersurfaces separate $\br$-space into infinitely-many, distinct connected components, i.e., the zeroth homotopy group $\pi_0$ comprises an infinite set. Each hypersurface may thus be viewed as a symmetry-protected, topological defect, which separates domains (in $\br$-space) where Landau levels are non-quasidegenerate. By varying $B$ alone (at fixed energy), one then transits between different domains; each penetration of a hypersurface is accompanied by a Landau-level quasidegeneracy. For the exactly-soluble Rashba model ($r_2{=}0$), these $\mathfrak{c}_{\infty}$-protected quasidegeneracies  are consistent with (and partially demystify)  the abundance of \textit{exact} level crossings in the spectrum of $H_R(\bK)$ [cf.\ \q{eq:Rashba-exact}]; such crossings are not expected from the Wigner-von Neuman non-crossing rule\cite{neumann2000behaviour}. The zero-radius sphere (that is, a point) has codimension three, as would be expected if symmetry were absent (cf. \s{sec:introducecodimension}). This point degeneracy can be understood from an elementary result in quantum mechanics: taking $\alpha{=}0$, $g_{s\perp}{=}2$ {and} $m{=}m_0$ in \q{rvectorrashba}, we recover the Zeeman-split Landau levels of a free electron gas without spin-orbit coupling; all levels (except the lowest) are spin-degenerate owing to the equality of the Zeeman and cyclotron energies\cite{landau2013course}.

The above discussion is reminiscent of topological semimetals whose corresponding Bloch Hamiltonians are energy-degenerate at Dirac-Weyl points\cite{wang2012dirac,wan2011topological} and/or line nodes\cite{burkov2011topological}. Both Dirac points (in 2D $\bk$-space, e.g., graphene\cite{neto2009electronic}) and line nodes (in 3D $\bk$-space) are attainable by tuning two wavenumbers; the stability of such codimension-two manifolds relies on crystallographic point-group symmetries. On the other hand, Weyl points in 3D $\bk$-space have codimension three and do not require any point-group symmetry for their stability. Just like Weyl points, the point degeneracy of $\A$ is associated with a nontrivial Chern number\cite{TKNN} (defined as an integral of the Berry curvature\cite{berry_quantal_1984} over a Gaussian surface surrounding the point)\footnote{The curvature is defined with respect to a line bundle formed from the nondegenerate eigenvector of $\cala$.} .

\begin{figure}
\includegraphics[width=1.0\columnwidth]{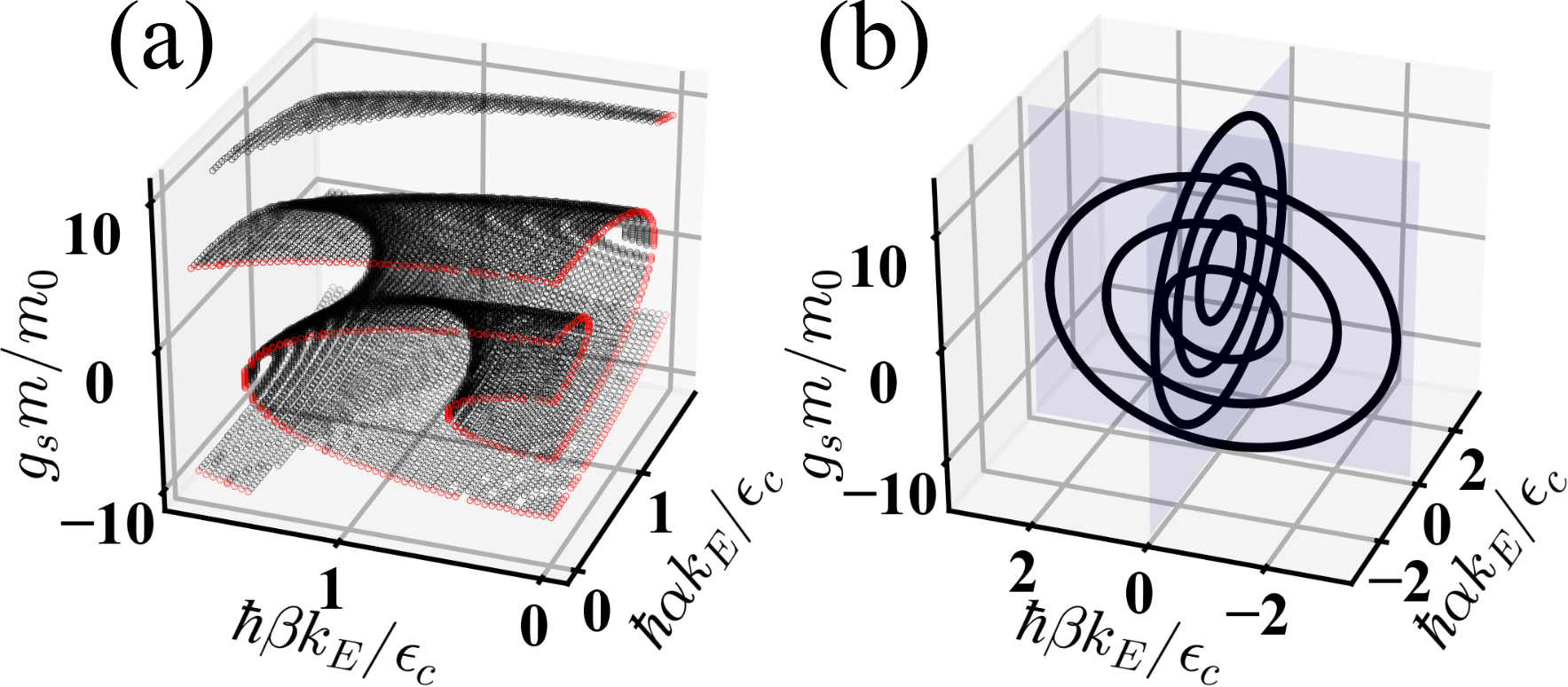}
\caption{Degeneracy manifolds of the Rashba-Dresselhaus model. (a) The degeneracy manifold of $\A{=}I$ is a path-connected hypersurface. Points in $\alpha=0$ or $\beta=0$ planes are colored in red. (b) illustrates the degeneracy manifold of $\A=-I$, which lie confined to  the $\alpha{=}0$ and $\beta{=}0$ planes (blue translucent planes). Each $\mathfrak{c}_{\infty}$-symmetric plane contains an infinite set of concentric, circular line degeneracies. Every circle (except for the smallest in either plane)  intersects two other circles in an orthogonal plane, such that the entire line node is path-connected. \label{fig:dgn}}
\end{figure}

\subsubsection{Rashba-Dresselhaus 2DEG with discrete rotational symmetry}\label{sec:disrot}

The Rashba model suffers from an oversimplification -- continuous rotational symmetry is not a symmetry in any crystallographic space group. It is possible for $\mathfrak{c}_{\infty}$-symmetric $\bk{\cdot}\bp$ models to approximate Bloch Hamiltonians (defined over the Brillouin torus) with a discrete, order-$N$ rotational symmetry (denoted $\mathfrak{c}_N$). How robust are the above-mentioned Landau-level quasidegeneracies  (over the $B$ axis) when $\mathfrak{c}_{\infty}$ symmetry is perturbatively reduced to $\mathfrak{c}_N$? For any  integer $N{\geq}2$, we find that $N{-}1$ of every $N$ quasidegeneracies are perturbatively stable. This general statement is proven in App. \ref{app:codimension}, and here we illustrate it for a specific model of $N{=}2$ -- we will reduce the $\mathfrak{c}_{\infty}$ symmetry of the Rashba model [cf.\ \q{eq:Rashba-Hamiltonian}] to  $\mathfrak{c}_2$  by adding the Dresselhaus spin-orbit interaction (with coupling parameter $\beta$): 
\begin{equation}
 H_{RD}(\bk)=\frac{{\hbar^2}k^2}{2m}+\hbar\alpha(k_{x}\sigma_{y}-k_{y}\sigma_{x})+\hbar\beta(k_{x}\sigma_{x}-k_{y}\sigma_{y}).\label{hamRD}
\end{equation}
{For the Rashba-Dresselhaus 2DEG, $\A$ depends on $\alpha,~\beta,~l,~E,~m$ only through three independent parameters:}%$\hbar\alpha k_E /\var_c$, $\hbar\beta k_E /\var_c$ and $g_sm/m_0$, let us consider the three-dimensional parameter space:
\e{\left( \f{\hbar\alpha k_{\sma{E}}}{\var_c},\f{\hbar\beta k_{\sma{E}}}{\var_c},g_{s\perp}\f{m}{m_0} \right){\in} \mathbb{R}^3. \la{parameter2}}  
$\A(\hbar\alpha k_{\sma{E}}/\var_c,\hbar\beta k_{\sma{E}}/\var_c,g_{s\perp}m/m_0){=}{\pm} I$ then determines a set of concentric circles in the $\beta{=}0$ plane, as per the $\mathfrak{c}_{\infty}$-symmetric case study in \s{sec:ctsrot}. Moving off this plane, we find that $\A{=}I$ is satisfied in a  neighborhood of $\beta{=}0$, i.e., $\A{=}I$ defines a hypersurface in $\mathbb{R}^3$ that is illustrated in \fig{fig:dgn}(a). On the other hand, $\A{=}{-}I$ is not satisfied in the neighborhood of $\beta{=}0$, hence the circles associated to $\A{=}{-}I$ remain isolated as line nodes, as illustrated in \fig{fig:dgn}(b). To recapitulate, we have found that one of every two Landau-level crossings (those labelled by odd $j$) destabilize due to the symmetry reduction.

To demonstrate how the $\A{=}I$ hypersurface derives from $\mathfrak{c}_2$ symmetry, consider  the effective Hamiltonian $\calh$ 
\e{\calh=\hbar\alpha (k_{x}\sigma_{y}{-}k_{y}\sigma_{x})+\hbar\beta (k_{x}\sigma_{x}{-}k_{y}\sigma_{y})-\f{g_{s\perp}}{2}\mu_{B}B\sigma_z,}
in a momentum-independent basis where the Pauli matrices correspond to spin operators. The two-fold rotational symmetry 
\e{\calh(\bk)=\sigma_z\calh(-\bk)\sigma_z, \;\: \bk(t)=-\bk\big(t+\f{T_c}{2}\big)\in \frako_0,}
implies that the propagator over the time interval $[T_c/2,0]$ is symmetry-related to that over $[T_c,T_c/2]$:
\e{\A=\A_{T_c\leftarrow T_c/2}\A_{T_c/2\leftarrow 0}=\sigma_z{\A}_{T_c/2\leftarrow 0}\sigma_z {\A}_{T_c/2\leftarrow 0}.\label{eq:sigmazconstraint}}
The tracelessness of $\calh$ (at each $\bk$) implies that  ${\A}_{\sma{T_c/2\leftarrow 0}}$  has unit determinant. Employing the canonical SU(2) parametrization for the half-loop propagator [Eq. (\ref{s3}) with ${\cal \bar{A}}$ replaced by ${\A}_{\sma{T_c/2\leftarrow 0}}$], we derive from \q{eq:sigmazconstraint} that
\e{
{\A}={\cal \bar{A}}(r_1,r_2,r_3)=\matrixtwo{1{-}2r_2(r_2{-}ir_1)}{2ir_2(r_3{+}ir_4)}{{-}2ir_2(r_3{-}ir_4)}{1{-}2r_2(r_2{+}ir_1)} \label{proprashba}
}
with $\sum_{j=1}^4r_j^2{=}1$. \q{proprashba} is a parametrization of the two-fold-symmetric propagator by the symmetric coordinates $\br{=}(r_1,r_2,r_3)$. It follows that $r_2{=}0$ is a necessary and sufficient condition for $\A{=}I$. The corresponding degeneracy manifold is a path-connected hypersurface, which separates $\br{=}(r_1,r_2,r_3)$-space into two distinct connected components. As proven in App. \ref{app:codimension}, this conclusion generalizes for any $N{\geq}2$: the degenerate manifolds of $\mathfrak{c}_N$-symmetric $\A$ separate the space of symmetric coordinates into $N$ distinct connected components. \fig{fig:dgn}(a) illustrates the $\A=I$ degenerate manifold {in the coordinates of  \q{parameter2} (which are distinct from the $\br$-coordinates in \q{proprashba}).} 

On the other hand, $\A{=}{-}I$ occurs if and only if $\br{=}(0,{\pm} 1,0)$. These two points are mapped to a path-connected line node confined to the $\alpha{=}0$ and $\beta{=}0$ planes [cf.\ \fig{fig:dgn}(b)]; the difference in codimension originates from the $\mathfrak{c}_{\infty}$ symmetry within these two planes (cf.\ \s{sec:ctsrot}), which was not accounted for in the $\mathfrak{c}_2$-symmetric analysis of \q{proprashba}.

%Let us relate \q{proprashba} to the Rashba-Dresselhaus model. For any choice of parameters for $\calh$, the time-ordered exponential of $\calh$ (over half a period) is uniquely defined. This defines the map $f:(\hbar\alpha k_{\sma{E}}/\var_c,\hbar\beta k_{\sma{E}}/\var_c,g_sm/m_0){\rightarrow}\br$. For a sufficiently regular $f$, the inverse image -- of a $d_s$-dimensional submanifold in $\br$-space -- must have  dimension that is greater or equal to $d_s$. In particular, the inverse image of the path-connected hypersurface ($r_2{=}0$) is a path-connected hypersurface in the space of $(\hbar\alpha k_{\sma{E}}/\var_c,\hbar\beta k_{\sma{E}}/\var_c,g_sm/m_0)$, as illustrated in \fig{fig:dgn}(a).

\subsection{Codimension reduction for 2DEG with distinct, symmetry-related II-Dirac points}\label{sec:rotsymmbreakdown}

\begin{figure}
\includegraphics[width=1.0\columnwidth]{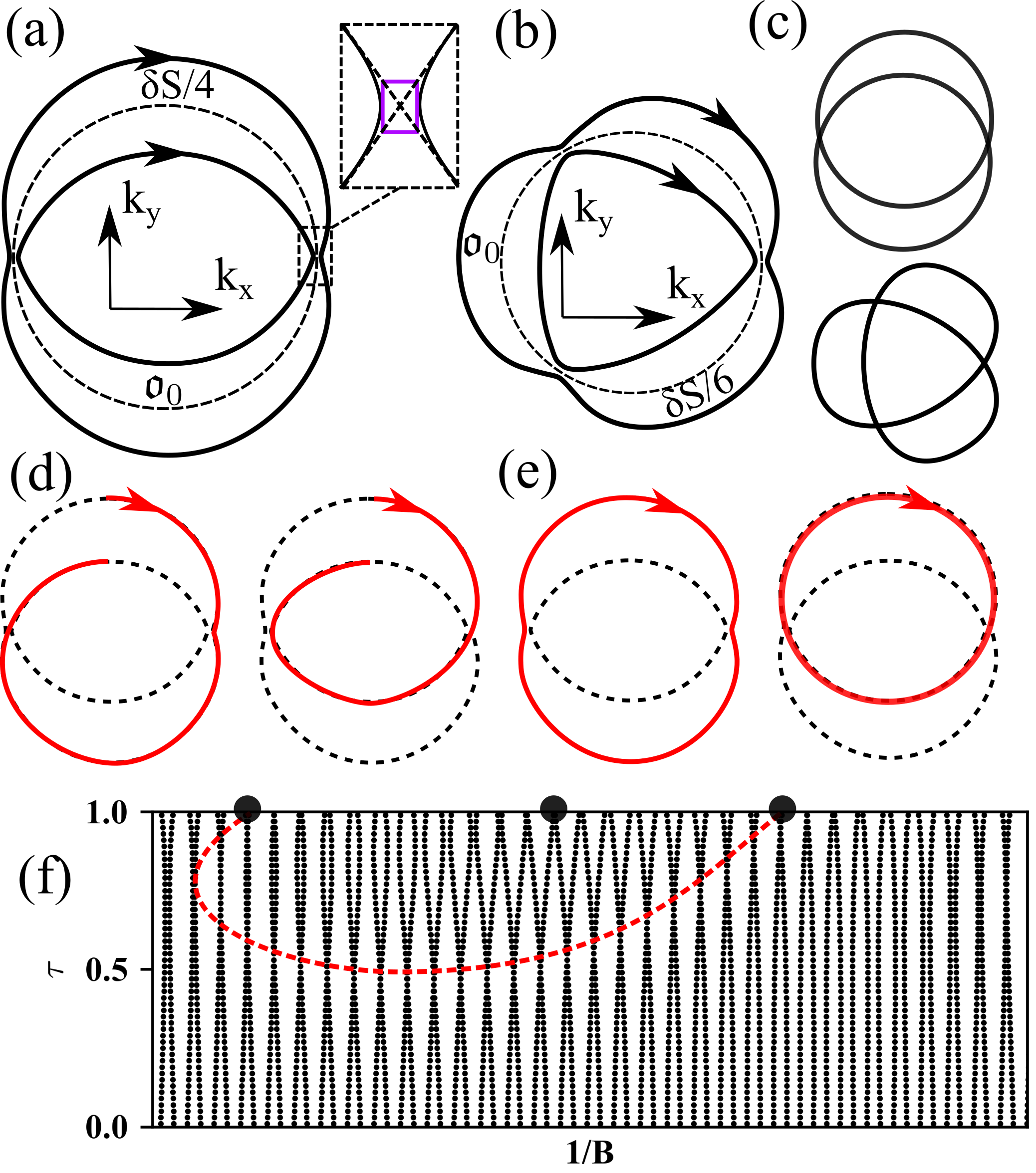}
\caption{(a-b) Cyclotron orbits for $\mathfrak{c}_2$-symmetric [(a)] and $\mathfrak{c}_3$-symmetric [(b)] breakdown. Top right inset of (a) presents an enlarged view of the breakdown junction, where the purple rectangle has area $S_{\sma{\square}}$. (c) Klein tunneling limit of $\mathfrak{c}_2$-symmetric (upper) and $\mathfrak{c}_3$-symmetric (lower) breakdown cyclotron orbits. (d) Feynman paths that switch between bands through  one tunneling event. (e) Feynman paths that preserve the band index through either zero or two tunneling events. (f) Schematic Landau-level dispersion for $\mathfrak{c}_3$ symmetric breakdown with respect to $\tau$ at a fixed energy. For the plotted interval of $1/B$, there are three quasidegenerate points (indicated by solid dots) in the  limit of zero Landau-Zener tunneling ($\tau=1$). Only two of three quasidegeneracies are stable upon inclusion of tunneling. That is to say, two of three  two points extend into quasidegenerate lines (indicated by red dashed lines) as $\tau$ is decreased from unity, while the third point is isolated, i.e., the quasidegeneracy is immediately lost as $\tau$ is decreased. At $\tau=0.5$, the two quasidegenerate lines eventually merge and annihilate.\label{fig:Cn-breakdown}}
\end{figure}

%one of them is destabilized by infinitesimal hybridization ($\tau<1$, indicated by green dashed lines) and two of them remains stable until they annihilate at $\tau=0.5$ (indicated by red dashed line)

In the adiabatic limit of two nearly-degenerate orbits, we have found (in \s{sec:introducecodimension}) that the codimension for $\cala{=}{\pm} I$ is unity. Here we investigate how this result changes as adiabaticity is relaxed by Landau-Zener tunneling -- that respects a discrete $N$-fold rotational symmetry ($\mathfrak{c}_N$). Since a II-Dirac cone is tilted in a special direction, a II-Dirac point cannot be the center of a rotational symmetry; the minimal model with $\mathfrak{c}_N$ symmetry thus requires $N$ distinct but symmetry-related II-Dirac points. For $N{=}2$, we will demonstrate that the codimension for $\cala{=}I$ (but not for $\cala{=}{-}I$) remains unity, in accordance with the symmetric-coordinate analysis of \q{proprashba}; the persistence of  Landau-level quasidegeneracies  (at finite tunneling probability) will be demystified in terms of the destructive interference of tunneling Feynman paths. For $N{=}3$, we will find that two of three Landau-level quasidegeneracies are perturbatively stable against tunneling, but even these will destabilize beyond a critical tunneling strength.

We begin with a minimal, $\mathfrak{c}_2$-symmetric model:
\e{H=\f{{\hbar^2}k^2}{2m_1}+ \bigg(\f{{\hbar^2}k^2}{2m_2}-\mu\bigg)\sigma_z +wk_y\sigma_x,\label{modelC2breakdown}}
with $0{<}m_1{\ll}m_2$ and positive $\mu$.  The $\mathfrak{c}_2$ symmetry manifests as $\sigma_z H(\bk)\sigma_z{=}H({-}\bk)$, and relates 
two Dirac points at
\e{\bar{k}_x = \pm\sqrt{2m_2\mu}/{\hbar},\;\bar{k}_y=0,\;\epsilon_{0}=\frac{m_2}{m_1}\mu.}
In the vicinity of either point, the  Hamiltonian has the linearized form 
\e{H_{\pm} =\pm {\hbar}\sqrt{\f{2\mu}{m_2}}\bigg(\frac{m_2}{m_1}+\sz\bigg)\delta k_x+w\delta k_{y}\sx+\epsilon_0, \label{linearHam2}}
from which we deduce that both Dirac points are  type-II if $m_1{<}m_2$; we further assume $m_1{\ll}m_2$ so that the nearly-degeneracy condition ($\delta S/S{\ll}1$) is satisfied. 

At energies close to the Dirac point ($E{-}\epsilon_0 {\ll} wm_2(2\epsilon_0/m_1)^{\sma{1/2}}$), there are two concentric orbits which (nearly) touch at both Dirac points, as illustrated in Fig.\ \ref{fig:Cn-breakdown}(a). For sufficiently weak field, interband tunneling is localized near either II-Dirac point, and occurs with the Landau-Zener probability:
\e{\rho^2=e^{-2\pi\bar{\mu}}, \; \bar{\mu}=\f{\big(Z_{\perp}^2+(m_1/m_2)^2(E-\epsilon_0)^2\big)l^2}{2w\sqrt{2E/m_1}}. \label{landauzenerc2}}
Equivalently, $\bar{\mu}{=}S_{\sma{\square}}l^2/8$, with $S_{\sma{\square}}$ the area of rectangle inscribed between the two hyperbolic arms [cf.\  Fig. \ref{fig:Cn-breakdown}(a)]. Elsewhere, an electron undergoes adiabatic (band-conserving) dynamics. Due to $\mathfrak{c}_2$ symmetry, the time-evolution propagator over the full cyclotron period is simply the square of the propagator for half the period: 
\e{\A=\left(\matrixtwo{\tau e^{i\bar{\varphi}}}{-\rho}{\rho}{\tau e^{-i\bar{\varphi}}}{\diagmatrix{e^{i\tilde{\lambda}_-}}{e^{i\tilde{\lambda}_+}}} \right)^2. \label{propC2breakdown}}
The half-period propagator may be taken from \qq{scattmat}{propinplanezeeman}, with  $\bar{\mu}$ here given by \q{landauzenerc2}, and  $\tilde{\lambda}_{\pm}{=}{\pm}l^2\delta S/4$ plus a field-independent geometric phase.

The necessary and sufficient condition for a degeneracy of $\cala$ is 
\begin{align}
\Delta:= \tilde{\lambda}_-{-}\tilde{\lambda}_{+}{+}2\bar{\varphi},\;\;
\tau\cos\f{\Delta}{2}= \begin{cases} 0, & {\A}=+I \\
                 \pm 1, & \A =-I\end{cases}.\label{c2breakdowndgncondition}
\end{align}
$\Delta$ is the differential phase between two Feynman paths  that involve a single tunneling event  over one cyclotron period, as illustrated in Fig. \ref{fig:Cn-breakdown}(d).
The adiabatic limit is given by $\tau{=}1$ and $\bar{\varphi}{=}0$, which leads to the satisfaction of \q{c2breakdowndgncondition} for $l^2$ equal to an integer multiple of $2\pi/\delta S$ (up to a field-independent offset). As noted in \s{sec:introducecodimension}, such a periodicity corresponds to a differential increment of a single, magnetic flux quantum. 

Deviating from the adiabatic limit, we deduce from \q{c2breakdowndgncondition} that $\A{=}I$ remains attainable by varying $B$, but $\A{=}{-}I$ is unattainable because $\tau{<}1$; this is consistent with the codimension analysis of \q{proprashba}. The conditions $\A{=}I$ and $\rho{>}0$ are interpreted as the destructive interference of  two, single-tunneling Feynman path [cf. Fig. \ref{fig:Cn-breakdown}(d)]. Concurrently, there is constructive interference of the Feynman loops involving zero and two tunneling events [cf. Fig. \ref{fig:Cn-breakdown}(e)]. The differential phase between outer and inner zero-tunneling (band-conserving) loops is simply 0 (mod $2\pi$), hence Landau levels are quasidegenerate. Such quasidegeneracies recur as $\Delta$ advances by $2\pi$, corresponding to a periodicity in $l^2$ of $4\pi/\delta S$ (or two flux quanta). This periodicity is not affected by the additional phase $2\bar{\phi}$ in $\Delta$, because $\bar{\phi}$ varies on the scale  $2\pi/S_{\sma{\square}}$\footnote{For $\bar{\varphi}(\bar{\mu})$ to vary $\sim 1$, the Landau-Zener parameter $\bar{\mu}$ must vary by $\sim 1$ (cf.\  Fig.\ 10 in Ref. \onlinecite{100p}). Moreover, $2\pi\barmu{:}{=} (2\pi/8) l^2S_{\sma{\square}}$\cite{AALG}}, which is much larger than $4\pi/\delta S$. To recapitulate, half the Landau-level quasidegeneracies (over the $B$ axis) are destabilized  by tunneling; the robustness of the other half relies on $\mathfrak{c}_2$ symmetry. There exists a complementary argument for the existence of Landau-level quasidegeneracies in the Klein-tunneling limit $(\tau{=}0)$ (cf. Fig. \ref{fig:Cn-breakdown}(c) upper panel): the two independent orbits are related by $\mathfrak{c}_2$ symmetry, leading to exactly-degenerate Landau levels.  

A completely different physical realization of two orbits linked by Landau-Zener tunneling was proposed in Ref. \onlinecite{yakovenko_angular_2006} and Ref. \onlinecite{pershoguba_effects_2015} -- for a bilayer {metal} subject to a tilted magnetic field. {In the absence of hybridization between the top and bottom layers, there are two decoupled orbits in each layer; in this example, the layers replace spin as a two-component degree of freedom. In the presence of hybridization,   Landau levels become layer-degenerate} only  at certain `magic angles'\footnote{These magic angles correspond to zeros of a Bessel function, as was first proposed by Yamaji.\cite{yamaji_angle_1989}} of the field orientation (a single parameter); this was experimentally confirmed in Ref.\ \onlinecite{gusev_interlayer_2008}. However, the role of $\mathfrak{c}_2${/spatial inversion} symmetry was not recognized in these works. 

For our last illustration, we consider a II-Dirac model with $\mathfrak{c}_3$  symmetry, as illustrated in Fig. \ref{fig:Cn-breakdown}(b). In the adiabatic regime ($\tau{\approx}1$), Landau-level quasidegeneracies are periodic in correspondence with a differential increment of one flux quantum; however, in the Klein-tunneling regime ($\tau{\approx}0$), there exists only one dominant orbit shaped like a three-fold-symmetric trefoil knot (cf. Fig. \ref{fig:Cn-breakdown}(c) lower panel) --  all Landau levels are nondegenerate with spacing determined by the effective mass of the trefoil; this property generalizes to $\mathfrak{c}_N$-symmetric II-Dirac models with odd $N$.   

To resolve this tension (for $N{=}3$), consider that the propagator $\A$ has a form analogous to \q{propC2breakdown} -- but with $(\cdot)^2$ replaced by $(\cdot)^3$, and with $\tilde{\lambda}_\pm{=}{\pm} l^2\delta S/4$ replaced by $\pm l^2\delta S/6$ (plus a field-independent geometric correction). The degeneracy condition is modified to $\tau\cos[(\tilde{\lambda}_+{-}\tilde{\lambda}_-)/2{+}\bar{\varphi}]{=}{\pm} 1,{\pm} 1/2$. With nonzero tunneling ($1/2{<}\tau {<}1$), only $\pm 1/2$ is attainable -- this implies that of every three degeneracies that exist in the adiabatic limit, only two are perturbatively stable. Beyond a critical field  (defined by $\tau{=}1/2$), all degeneracies coalesce pairwise (on the $B$ axis) and annihilate. This behaviour is illustrated in Fig. \ref{fig:Cn-breakdown}(f).

We conclude this section with a general result that is valid for any integer $N{\geq}2$: of every $N$ quasidegeneracies that exist in the adiabatic limit, $N{-}1$ of them are perturbatively stable against tunneling that respects  $\mathfrak{c}_N$ symmetry. We prove this in App. \ref{app:codimension}.

\subsection{The ten-fold table for symmetry constrained codimensions}\label{sec:tenfold}

\begin{table*}
\begin{tabular}{|c|c|c|c|c|c|c|}
\hline 
action on orbit & action on $\boldsymbol{k}$  & $u$ & $s$  & class label & $g$  & codimension\tabularnewline
\hline 
\multirow{5}{*}{$|g\circ\mathfrak{o}_{0}|=|\mathfrak{o}_{0}|$} & \multirow{2}{*}{$\forall\boldsymbol{k}\in\mathfrak{o}_{0},g\circ\boldsymbol{k}=\boldsymbol{k}$} & \multirow{2}{*}{0} & \multirow{2}{*}{0} & \multirow{2}{*}{I-1} & $\breve{g}\propto I$  & 3\tabularnewline
\cline{6-7} 
 &  &  &  &  & $\breve{g}\not\propto I$  & 1\tabularnewline
\cline{2-7} 
 & \multirow{3}{*}{$\exists\boldsymbol{k}\in\mathfrak{o}_{0},g\circ\boldsymbol{k}\neq\boldsymbol{k}$} & 0 & 0 & II-1 & - & 1\tabularnewline
\cline{3-7} 
 &  & \multirow{2}{*}{1} & \multirow{2}{*}{1} & \multirow{2}{*}{II-2} & $g^{2}=1$  & 2\tabularnewline
\cline{6-7} 
 &  &  &  &  & $g^{2}=-1$  & 0\tabularnewline
\hline 
\multirow{2}{*}{$|g\circ\mathfrak{o}_{0}|\neq|\mathfrak{o}_{0}|$ } & \multirow{2}{*}{-} & 0 & 0 & III-1 & - & 3\tabularnewline
\cline{3-7} 
 &  & 1 & 1 & III-2 & - & 3\tabularnewline
\hline 
\end{tabular}
\caption{Symmetry-constrained codimensions for nearly-degenerate orbits, which fall into five symmetry classes (as numbered in the first column). This table is also applicable to exactly-degenerate orbits, which fall into ten symmetry classes -- five here, and five in  \tab{table:codimension-exactlydegen}. The second and third columns define how the symmetry acts on the cyclotron orbit. The next two columns  inform us if the mapping $\bk{\rightarrow}g{\circ}\bk$ [cf.\ \q{gmapk}]  preserves ($u{=}0$) or inverts ($u{=}1$) the orientation of the orbit, and if $g$ preserves ($s{=}0$) or inverts ($s{=}1$) the arrow of time. The sixth column specifies additional conditions -- on the representation of $g$ -- which must be imposed to uniquely determine the codimension. $\breve{g}$ is defined as the diagonalized, unitary matrix representation of $g$ in the degenerate subspace of $\hat{H}_0$. For the reader's convenience, we present some commonly-encountered matrix representations for exactly-degenerate orbits in Tab.\ \ref{table:sewing-matrix}. The last column lists the symmetry-constrained codimensions [cf. Sec. \ref{sec:introducecodimension}]. \label{table:codimension-nearlydegen}}
\end{table*}
%the two-by-two propagator $\A$ -- for the five symmetry classes which apply to both nearly- and exactly-degenerate orbits. . 

\begin{table*}
\begin{tabular}{|c|c|c|c|c|c|c|}
\hline 
action on orbit  & action on $\boldsymbol{k}$  & $u$ & $s$  & class label & $g$  & codimension\tabularnewline
\hline 
\multirow{5}{*}{$|g\circ\mathfrak{o}_{0}|=|\mathfrak{o}_{0}|$} & \multirow{2}{*}{$\forall\boldsymbol{k}\in\mathfrak{o}_{0},g\circ\boldsymbol{k}=\boldsymbol{k}$} & \multirow{2}{*}{0} & \multirow{2}{*}{1} & \multirow{2}{*}{I-2} & $g^{2}=1$  & 1\tabularnewline
\cline{6-7} 
 &  &  &  &  & $g^{2}=-1$  & 3\tabularnewline
\cline{2-7} 
 & \multirow{3}{*}{$\exists\boldsymbol{k}\in\mathfrak{o}_{0},g\circ\boldsymbol{k}\neq\boldsymbol{k}$} & 0 & 1 & II-3 & - & 1\tabularnewline
\cline{3-7} 
 &  & \multirow{2}{*}{1} & \multirow{2}{*}{0} & \multirow{2}{*}{II-4} & $\breve{g}\propto\sigma_{z}$  & 2\tabularnewline
\cline{6-7} 
 &  &  &  &  & $\breve{g}\propto\sigma_{z}$  & 0\tabularnewline
\hline 
\multirow{2}{*}{$|g\circ\mathfrak{o}_{0}|\neq|\mathfrak{o}_{0}|$ } & \multirow{2}{*}{-} & 0 & 1 & III-3 & - & 3\tabularnewline
\cline{3-7} 
 &  & 1 & 0 & III-4 & - & 3\tabularnewline
\hline 
\end{tabular}
\caption{Symmetry-constrained codimensions for five of ten symmetry classes of exactly-degenerate orbits; the other five are described in \tab{table:codimension-nearlydegen}. We employ the same notations as in the caption of Tab. \ref{table:codimension-nearlydegen}. In class II-4, $\breve g$ is evaluated at $g$-invariant $\bk$ ($g\circ\bk=\bk$). For exactly-degenerate orbits in class I-2, II-3 and II-4, the two-by-two matrix propagator $\A$ has unit determinant, assuming that spin-orbit coupling can be turned off continuously without changing the degeneracy of the orbit.\cite{topoferm}\label{table:codimension-exactlydegen}}
\end{table*}

Generalizing our rotational-symmetric case studies in
Sec. \ref{sec:singleparameterrashba}-\ref{sec:rotsymmbreakdown}, we now present the symmetry-constrained codimensions for all ten symmetry classes of orbits\cite{topoferm}. Depending on the class, the codimension is either $0,1,2$ or $3$. We begin by briefly reviewing the ten-fold symmetry classification of orbits in \s{sec:reviewtenfold}. We then study in Sec. \ref{sec:codimquasideg} five of the classes which is relevant to nearly-degenerate orbits. The remaining five classes are relevant to exactly degenerate orbits, which we study in Sec. \ref{sec:codimexactdeg}.

\subsubsection{Ten-fold classification of orbits}\label{sec:reviewtenfold}

Any symmetry $g$ of a field-independent, translation-invariant Hamiltonian $\hat{H}$ acts on space-time as
\begin{align}
&g: \vectwo{\br}{t} \rightarrow \matrixtwo{\check{g}}{0}{0}{ (-1)^{s(g)}}\vectwo{\br}{t}+\vectwo{\bdelta}{0}, \label{definesg}
\end{align}
with $s(g){=}0$ (resp.\ $1$) if $g$ preserves (resp.\  reverses) the arrow of time. $\check{g}$ is the matrix representation of $g$ in real space, and the corresponding action of  $g$ on the crystal wavevector is  
\e{\bk \rightarrow g\circ\bk := (-1)^{s(g)}\check{g}\bk. \label{gmapk}}
Given $\hat{H}$ and the orientation of the magnetic field, electrons follow orbits (denoted $\frako$) that are confined to field-orthogonal $\bk$-planes.  Given an orbit $\frako$ in a $\bk$-plane, we restrict our attention to the subgroup of $\hat{H}$ that preserves the plane. Any element of this subgroup maps the orbit to itself (denoted as $|g\circ\frako|{=}|\frako|$) or to a distinct orbit ($|g\circ\frako|{\ne}|\frako|$). In the former mapping, we may further distinguish two cases: (a)  $g$ maps every $\bk$ on the orbit to itself, as exemplified by the composition $T\inv$ of time reversal ($T$) and spatial inversion ($\inv$)  symmetries. (b) There exists at least one $\bk$ on the orbit that is mapped to a distinct wavevector, e.g., time-reversal symmetry relates two wavevectors ($\bk$ and ${-}\bk$) on an orbit that encircles a time-reversal-invariant point. 

Lastly, we distinguish  between mappings [\q{gmapk}] that preserve  ($u(g){=}0$) or invert ($u(g){=}1$) the orientation of the orbit. For example, a rotation with axis parallel to the field is orientation-preserving, while a reflection with axis orthogonal to the field is orientation-inverting. The combination of all distinguishing criteria leads to   a ten-fold classification of orbits $\frako_0$ with $g$ symmetry, as was first presented in Ref. \onlinecite{topoferm} and is summarized by the ten numbered rows in Tab. \ref{table:codimension-nearlydegen}-\ref{table:codimension-exactlydegen}.

\subsubsection{Symmetry-constrained codimension for nearly-degenerate orbits}\label{sec:codimquasideg}

In the particular context of nearly-degenerate bands, $\hat{H}{=}\hat{H}_0{+}\delta \hat{H}$ is the sum of a zeroth-order Hamiltonian (with spin-degenerate bands) and a symmetry-lowering perturbation.  Given a zeroth-order orbit $\frako_0$ of $\hat{H}_0$, we  are interested in the subgroup of $\hat{H}$ that  maps $\frako_0$ to itself or to a distinct zeroth-order orbit. 

Of the ten classes described in \s{sec:reviewtenfold}, only five   are relevant to quasidegenerate orbits (which are split with differential area  $\delta S$ that is much less than the area of $\frako_0$).
To identify these five, recall that $\lambda_a$ in our quantization rule is determined by the propagator $\A$ [cf.\ \q{eq:prop}], which is a time-ordered exponential of an effective Hamiltonian $\H{=}\delta\epsilon{+}B \mathcal{M}$ [cf.\ \q{eq:H1}]; $\H$ perturbatively encodes $\delta H$ and the Zeeman interaction, for an electron travelling over  $\frako_0$. Since $g$ is  a symmetry of  $\delta \hat{H}$, $\delta\epsilon$ is invariant under action of $g$. On the other hand, $\mathcal{M}$ is the field-parallel component of a pseudovector that is odd under time reversal\cite{sakurai1995modern}, hence changes under $g$ to  $(-1)^{u+s}\mathcal{M}$ (plus a gauge-dependent term\cite{100p} which is not essential to the present argument). Only with $u{+}s{=}0$ or $2$ would both terms in $\H$ transform consistently (invariantly) under $g$ -- this identifies the five classes in Tab.\ \ref{table:codimension-nearlydegen}.

For these five classes, we list their corresponding \textit{symmetry-constrained codimensions} in the right-most column of Tab.\ \ref{table:codimension-nearlydegen}; these numbers are derived in  App.\ \ref{app:codimension}. Recall that the codimension is the number of independently-tunable parameters needed to attain a degeneracy for $\A$. Each listed codimension is robust to any perturbation of the Hamiltonian that preserves the corresponding symmetry (specified by the row) and the homotopy class\cite{100p} of the zeroth-order orbit. (An example of a change in homotopy class is given by a deformation of a circular orbit into a figure of eight.) 

Of the five classes only classes I-1, II-1 and II-2 potentially have symmetry-constrained codimensions that are reduced from three: classes I-1, II-1 and II-2 have in common that  $|g\circ\frako_0|{=}|\frako_0|$, and therefore the propagator $\A$ [cf.\ \q{eq:prop}] is self-constrained. To clarify, a self-constraining symmetry relates the propagator $\A$ [cf.\ \q{eq:prop}] to itself or to its inverse $\A^{-1}$, depending respectively on whether $u{=}0$ or $1$.

Class I-1 groups together spatial symmetries that preserve the arrow of time and map every $\bk$ on $\frako_0$ to itself. The simplest example of such a symmetry ($g$) is a reflection symmetry with axis parallel to the field, and of  $\frako_0$ that is confined to a $g$-invariant $\bk$-plane. To uniquely determine the symmetry-constrained codimension in class I-1, one must specify if the nearly degenerate bands belong to identical or distinct representations of $g$. The symmetry-constrained codimension is reduced to unity in the latter (and only the latter) case, due to the absence of level repulsion between distinct representations of $g$. 

Class II-1 corresponds to spatial symmetries that preserve the arrow of time and the orientation of $\frako_0$, as exemplified by spatial rotational symmetry with rotational axis parallel to the field. Another example is given spatial inversion symmetry, for $\frako_0$ that is confined to an inversion-invariant plane: $k_z=0$ or $\pi$. The symmetry-constrained codimension here is unity, as has been modelled in Sec.\ \ref{sec:singleparameterrashba}-\ref{sec:rotsymmbreakdown}. For a fixed orbit $\frako_0$ in class II-1, there may exist different degeneracy manifolds with different codimensions; in such case we define the symmetry-constrained codimension to be the minimal codimension among all manifolds.\footnote{For a fixed orbit in class I-1 or II-2, all degeneracy manifolds have the same codimension.} For example, we have found that two-fold rotational symmetry reduces the codimension of the $\A{=}{+}I$ manifold (to unity), but not the codimension of the $\A{=}{-}I$ manifold [cf.\ \fig{fig:dgn} and \q{proprashba}]. The rotation-constrained codimension is then unity, and hence the entry $1$ in the right-most column for class II-1.

Symmetries in class II-2 reverse both the arrow of time and the orientation of the orbit. The symmetry-constrained codimension is $2$ or $0$, depending respectively on whether the antiunitary representation of $g$ squares to $+1$ or $-1$. The former is exemplified by the composition of time reversal with a reflection symmetry (with reflection axis orthogonal to the field); this is a symmetry of the Rashba-Dresselhaus Hamiltonian [cf.\ \q{hamRD}]:
\e{ g H_{RD}(k_x,k_y)g^{-1} = H(k_y,k_x),\;\; g = e^{i\sigma_z\pi/4}K,}
with $K$ implementing complex conjugation. A consequence of this symmetry is that any of the quasidegeneracy lines in Fig. \ref{fig:dgn}(b) is robust -- a property we had previously (and consistently) deduced from continuous rotational symmetry. 

An example of $g^2{=}-1$ is given by
the composition $T\mathfrak{r}_{x,\boldsymbol{c}/2}$ of time reversal with a glide symmetry; $\mathfrak{r}_{x,\boldsymbol{c}/2}$ is itself the composition of a reflection (with axis orthogonal to the field) and half a lattice translation (parallel to the field). The propagator for an orbit in the Brillouin-zone boundary ($k_z{=}\pi$ plane) would then be Kramers-degenerate -- the symmetry-constrained codimension vanishes.

Finally, we point out that the Rashba 2DEG in a tilted magnetic field, as studied in \ref{sec:inplanezeeman}, also falls into class II-2  owing to the symmetry:
\begin{equation}
\sigma_z K H_{RZ}(k_x, k_y) K \sigma_z=H_{RZ}(k_x, -k_y).
\end{equation}
Consequently, the degeneracy manifolds are lines in the parameter space $(\alpha, B_\parallel, g_{s\perp})$. Numerically, we have determined that these lines lie solely within the $B_\parallel=0$ and $\alpha=0$ planes. At any point within either plane,  the Landau levels are exactly solvable. For $B_\parallel=0$, $H_{RZ}$ reduces to the Rashba Hamiltonian and the degeneracy manifolds form a series of concentric circles [illustrated in Fig. \ref{fig:dgn}(b) with $\beta=0$]. For $\alpha=0$, the degeneracy manifolds occur at
\begin{equation}
    \f{g_s}{4} \frac{ m}{ m_0}\text{sec}(\theta)=n,\;\;(n\in\mathbb{Z}),
\end{equation}
with $\theta$ the tilt angle: $\cos(\theta)=B_\perp/\sqrt{B_\perp^2+B_\parallel^2}$. 

\subsubsection{Symmetry-constrained codimension for exactly-degenerate orbits}\label{sec:codimexactdeg}

This work has mainly focused on the Landau quantization of nearly-degenerate bands, which are eigenstates of the field-independent Hamiltonian $\hat{H}_0{+}\delta \hat{H}$. In this section we focus on higher-symmetry solids where the degeneracy-lifting perturbation $\delta \hat{H}$ is disallowed. $\hat{H}_0$ may or may not include spin-orbit coupling; in the former case, a crystalline point-group symmetry (e.g., $T\inv$ symmetry) is required for energy bands of $\hat{H}_0$ to be two-fold degenerate at each $\bk$.

Given $\hat{H}_0$ and the orientation of the field, electrons of opposite spin follow exactly-overlapping orbits (denoted $\frako_0$) which are confined to field-orthogonal $\bk$-planes. Such overlapping orbits will be referred to as \textit{exactly-degenerate}, in contrast with quasidegenerate orbits that are split on the $\bk$-plane. Despite the exact degeneracy of orbits, the corresponding Landau levels are generically split by the Zeeman interaction $B\calm$. We may therefore ask, in complete analogy: how many tunable parameters are required to attain a Landau-level quasidegeneracy for exactly-degenerate orbits? 

We answer this question in nearly the same vein as for nearly-degenerate orbits: the quantization rule for exactly-degenerate orbits\cite{topoferm} has the same form as \qq{eq:rule}{berryconn} with $\delta \var{=}\delta S{=}0.$ An eigenvalue-degeneracy of $\A$ would also imply a Landau-level quasidegeneracy, since the analysis of \s{sec:relatedegeneracies} applies to the special case $\delta S{=}0$. 

There is, however, a difference in the possible symmetry classes that apply to exactly-degenerate vs quasidegenerate orbits.  In \s{sec:codimquasideg}, we restricted ourselves to symmetries  that map the magnetic moment  $\calm$ to itself (plus a gauge-dependent term\cite{100p}), so that both terms in $\calh{=}\delta \var{+}B\calm$ transform consistently as scalars. In the present context, this restriction is void because $\delta\var{=}0$, hence we \textit{also} allow for symmetries that invert the sign of $\calm$. 

Some $\calm$-inverting symmetries also exhibit reduced symmetry-constrained codimensions, as listed in the five classes of \tab{table:codimension-exactlydegen}. As a case in point, class II-2 symmetries invert time and $\calm$, and map every wavevector (in $\frako_0$) to itself. The corresponding codimension is three (resp.\ one) if $g$ squares to $-1$ (resp.\ $+1$). The former is exemplified by $T\inv$ symmetry, and the latter by $T\rot_{2z}$ (the composition of time reversal with a two-fold rotation about the field direction). For an exactly-degenerate orbit $\frako$ confined to a $\rot_{2z}$-invariant plane, the associated propagator $\A$ is unitarily equivalent to a special orthogonal matrix\cite{100p,alexandradinata_berry-phase_2016} with a single, tunable angle of rotation -- this provides an intuitive explanation for the unit codimension.

The simplest example of codimension reduction by a $\calm$-preserving symmetry is that of continuous spin rotation in class I-1. This is a symmetry of any solid with negligible spin-orbit coupling, and results in the absence of Landau-level repulsion between different spin species -- this implies codimension is 1 [cf.\ class I-1 in \tab{table:codimension-nearlydegen}, with non-identical symmetry eigenvalues].

The detailed modelling of all ten classes of exactly-degenerate orbits is left to future work. We conclude this section by commenting on the tunable experimental parameters that are relevant to exactly-degenerate orbits. The absence of $\delta\var$  implies that the eigenvalues of $\cala$ are independent of the magnitude of $B$ [cf.\ \q{eq:Rashba-lambda} with $\delta S{=}0$]. However, the generic anisotropy of the Zeeman interaction suggests that we may sweep the orientation of $B$; the generic energy-dependence of $\calm$ suggests that we may tune the Landau-level index, e.g., by varying a bias voltage in tunneling spectroscopy.

\begin{table}
\begin{tabular*}{\columnwidth}{c@{\extracolsep{\fill}}ccc}
\hlineB{2.0}
class No. & symmetry  & origin of degeneracy & $\breve{g}\propto$ \\
\hline 
\multirow{2}{*}{1} & \multirow{2}{*}{$\mathfrak{r}_z$, $\mathfrak{r}_{z,\boldsymbol{a(b)}/2}$} & $T\mathfrak{i}$, spin SU(2) & $\sigma_z$ \\
\cline{3-4}
 & & $T\mathfrak{c}_{2z,\boldsymbol{c}/2}$ & $I$ \\
\hline
\multirow{3}{*}{5} & $\mathfrak{r}_{x(y),\boldsymbol{c}/2}$ & $T\mathfrak{i}$, spin SU(2), $T\mathfrak{c}_{2z,\boldsymbol{c}/2}$ & $\sigma_z$ \\
\cline{2-4}
& $\mathfrak{r}_{x(y)}$,$\mathfrak{r}_{x(y),\boldsymbol{b(a)}/2}$, & $T\mathfrak{i}$, spin SU(2) & $\sigma_z$ \\
\cline{3-4}
& $\mathfrak{c}_{2x(y)}$, $\mathfrak{c}_{2x(y),\boldsymbol{b(a)}/2}$ & $T\mathfrak{c}_{2z,\boldsymbol{c}/2}$ & $I$\\
\hlineB{2.0}
\end{tabular*}
\caption{Matrix representations of reflections (e.g., $\mathfrak{r}_x$), rotations (e.g. $\mathfrak{c}_{2y}$), glide reflections (e.g., $\mathfrak{r}_{x,\ba/2}$) and screw rotations (e.g., $\mathfrak{c}_{2y,\ba/2}$). The latter two nonsymmorphic elements involve  half a lattice translation in the $x,~y,$ or $z$ direction (labeled by $\boldsymbol{a,~b,~c}$ respectively). The $z$ axis of our Cartesian coordinate system is aligned parallel to the  magnetic field. This table is applicable to exactly-degenerate orbits whose degeneracy is protected by either spacetime inversion symmetry ($T\mathfrak{i}$), the composition of time reversal and nonsymmorphic two fold rotation symmetry ($T\mathfrak{c}_{2z,\boldsymbol{c}/2}$) [for orbits confined to the $k_z=\pi$ plane] or spin SU(2) symmetry (i.e., negligible spin-orbit coupling). \label{table:sewing-matrix}}
\end{table}

\section{Quantum-oscillatory phenomena for exactly- and nearly-degenerate orbits}\label{sec:qo}

In \s{sec:qtznrules} we have presented a quantization rule [\qq{eq:rule}{berryconn}] that is applicable to both nearly-degenerate ($0{<}\delta S{\ll}S$) orbits and exactly-degenerate ($\delta S{=}0$) orbits. This section explores the implication of this rule for quantum-oscillatory phenomena of the Shubnikov-de Haas (SdH)\cite{SdH} and de Haas-van Alphen (dHvA)\cite{dHvA} types. In \s{sec:quantosc_equidis}, we present a generalized Lifshitz-Kosevich formula\cite{lifshitz_kosevich,lifshitz_kosevich_jetp} that inputs our rule and outputs quantum oscillations. Applying this formula in the high-temperature regime, we will prove that  destructive interference of the fundamental quantum-oscillatory harmonic may  be tuned by a single parameter in any symmetry class; this parameter may be the magnitude ($B$) of the magnetic field for quasidegenerate orbits, but not for exactly-degenerate orbits. \s{sec:quantosc_quasideg} focuses on the low-temperature SdH effect for nearly-degenerate orbits in symmetry classes I-1 and II-1 [cf.\ \tab{table:codimension-nearlydegen}], wherein we predict a smooth crossover from period-doubled to -undoubled oscillations upon variation of $B$ (at fixed particle density).

\subsection{Single-parameter tunability for the destructive interference of the fundamental harmonic}\label{sec:quantosc_equidis}

Lifshitz-Kosevich formulae\cite{lifshitz_kosevich,lifshitz_kosevich_jetp} allow for a harmonic analysis of quantum oscillations, and have previously been derived for  exactly-degenerate orbits in symmetry class I-1\cite{rothmag}, with the generalization to all ten symmetry classes in Ref. \onlinecite{topoferm}. A simple generalization of the derivation in Ref. \onlinecite{topoferm} extend the utility of these existing formulae to nearly-degenerate orbits. For example, the oscillatory component of the magnetization of 2D metals has the form
\begin{equation}
\delta M=-\frac{1}{2\pi}\frac{k_BT}{BS^{-1}}\sum_{a=1}^2\sum_{r=1}^{\infty}e^{-\f{r\pi}{\omega_c\tau_{\sma{D}}}} \frac{\text{sin}[r(l^2S{+}\lambda_a{+}\gamma)]}{\text{sinh}(2\pi^2rk_BT/\var_c)},\label{eq:LK}
\end{equation}
where the argument of the sine function involve quantities from our quantization rule [\qq{eq:rule}{berryconn}]; $T$ is temperature and $\tau_{\sma{D}}$  the Dingle scattering lifetime\cite{Dingle_collisions}. All the quantities on the right-hand side of Eq.\ (\ref{eq:LK}) are evaluated at the chemical potential. [\q{eq:LK} corrects the analogous formula in Ref.\ \onlinecite{topoferm} by a factor of half.]

At high temperature ($kT{\gg}\var_c$) and/or with strong disorder ($\omega_c\tau_{\sma{D}}{\ll}1$), the dominant contribution to $\delta M$ is from the fundamental harmonic ($r{=}1$), which sums two sine functions with generically distinct phase offsets $\lambda_{1,2}$. This summation (or interference) is constructive if $\delta \lambda{=}|\lambda_1{-}\lambda_2|{\equiv}0$, and destructive if $\delta \lambda{\equiv}\pi$  ($\equiv$ denotes an equivalence modulo $2\pi$). In the latter case, only even-$r$ harmonics remain.

We have shown in \s{sec:llquasideg} that  the number of tunable parameters required to attain  $\delta \lambda{\equiv}0$ ranges from $0$ to $3$; this number is a symmetry-dependent measure of Landau-level repulsion [cf.\ \tab{table:codimension-nearlydegen} and \tab{table:codimension-exactlydegen}]. As illustrated in the right-most panel of Fig 7(b), $\delta \lambda{\equiv}\pi$ is the condition  for Landau levels to be quasi-equidistant  with spacing ${\approx}\var_c/2$; since no level repulsion need be overcome,
we expect $\delta \lambda{\equiv}\pi$ is attainable with at most one tunable parameter (in any symmetry class). To prove this, we return to the canonical parametrization of the SU(2) component (${\cal \bar{A}}$) of $\A$ [cf.\ \q{s3}]. The necessary and sufficient condition for $\delta \lambda{\equiv}\pi$ is that ${\cal \bar{A}}(r_1,r_2,r_3)$ has eigenvalues $+i$ and $-i$; this is equivalent to setting $r_1{=}0$ in \q{s3} -- a condition on a single parameter.\footnote{For an alternative proof,  $\delta \lambda{\equiv}\pi$ is equivalent to ${ \bar{\cal A}}{=}Ve^{i\sigma_z\pi/2}V^{-1}$ for some unitary $V\in U(2)$. The manifold associated to $\delta \lambda{\equiv}\pi$ is then the space of unitary matrices  that do not commute with $e^{i\sigma_z\pi/2}$. The dimension of this manifold is the number of linearly-independent Hermitian generators (two) for unitaries that do not commute with $e^{i\sigma_z\pi/2}$. The codimension is then obtained from deducting two from the total number (three)  of linearly-independent Hermitian generators  for U(2). This approach of determining codimension is similar to that in Ref. \onlinecite{holler_non-hermitian_2018}.} 

For exactly-degenerate orbits, the single parameter to attain destructive interference might be the chemical potential (as illustrated for the quantum oscillations of Na$_3$Bi in Ref. \onlinecite{topoferm}) or the orientation of the magnetic field; however $\lambda_a$ is independent of the field magnitude ($B$) to the accuracy\cite{rothmag,fuchs_landau_2018,gao_zero-field_2017,fischbeck_review} of our theory, as explained in \s{sec:codimexactdeg}. 

%Going beyond the accuracy of our theory, one may account for linear-in-$B$ corrections to $\lambda_a$ at stronger fields, however a quantitative theory for such corrections only exists (thus far) for non-degenerate orbits\cite{rothmag,fuchs_landau_2018,gao_zero-field_2017,fischbeck_review}.

For nearly-degenerate orbits, $\lambda_a$ is expandable as a Laurent series in $B$ with leading term $1/B$ [cf.\ \q{lambdasmallB}], hence $B$ can be tuned to attain destructive interference. The slow variation of $\lambda_a$ compared to the action $l^2S$ [cf.\ \app{sec:proofLLquasideg}] implies a beating pattern for the fundamental harmonic. In the weak-field, adiabatic regime, the nodes of this pattern occur with period $2\pi/\delta S$ in $l^2$ [cf.\ \s{sec:introducecodimension}]; in the non-adiabatic regime, this periodicity is lost due to field-induced hybridization of orbits [cf.\ \q{lambdasmallB}]. 

Such beatings have been observed for semiconductor heterostructures\cite{das_evidence_1989,hu_zero-field_1999,wilde_inversion-asymmetry-induced_2009} and inversion-asymmetric metals\cite{terashima_fermi_2008,onuki_chiral-structure-driven_2014,maurya_splitting_2018}. In most experiments, the beating has been used to infer the magnitude of the spin-orbit coupling\cite{das_evidence_1989,onuki_chiral-structure-driven_2014,maurya_splitting_2018}; some of these experiments\cite{das_evidence_1989} paid attention to  the aperiodicity of the beating, and attributed it phenomenologically to a $B$-dependent splitting of the two orbits, which have distinct frequencies $f_1$ and $f_2$ in $1/B$. In our language, $|f_1-f_2|=B\delta\lambda/2\pi$, with the leading term of $\delta \lambda(E,B)$ proportional to $1/B$.

Our contribution is a quantitative theory of the $B$ dependence of $\delta\lambda$ -- we have related the subleading term  $({\sim}B^0)$ to a generalized Zeeman interaction, and higher-order terms to a competition between spin-orbit coupling and the Zeeman interaction [cf.\ \q{eq:Rashba-lambda}]. As one application, one may extract the intervals (in $1/B$) between nodes for the Rashba 2DEG [by imposing $\delta \lambda{\equiv}\pi$ in \q{eq:Rashba-lambda}] and for the Rashba-Dresselhaus 2DEG (see the subsequent \s{sec:quantosc_quasideg}).

%In some experiments\cite{das_evidence_1989}, it has been recognized that the oscillation frequency in $1/B$, denoted as $f$, is weakly dependent on $B$ itself; 
%the possibly aperiodic beatings has been interpreted as a $B$ dependent oscillation frequency difference $\delta f$ between the two splitted orbits. In our language, the frequency difference is related to $\delta\lambda$ as $\delta f=B\delta\lambda/2\pi$. It has been appreciated that the leading-order term of $\lambda$ (${\sim}1/B$) originates from the zero-field splitting of orbits; this has been applied to estimate the strength of the spin-orbit coupling\cite{das_evidence_1989,onuki_chiral-structure-driven_2014,maurya_splitting_2018}. 

\begin{figure}
\includegraphics[width=1.0\columnwidth]{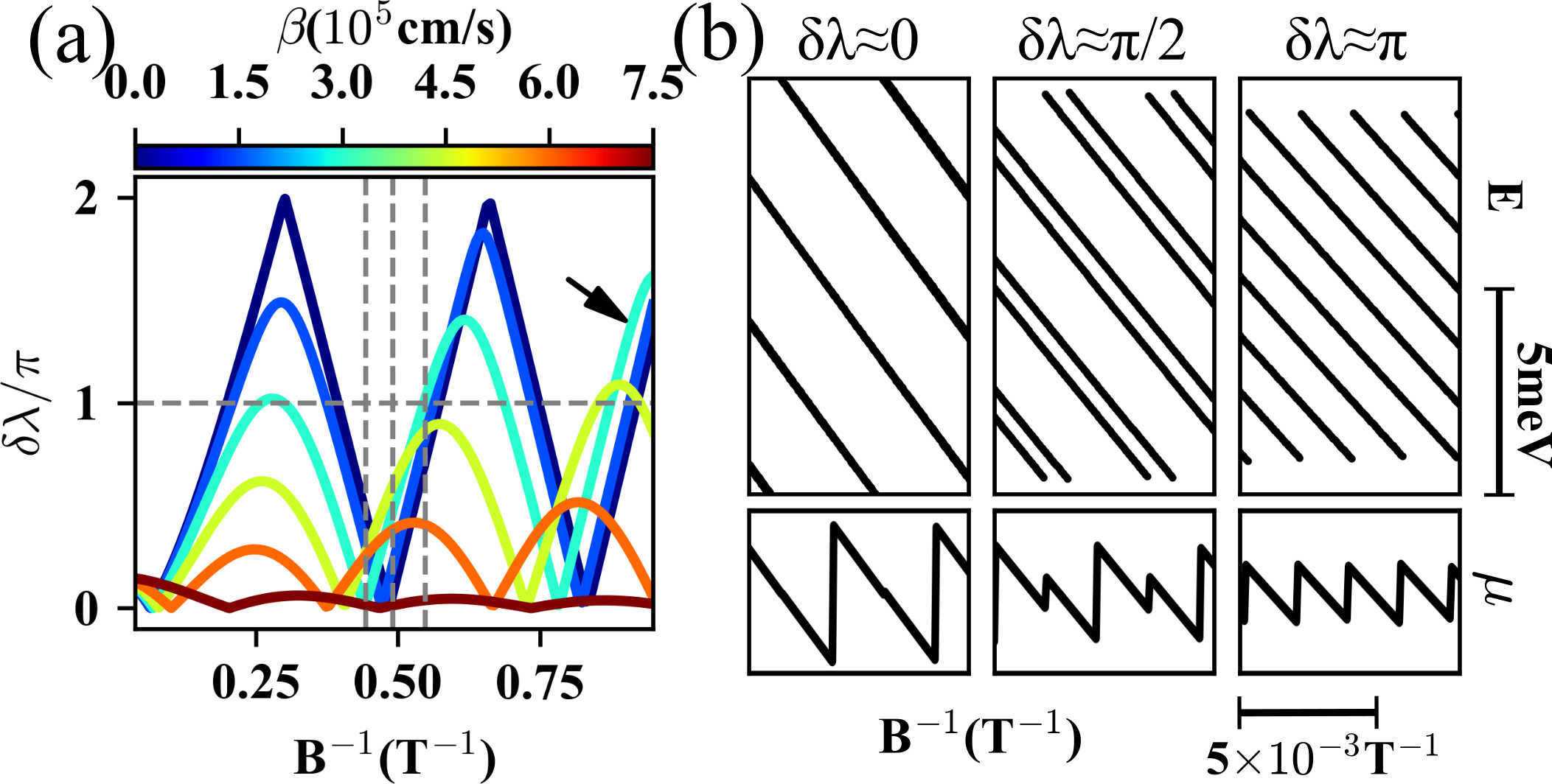}
\caption{Quantum oscillations in the Rashba-Dresselhaus 2DEG. (a) Evolution of $\delta\lambda$ with respect to the inverse field for different values of $\beta$. The horizontal dashed line indicates $\delta\lambda=\pi$. For $\beta=3.0\times 10^{5}$cm/s, we have plotted -- in the upper panel of (b) -- the Landau levels at three $B$ values, as indicated by the vertical dashed lines in (a). The evolution of $\delta \lambda$ for $\beta=3.0\times 10^{5}$cm/s is given by a cyan-colored line indicated by a black arrow in (a); the three chosen $B$ values correspond approximately to  $\delta\lambda{=}0$, $\delta\lambda{=}\pi/2$ and $\delta\lambda{=}\pi$ respectively. Lower panel of (b) illustrates the saw-tooth oscillation of the chemical potential (at fixed electron density) corresponding to the upper panel. Other parameters chosen in (b): $m{=}0.076m_0$, $\alpha{=}7.5\times10^{5}$cm/s and $E=0.4$eV.
\label{fig:qo}}
\end{figure}

\subsection{Shubnikov-de Haas effect for nearly-degenerate orbits in symmetry class I-1 and II-1}\label{sec:quantosc_quasideg}

Let us focus on nearly-degenerate orbits for which a single parameter ($B$) can be tuned to attain $\delta \lambda{\equiv}0$ (equivalently, to attain a Landau-level quasidegeneracy). Such orbits with unit codimension lie in symmetry classes I-1 and II-1; in class I-1 further assumptions must be made about the symmetry representation, as explained in \s{sec:codimquasideg} and summarized in \tab{table:codimension-nearlydegen}.

We propose a signature for tunable quasidegeneracies in the  Shubnikov-de Haas effect at fixed particle density $n_e$, low temperature ($k_BT{\ll}\var_c$) and weak disorder ($\omega_c\tau_{\sma{D}}{\gg}1$).
In this regime, it is known that minima of the longitudinal conductivity occur at discrete fields (denoted $B_{\nu}$)  where the filling is integer-valued\cite{vinter_resolution_1980}; the period of quantum oscillations is determined by the total density $n_e$ through $1/B_j{-}1/B_{j+1}{=}e/n_eh$. As $B$ increases through $B_j$, a Landau level is completely depopulated, leading to a periodic drop in the chemical potential ($\mu$) by the energy gap  between adjacent Landau levels, as illustrated in \fig{fig:qo}(b). Where $\delta\lambda{\equiv}\pi$, Landau levels are  quasi-equidistant, hence adjacent minima (of conductivity) should be identically low. Where $\delta\lambda{\equiv}0$, every consecutive gap is small compared to $k_BT$ and $h/\tau_{\sma{D}}$, hence every other  minimum (of conductivity) vanishes\cite{bychkov_oscillatory_1984}. $0{\leq}\delta\lambda{\leq}{\pi}$ thus characterizes a smooth crossover between period-doubled and -undoubled oscillations, as representatively illustrated in \fig{fig:qo}(b); the amplitude of oscillations in the undoubled regime is roughly half the amplitude in the doubled regime.

Our contribution is the exhaustive  identification of symmetry classes (I-1 and II-1) for which such a crossover will occur. In particular, we are proposing that the crossover occurs for the Rashba-Dresselhaus 2DEG [cf.\ \q{hamRD}], whose nearly-degenerate orbits fall into class II-1. For this model,  Fig.\ \ref{fig:qo}(a) plots $\delta \lambda$ vs $1/B$ for various choices of the Dresselhaus coupling $\beta$. As expected from the codimension analysis of \s{sec:disrot}, half of the degeneracies ($\delta\lambda {\equiv}0{\equiv}2\pi$) are lifted by introducing nonzero $\beta$.

\section{Summary, discussion and outlook}\label{sec:discussion}

We have extended the semiclassical theory of Landau quantization to describe nearly-degenerate cyclotron orbits, i.e., orbits which are split in $\bk$-space with a differential area ($\delta S$) that is much less than their average area. The quantization rule for two nearly-degenerate orbits is summarized in \qq{eq:rule}{berryconn}, and may be applied to spin-degenerate orbits which are split by spin-orbit coupling or by breaking of a crystalline point-group symmetry. The former type of splitting is exemplified by  our case study of the Rashba-Dresselhaus 2DEG with an arbitrarily-oriented field.

The subleading phase corrections ($\lambda_1,\lambda_2$) in our quantization rule  [\q{eq:rule}] encode the dynamical correction due to the splitting of  orbits, as well as a generalized two-band Zeeman interaction. If $\lambda_1{=}\lambda_2$ mod $2\pi$ for specific values of energy $E$ and magnetic field $B$, there exist -- in close proximity to $E$ and $B$ --    paired Landau levels which are either exactly pseudospin-degenerate, or nearly pseudospin-degenerate in the sense of \qq{llquasideg}{llquasidegB}. Such paired Landau levels have been referred to as \textit{quasidegenerate}, and are illustrated in Fig. \ref{fig:LL}.

For each of the five symmetry classes of nearly-degenerate orbits, we have determined the number of tunable real parameters needed to attain a Landau-level quasidegeneracy; these numbers may be viewed as symmetry-dependent measures of Landau-level repulsion. We have referred to these numbers as \textit{symmetry-constrained codimensions}. Depending on the symmetry class and representation, the codimension ranges from 0, 1, 2 to 3, as summarized in the right-most column of Tab.\ \ref{table:codimension-nearlydegen}- \ref{table:codimension-exactlydegen}.

If the codimension is unity, it is possible (for nearly-degenerate orbits) to tune the magnitude of the field to attain Landau-level quasidegeneracies; the implications for this tunability -- in the Shubnikov-de Haas effect of two-dimensional metals -- are discussed in \s{sec:quantosc_quasideg}. If the codimension is two, an additional parameter is needed: e.g., the orientation of the field\cite{yakovenko_angular_2006}, or the Landau-level index (tuned through the bias voltage in tunneling spectroscopy\cite{Sangjun_Cd3As2,Ilija_SnTe,kushwaha_Dirac}). While we have mainly exemplified our theory for 2DEGs or two-dimensional metals, we remark that for the Landau levels of three-dimensional solids, the crystal wavevector parallel to the field provides yet another tunable parameter.

We highlight a result from  Tab.\ \ref{table:codimension-nearlydegen} that is widely applicable: the codimension is unity for orbits having either a spatial inversion or rotation symmetry (class II-1). Both symmetries are commonly found in solids. From a geometric perspective, the quasidegeneracies (in this symmetry class) lie on hypersurfaces in the space of tunable parameters. From a topological perspective, each hypersurface is a domain wall that separates distinct connected components, and there are $N$ connected components for orbits with an order-$N$ rotational symmetry; $N=2$ in the case of inversion symmetry. This topological distinction between rotational symmetries of different orders has the following implication: if the rotational symmetry is reduced in order, a specific subset of the quasidegeneracies remain stable, as explained in \s{sec:singleparameterrashba} and exemplified for the Rashba-Dresselhaus 2DEG.  For further motivation in an entirely different context, a reduction in rotational symmetry   occurs also in a phase transition to a nematic Fermi fluid\cite{fradkin_nematic_2010}; it may be possible to utilize  the selective stability of Landau-level quasidegeneracies as an experimental indicator for nematic phase transitions. 

Our quantization rule [\qq{eq:rule}{berryconn}] may be applied, as a special case, to \textit{exactly-degenerate orbits} -- spin-differentiated orbits which overlap exactly in $\bk$-space. There are ten possible symmetry classes for exactly-degenerate orbits -- five more than for near-degenerate orbits. For each of this ten, we have also determined their symmetry-constrained codimensions in Tab.\ \ref{table:codimension-nearlydegen}-\ref{table:codimension-exactlydegen}. We remark on one further distinction from nearly-degenerate orbits --  the magnitude of the field becomes less useful as a tuning parameter, however the field orientation and Landau-level index remain feasible options.

%\red{Symmetry constrained codimension is the number of parameters needed to find $\lambda_1=\lambda_2 \mod 2\pi$, which is generally not a solution of our quantization rule. If one more parameter is available, the simultaneous solution of quantization rule and $\lambda_1=\lambda_2 \mod 2\pi$ can generally found, as illustrated in Fig. \ref{fig:LL}. These solutions are exact degeneracies in the sense of first order semiclassical approximation. It is possible that these solutions are exact even beyond semiclassical approximation, for example when the two crossing Landau levels belongs to the different representation of a symmetry operation. A general answer to whether these solutions are always exact degeneracies beyond semiclassical approximation is left for further investigation.}

Our codimension analysis is applicable to contexts beyond that of Landau-level repulsion. By keeping only the Berry term in \q{eq:H1}, Eq.\ ({\ref{eq:prop}}) reduces to the Wilson loop of the non-abelian Berry gauge field\cite{wilczek_appearance_1984}, i.e., the matrix representation of holonomy.  Any symmetric Wilson loop can be classified according to Tab.\ \ref{table:codimension-nearlydegen}-\ref{table:codimension-exactlydegen}; the codimensions for an eigenvalue-degeneracy of the Wilson loop are also identical to those listed in Tab.\ \ref{table:codimension-nearlydegen}-\ref{table:codimension-exactlydegen}. Our work thus also explains the existence of robust crossings in the Wilson-loop spectra, which have been widely applied in topological band theory. For example, the existence of such crossings have been used to diagnose the presence of a 3D Dirac point protected by rotational symmetry\footnote{This has been observed numerically in \onlinecite{gresch_z2pack:_2017}, and the group-theoretic explanation may be found in Sec. VI-D-7 of \onlinecite{100p} or \onlinecite{li_topological_2017}}, as well as to diagnose fragile topological phases with rotational symmetry\cite{bouhon_wilson_2018,bradlyn_disconnected_2018}.

Symmetry does not only constrain the level repulsion between $\lambda_1$ and $\lambda_2$ -- in certain classes, it constrains their sum. In our previous work on exactly-degenerate, spin-differentiated orbits\cite{100p,topoferm}, we have derived a zero-sum rule ($\lambda_1{+}\lambda_2{\equiv}0$) in classes  I-1, II-3 and II-4, with the assumption that spin-orbit coupling can be continuously tuned to zero without changing the degree of degeneracy; this zero-sum rule implies a quantized phase offset for the fundamental harmonic of quantum oscillations\cite{topoferm}. Here, we report that the zero-sum rule applies \textit{also} to nearly-degenerate orbits in classes I-1, II-3 and II-4; this may be derived by combining the symmetry analysis of Ref.\ \onlinecite{topoferm} with the tracelessness of $\delta\var$ [cf.\ \q{eq:H1}]. 

%Although these three symmetry classes do not impose constraints on nearly-degenerate orbit, due to the traceless spin splitting Hamiltonian matrix, $\lambda_1+\lambda_2=0$ remains true for nearly-degenerate orbits.

With the inclusion of nonsymmorphic space-group symmetries, it is not uncommon to find energy bands which are (nearly) four-fold degenerate\cite{michel_elementary_2001,wang_hourglass_2016,bradlyn_topological_2017}. This motivates a generalization of our rule -- for any number of nearly-degenerate orbits -- that is presented in \app{app:quantizationruleproof}. The symmetry-constrained codimensions in the more general case is left to future investigations. 

\begin{acknowledgments}

We are grateful to Nicholas Read and Judith H\"oller for clarifying notions related to codimensions of eigenvalue-degeneracies, and to Xi Dai for an instructive discussion on the Zeeman interaction in the context of breakdown. We thank Victor Yakovenko for  educating us on `magic-angle' magnetoresistance oscillations, and  Eduardo Fradkin for his expertise on nematic Fermi fluids.  We acknowledge support from the Ministry of Science and Technology of China, Grant No.\ 2016YFA0301001 (CW and WD), the National Natural Science Foundation of China, Grants No.\ 11674188 and 11334006 (CW and WD), the Beijing Advanced Innovation Center for Future Chip (CW and WD), the Yale Postdoctoral Prize Fellowship (AA) and NSF DMR Grant No.\ 1603243 (LG). This work was performed in part at Aspen Center for Physics, which is supported by National Science Foundation grant PHY-1607611. This work was supported in part by the Gordon and Betty Moore Foundation EPiQS Initiative through Grant No. GBMF4305 at the University of Illinois.\\
\end{acknowledgments}

\appendix

The appendices are organized as follows: in App. \ref{app:quantizationruleproof}, we prove the quantization rule of \qq{eq:rule}{berryconn}. The Landau levels derived from this quantization rule are then compared with numerically-exact Landau levels in App. \ref{app:approximatevsexact}, for the Rashba-Dresselhaus 2DEG in a magnetic field.  In App. \ref{sec:proofLLquasideg}, the existence of Landau-level quasidegeneracies is proven. In App. \ref{app:codimension}, we derive the symmetry-constrained codimensions of Landau-level quasidegeneracies, in all symmetry classes of cyclotron orbits. 

%utilizing a mathematical result proved in App. \ref{app:makinggoffdiagonal}.

%The numerical diagonalization method is reviewed in App. \ref{sec:numerical}.

\section{Proof of quantization rule for the Landau levels of nearly-degenerate bands \label{app:quantizationruleproof}}

In this appendix, we derive the quantization rule for the Landau levels of nearly-degenerate energy bands. We assume the energy bands are nearly $D$-fold degenerate at generic wavevectors. The Landau quantization of such bands results in the quantization rule of  \qq{eq:rule}{berryconn} for $D=2$; these equations will be generalized (below) for $D>2$. Although less common than $D=2$, $D>2$ is relevant in  materials with nonsymmorphic symmetries. For example in screw- and time-reversal-invariant solids with weak spin-orbit coupling, bands in a high-symmetry plane (e.g., $k_z=\pi$ for the screw $\mathfrak{c}_{2z,\boldsymbol{c}/2}$)  are nearly four-fold degenerate.

The quantization rule is derived from the WKB solution of an effective Schr\"odinger equation. During the derivation, it is important to control the error in our approximation scheme, which we use big-O and little-o notations to quantify. They are defined as: 
\begin{itemize}
    \item 
    \begin{equation}
        f(\boldsymbol{x})\sim\text{O}(\{g_i(\boldsymbol{x}), i\in 1..N\})
    \end{equation}
    as $\boldsymbol{x}\to\boldsymbol{0}$ if and only if $\exists C>0$ and $\exists \delta>0$ such that for all $\boldsymbol{x}$ with $||\boldsymbol{x}||_\infty<\delta$,
    \begin{equation}
        |f(\boldsymbol{x})|<C\text{max}\{|g_i(\boldsymbol{x})|, i\in 1..N\};
    \end{equation}
    \item 
    \begin{equation}
        f(\boldsymbol{x})\sim\text{o}(\{g_i(\boldsymbol{x}), i\in 1..N\})
    \end{equation}
    as $\boldsymbol{x}\to\boldsymbol{0}$ if and only if $\forall \epsilon>0,~\exists \delta>0$ such that for all $\boldsymbol{x}$ with $||\boldsymbol{x}||_\infty<\delta$,
    \begin{equation}
        |f(\boldsymbol{x})|<\epsilon\text{max}\{|g_i(\boldsymbol{x})|, i\in 1..N\}.
    \end{equation}
\end{itemize}
Here, $||\boldsymbol{x}||_\infty=\text{max}\{|x_i|\}$ is the uniform norm of the vector $\boldsymbol{x}$.

As stated in \s{sec:qtznrules}, our starting point is a decomposition of the Hamiltonian $\hat{H}$ into $\hat{H}_0$ and a perturbation $\delta\hat{H}=\eta\hat{H}_1$, with $\eta$ a dimensionless small parameter. We assume that $\eta\hat{H}$ perturbatively breaks the  $D$-fold energy degeneracy (at generic wavevectors) of $\hat{H_0}$. For 2-fold nearly-degeneracy, $\eta$ can be chosen as $\delta S/S$ in the main text.

For an approximate, semiclassical solution to the time-independent Schr\"odinger equation, we employ the effective Hamiltonian formalism \cite{rotheffham,100p}.  The effective Hamiltonian is expressed in a basis of a field-augmented Bloch functions $\{\tilde{\psi}_{n\boldsymbol{k}}\}_{n=1}^D$,\cite{rotheffham} which are modifications of Bloch functions that span the  energy-degenerate subspace of $\hat{H}_0$. Expanding an energy eigenfunction $\phi$ in this basis ($\phi=\sum_{n=1}^D\sum_{\boldsymbol{k}}f_{n\boldsymbol{k}}\tilde{\psi}_{n\boldsymbol{k}}$), the Schr\"odinger equation effectively becomes
\begin{equation}
(\mathfrak{H}(\boldsymbol{K})-E)\boldsymbol{f}_{\boldsymbol{k}}=0,\label{eq:schrodinger}
\end{equation}
with $\mathfrak{H}(\boldsymbol{K})$ (the effective Hamiltonian) a $D$-by-$D$ matrix differential operator. 

In general, the effective Hamiltonian has an asymptotic expansion in powers of the  parameters $a/l^2 (\propto B)$ and $\eta$, which are both assumed small with finite ratio $l^2\eta/a^2$; $a$ here is a crystalline lattice period, which we henceforth set to one for convenience.
To the leading order in $l^{-2}$ and $\eta$, the effective Hamiltonian  is the Peierls-Onsager Hamiltonian: $\mathfrak{H}_{0}(\boldsymbol{K}):=\epsilon(\boldsymbol{K})$. This is just the Weyl-symmetric Peierls substitution $\bk \rightarrow \bK$ for the dispersion ($\epsilon(\boldsymbol{k})$) of the $D$-fold degenerate band of $\hat{H}_0(\bk)$, and $\boldsymbol{K}=\boldsymbol{k}+(e/\hbar)\boldsymbol{A}(i\nabla_{\boldsymbol{k}})$ here are the kinematic quasimomentum operators.

The subleading term $\mathfrak{H}_{1}(\boldsymbol{K})$ (denoted $\H$ in the main text) of the effective Hamiltonian has two additive components: the first is obtained by Peierls substitution of the generalized Zeeman interaction (in the absence of $\delta \hat{H}$), which includes the Zeeman coupling to orbital and spin moments, as well as a geometric Berry contribution. The second component is the Peierls-substitution of   $\delta \epsilon(\bk)$, which is the projection of $\hat{H}_1$ to the $D$-fold degenerate bands of $\hat{H}_0$. In combination,
\begin{equation}
\effH_1(\bk) = \eta\delta \epsilon(\bk)+B(M_{z}-g_s\mu_{B}s_{z}/\hbar+e\epsilon_{\alpha\beta}\mathfrak{X}_{\beta}v_{\alpha}). \label{effham1}
\end{equation}
$M_z$, $s_z$, $\mathfrak{X}$ and $v$ in Eq. (\ref{effham1}) should be evaluated with wavefunctions in the degenerate subspace of $\hat{H}_0$. The above equation presupposes that the  magnetic field $B$ lies in the $-z$ direction.

Employing the Landau gauge for Eq. (\ref{eq:schrodinger}) with $k_x$ a good quantum number, we seek a semiclassical solution in the WKB approximation. To the leading order (i.e., with $\mathfrak{H}_0(\bK)$ only), solutions of Eq. (\ref{eq:schrodinger}) are Zil'berman functions\cite{zilberman} labelled by the wavevector $k_x^0$:
\begin{equation}
g_{\bk}^\nu=\frac{1}{\sqrt{|v_x^\nu|}}e^{il^2k_x^0k_y}e^{-il^{2}\int k_x^\nu dk_y}\delta_{k_x^0 k_x},\la{zilberman}
\end{equation}
where $k_x^\nu$ is a function of $k_y$ satisfying   $\epsilon(k_x^\nu,k_y)=E$. The multiple solutions to $\epsilon(k_x^\nu,k_y)=E$ are indexed by $\nu$. Quantities   with the superscript $\nu$ (e.g., $v_x^\nu(\bk)$ in \q{zilberman}) should be evaluated at  $\bk = (k_x^\nu, k_y)$.

%The solution to the  effective Schr\"odinger equation is defined by imposing a hard-wall boundary condition [on $f_{\bk} = \sum_\nu g_{\bk}^\nu$] at the classical turning points. In this manner we obtain  the  quantization rule for cyclotron motion.

Similar to what is done in Ref. \onlinecite{100p}, we seek the solution to the Schr\"odinger equation using the following multicomponent wave function ansatz
\begin{equation}
\boldsymbol{f}^\nu=\A^\nu\boldsymbol{g}^\nu,\label{wkb-wf}
\end{equation}
where 
\begin{equation}
g_a^\nu=c_{a}^\nu\frac{1}{\sqrt{|v_x^\nu|}}e^{ik^{0}_{x}k_{y}l^{2}}e^{-il^{2}\int k_{x}^\nu dk_{y}}\delta_{k^{0}_{x}k_{x}},~a\in{1..D}
\end{equation}
and $c_a^\nu$ is $\bk$ independent. $\A^\nu$ is a $k_y$ dependent square matrix with the assumption $\A_{ab}^\nu$ is of order $\order(1, \eta l^2):=\order(\{1,\eta l^2\})$. Following section V A 2 of Ref. \onlinecite{100p},:
\e{
[\effH_{0}(\boldsymbol{K})]_{ab}\A_{bc}^\nu g_{c}^\nu=&\A_{ac}^\nu\epsilon_0(\bK)g_{c}^\nu+i\hbar l^{-2} v_{x}^\nu(\partial_{y}\A_{ac}^\nu)g_{c}^\nu\nonumber\\
&+\text{o}(l^{-2}, \eta),
}
and
\begin{equation}
[\effH_{1}(\bK)]_{ab}\A_{bc}^\nu g_{c}^\nu=[\effH_{1}]_{ab}\A_{bc}^\nu g_{c}^\nu+\text{o}(l^{-2}, \eta).\label{effH1onA}
\end{equation}
One term omitted in Eq. (\ref{effH1onA}) in $\text{o}(l^{-2},\eta)$ is $i l^{-2}(\partial_{y}A_{ac}^\nu)g_{c}^\nu d\effH_1/dk_x$. This omission can be justified if terms in Eq. ({\ref{effham1}}), including the change of velocity due to $\delta\hat{H}$, are not anomalously large.

Sch\"rodinger equation then reads 
\begin{equation}
i\hbar l^{-2} v_{x}^\nu (\partial_{y}\A_{ac}^\nu)  g_{c}^\nu+[\effH_{1}]_{ab}\A_{bc}^\nu g_{c}^\nu=\text{o}(l^{-2}, \eta).
\end{equation}
For the above equation to hold for any $\mathbf{c}$,
\begin{equation}
\hbar\partial_{y}\A_{ac}^\nu=il^{2}[v_{x}^\nu]^{-1}[\effH_{1}]_{ab}\A_{bc}^\nu+{\text{o}(1, \eta l^2)}.
\end{equation}
The solution to this differential equation is a path-ordered exponential:
\begin{equation}
\A^\nu=\overline{\exp}[il^{2}\int\frac{\effH_{1}}{\hbar v_{x}^\nu}dk_{y}]+{\text{o}(1, \eta l^2)},
\end{equation}

%Eq. ({\ref{eq:prop}}) can be obtained by solving boundary conditions at turning points. 

By imposing a hard-wall boundary condition at classical turning points, and requiring that the wavefunction in Eq. (\ref{wkb-wf}) be single-valued,\cite{100p} we obtain the quantization rule  stated in \qq{eq:rule}{berryconn}; in \q{eq:rule}, $a$ now runs from $1$ to $D$. Due to the geometric Berry term, Eq. ({\ref{eq:prop}}) transforms covariantly under basis transformations within the degenerate subspace of $\hat{H}_0$ [See Eq. (133) of Ref. \onlinecite{100p}], and as a consequence $\{\lambda_a\}_{a=1}^D$ is well-defined modulo $2\pi$.

Since $\lambda_a$ is the eigenphase of $\A$ over a full cyclotron orbit, it carries the same uncertainty $\text{o}(1, \eta l^2)$. For example, in the Rashba model with an out-of-plane field, a comparison with the exact solution shows that our quantization rule misses terms of the order $\eta^2 l^2$; precisely, the missed terms are each proportional to  $l^2\delta S^2/S$ in Eq. (\ref{eq:Rashba-exact})]. These missed terms are small under our assumption of small $\eta$ and small $l^{-2}$ with finite ratio $l^2\eta$. We remark that $l^2\delta S^2/S\ll 1$, along with the standard semiclassical condition, sets a double-sided constraint on the magnetic field: $S^{-1} \ll l^2 \ll S/(\delta S)^2$. 

%The quantization rule thus requires an extra condition $l^2\delta S^2/S\ll 1$, which is an upper bound for Landau level index: $n\ll(S/\delta S)^2$.

Besides the effective-Hamiltonian approach described here, semiclassical wave packet theory has been developed for multiple coupled bands\cite{culcer_coherent_2005}. It is possible that quantization rule [of \qq{eq:rule}{berryconn}] can be alternatively derived by a `requantization' of the wave packet theory\cite{xiao_berry_2010}; we leave this for future investigation.  

\section{Comparison of numerically-exact Landau levels with our approximation scheme}\label{app:approximatevsexact}

\begin{figure}
\includegraphics[width=1.0\columnwidth]{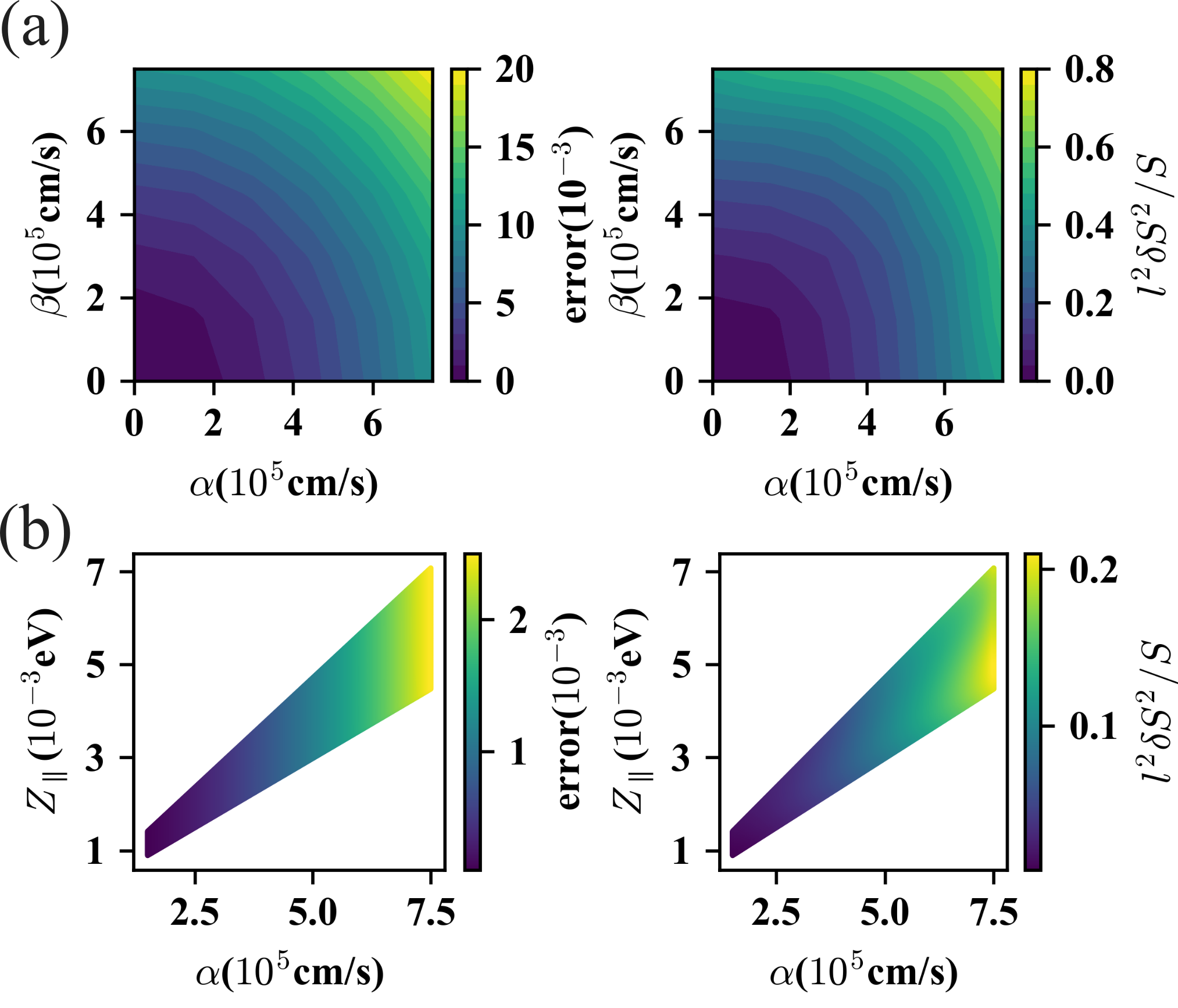}
\caption{{Comparison of quantization rule and exact diagonalization. (a) For various spin-orbit parameters $(\alpha,\beta)$ of the Rashba-Dresselhaus model, we plot in left panel $\text{error}:=|\epsilon_\text{rule}-\epsilon_\text{exact}|/\var_c$ for a single Landau level that lies closest to energy $E=0.4eV$, where $\epsilon_\text{rule}$ is determined by Eq. (\ref{eq:rule}) and $\epsilon_{\text{exact}}$ is obtained by numerical diagonalization of the Peierls-Onsager Hamiltonian $H_{RD}(\bK)$ plus the Zeeman coupling to spin. We also plot in right panel $l^2\delta S^2/S$, which has a strong positive correlation with the error. (b) For various parametrizations ($\alpha$, $Z_\parallel$) of $H_{RZ}$, we plot in the left panel the error for a single Landau level that is closest to $\epsilon_0$ [as defined in Eq. (\ref{whereisdiracpoint})] $\epsilon_\text{rule}$ is determined by Eq. (\ref{eq:rule}) and $\epsilon_{\text{exact}}$ is obtained by numerical diagonalization of the Peierls-Onsager Hamiltonian $H_{RZ}(\bK)$ plus the Zeeman coupling to spin. The corresponding $l^2 \delta S^2/S$ is plotted in the right panel. For numerical convergence\cite{RZfignote}, we have presented data only in a strip [within ($\alpha$, $Z_\parallel$)-space] where $0.4{<}\epsilon_0{<}1.0$(eV).Parameters for (a): $m=0.076m_0$, $l=200$\AA~and $g_{s\perp}=2$; parameters for (d): $m=0.076m_0$, $l=100$\AA~and $g_{s\perp}=2$.}\label{fig:error}}
\end{figure}

In this appendix, we compare the exact Landau levels obtained from numerical diagonalization (with standard techniques reviewed in App. \ref{sec:numerical}) to the approximate Landau levels obtained from our quantization rule, as summarized in \qq{eq:rule}{berryconn}. The comparison shall be made for the Rashba-Dresselhaus 2DEG in an out-of-plane field [see Fig.\ \ref{fig:error}(a)], and for the Rashba 2DEG in a tilted field [see Fig.\ \ref{fig:error}(b)]. In both cases, a closed-form analytic expression for the Landau-level spectrum is not known. We quantify the error of our quantization rule by $\text{error}:=|\epsilon_\text{rule}-\epsilon_\text{exact}|/\var_c$ for different parameter choices in both models. The semiclassical ($l^2 S\gg 1$) and near-degeneracy conditions ($\delta S/S\ll 1$) are satisfied within the parameter range in Fig. \ref{fig:error}. Just as for the Rashba 2DEG in an out-of-plane field [cf. Eq.(\ref{eq:Rashba-exact})], the accuracy of our rule improves with decreasing parameter $l^2\delta S^2/S$,  suggesting that this parameter gives the dominant contribution to the error of the quantization rule [cf. the discussion in App. \ref{app:quantizationruleproof}].

\subsection{Method of numerical diagonalization}\label{sec:numerical}

Numerical diagonalization relies on  $[K_x,K_y]=il^{-2}$ being algebraically similar  to the commutation relation  of position and momentum (in quantum mechanics with quantum parameter $\hbar$). Analogously, $K_x$ and $K_y$ can expressed with creation and annihilation operators
\begin{align}
\begin{split}
K_x=&l^{-1}(a+a^\dagger)/\sqrt{2},\\
K_y=&-il^{-1}(a-a^\dagger)/\sqrt{2}.
\end{split}
\end{align}
Performing the above substitution for  $H_R(\bK)$ (with the Zeeman coupling to spin),
\begin{align}
\begin{split}
H_{R}(\boldsymbol{K})=&\frac{\hbar^2}{ml^2}(a^{\dagger}a+\frac{1}{2})
\\&+\sqrt{2}\hbar\alpha l^{-1}\left(\begin{array}{cc}
0 & -ia^{\dagger}\\
ia & 0
\end{array}\right)
-\frac{g_{s\perp}\hbar^2\sigma^z}{4m_0l^2}.
\end{split}\label{eq:Rashba-ho}
\end{align}
Eq. (\ref{eq:Rashba-ho}) have solutions in the form of $(c_1|n+1\rangle,c_2|n\rangle)$ where $c_1$ and $c_2$ are some constants and $|n\rangle$ is an eigenstate of $a^\dagger a$ with eigenvalue $n$. Any Hamiltonian $H(\bK)$ can be numerically diagonalized with a  cutoff for the index $n$, which we assume to be large.

%\subsection{Rashba 2DEG in a tilted field}\la{app:rashba2DEGtilted_comparenumerics}

%In \s{sec:inplanezeeman}, we focused on a parameter regime ($\hbar\alpha k_{\sma{E}}{\gg}\var_c$) of the Rashba 2DEG (in tilted field), where the nearly-degenerate spin-split orbits are negligibly hybridized by the Zeeman interaction -- except near a type-II Dirac point. 

\section{Existence of Landau-level quasidegeneracy\label{sec:proofLLquasideg}}

For our quantization rules to be consistent, the following inequalities are necessary:
\e{ |l^2S| \gg |\lambda_a|,\, \left|l^2\p{S}{E}\right| \gg \left|\p{\lambda_a}{E}\right|,\, \left|S\right| \gg \left|\p{\lambda_a}{l^2}\right|.} 
Due to the latter two inequalities, we shall refer to $\lambda_{a}$ as \textit{slowly-varying} (with respect to $E$ and $B$), in comparison with the rapidly-varying $l^2S$. 

To appreciate the difference in scales, consider that for $\lambda_a$ to change by $2\pi$ due to a variation $l^2$ (at fixed energy), we estimate from \q{phaseindependentorbit} that $l^2$ must change on the scale of $2\pi/\delta S$; this scale is much greater than the period of quantum oscillations ($T_{l^2}{:}{=}2\pi/S$), by our assumption of near degeneracy. The above estimate was carried out in the weak-hybridization regime, where the Zeeman interaction is much weaker than the spin-orbit-splitting of energy bands. In the strong-hybridization regime, $\lambda$ is independent of $l^2$ to the accuracy of our theory.

Analogously, for $\lambda_a$ to vary by $2\pi$ due to a variation of $E\rightarrow E+\delta E$ (at fixed field), we estimate in the weak-hybridization regime [cf.\ \q{phaseindependentorbit}] that $\delta E {\sim} 2\pi/l^2 (\partial \delta S/\partial E)$, which is generically much larger than the cyclotron energy $\epsilon_c{=}2\pi/l^2 (\partial  S/\partial E)$. In the strong-hybridization regime, $\lambda_a$ depends on $E$ solely through the slow variation of $\calm$.  In the absence of symmetry, we estimate 
\e{\p{\lambda_{a}}{E}=\order\left(\f{g_s}{m_0}\p{m}{E},\f{\partial^2S/\partial E^2}{\partial S/\partial E} \right)\ll \f{2\pi}{\var_c}, \la{dlambdadE}} 
where $m_0$ is the free-electron mass, and $m{=}\hbar^2(\partial S/\partial E)/2\pi$ the effective mass. The first argument of $\order$ originates from the energy dependence of the effective $g$-factor; the second argument accounts for possible non-analyticities in the area of $\frako_0$, which may originate from saddle- or II-Dirac points in the zeroth-order Hamiltonian $\hat{H}_0$.

Armed with the above inequalities, we now demonstrate that, if a degeneracy of $\cala$ [$e^{i\lambda_+}{=}e^{i\lambda_-}$] occurs at $(\bar{E},\bar{B})$, there exists two Landau levels in close proximity to $(\bar{E},\bar{B})$ which satisfy the nearly-degeneracy conditions defined in \qq{llquasideg}{llquasidegB}.

To begin, it is useful to define
\e{ \calq_{\pm}(E,B):= \f1{2\pi}\left(l^2S +\lambda_{\pm} -\gamma \right)}
such that the quantization rule \q{eq:rule} is satisfied if $\calq_{\pm}(E,B){\in}\Z$. 
Generically, either of $m_{\pm}:=\calq_{\pm}(\bar{E},\bar{B})\in \R$ is not integer-valued, but 
\e{\big(\calq_{+}-\calq_{-}\big)\bigg|_{\bar{E},\bar{B}}=\f{\lambda_+-\lambda_-}{2\pi}\bigg|_{\bar{E},\bar{B}} \in \Z. \label{diffinteger}} 
Let $n_{\pm}$ be the closest integer to $m_{\pm}$ (this implies $|n_{\pm}-m_{\pm}|<1$).  From \q{diffinteger}, we deduce that $n_+-m_+$ is equal to $n_--m_-$; this quantity is henceforth denoted as $r:= n_{\pm}-m_{\pm}$, and  $\calq_{\pm}(\bar{E},\bar{B}){+}r{\in} \Z$. 

Let us first tackle \q{llquasideg}. We define $E_{\pm}$ such that $\calq_{\pm}(E_{\pm},\bar{B}){=}\calq_{\pm}(\bar{E},\bar{B}){+}r{\in} \Z$. Since  $\lambda_{\pm}$ is a slowly varying function of $E$ relative to $l^2S$, $|E_{\pm}{-}\bar{E}|/\var_c{\approx} |r|{<}1$, with the cyclotron energy $\var_c=2\pi/l^2|\partial S/\partial E|$ evaluated at $(\bar{E},\bar{B})$. Let us denote $O':=\partial O/\partial E$ evaluated at $(\bar{E},\bar{B})$. We estimate
\e{\f{E_+-E_-}{\var_c} \approx &\; \f{r}{\var_c}\left(\f1{\calq'_+}-\f1{\calq'_-}\right)\approx  \f{r}{2\pi}\var_c \left( \lambda'_- -\lambda'_+\right). \label{estimate}}
Combining \q{estimate} with the above estimates for $\lambda'_{\pm}$, we derive \q{llquasideg}.  
%From \s{sec:qtznrules}, we know that for sufficiently weak field ($\var_c\ll \Delta$), $\lambda_{\pm}'\approx \pm l^2\delta S'/2$. For $\var_c\gg \Delta$, $\lambda_{\pm}'$ is determined by the energy dependence of the field-dependent terms in the effective Hamiltonian $\calh$.  

Let us next tackle \q{llquasidegB}. It is most convenient at this point to change variables from $B\rightarrow l^2$ (the square of the magnetic length). Let us  define   $l^2_{\pm}$ such that $\calq_{\pm}(\bar{E},l^2_{\pm}){=}\calq_{\pm}(\bar{E},\bar{l}^2){+}r{\in} \Z$. We denote  $\dot{O}:=\partial O/\partial(l^2)$ evaluated at $(\bar{E},\bar{B})$, and estimate
\e{\f{l^2_+-l^2_-}{T_{l^2}} \approx  \f{r }{2\pi}T_{l^2} \left( \dot{\lambda}_- -\dot{\lambda}_+\right). \label{estimateB}}
Combining \q{estimateB} with the above estimates for $\dot{\lambda}_{\pm}$, we derive \q{llquasidegB}.

%The order-of-magnitude estimate is obtained in the limit $l^2\delta S \gg 1$, where $\lambda_{\pm}\approx \pm l^2\delta S/2$ [cf.\ \q{phaseindependentorbit}]. In the opposite limit $l^2\delta S \ll 1$, $\lambda_{\pm}$ is independent of $B$ within the accuracy of our semiclassical theory.    

\section{Codimension of Landau level quasidegeneracy}\label{app:codimension}

\begin{table*}[t]
\begin{tabular*}{2\columnwidth}{c@{\extracolsep{\fill}}ccccccc}
\hlineB{2}
class label & $u$ & $s$ & symmetry constraint & $\det{\A}$ & $\breve{g}$ & codimension \\
\hline
I-1 & 0 & 0 & $\A=\breve{g}\A\breve{g}^{-1}$ & - & $\breve{g}\propto I$ & 3  \\
&  &  &  & - & $\breve{g} \not\propto I$ & 1  \\
I-2 & 0 & 1 & $\A=\breve{g}\A^*\breve{g}^{-1}$ & 1 & $(\breve{g}K)^2=I$ & 1 \\
&  &  &  & 1 & $(\breve{g}K)^2=-I$ & 3 \\
&  &  &  & -1 & $(\breve{g}K)^2=I$ & $\infty$ \\
&  &  &  & -1 & $(\breve{g}K)^2=-I$ & $\times$ \\
II-1 & 0 & 0 & $\A=\breve{g}(g^{L-1}\circ\bk_0) \A_{1/L}^{L}$ & - & - & 1 \\
II-2 & 1 & 1 & $\A=\breve{g}\A^T\breve{g}^{-1}$ & - & $(\breve{g}K)^2=I$ & 2 \\
& & & & - & $(\breve{g}K)^2\ne I$ & 0 \\
II-3 & 0 & 1 & $\A=\breve{g}(g^{L-1}\circ\bk_0) (K^i\A_{1/L}K^i)^L$ & - & $L=N$ & 1 \\
& & & & 1 & $L\ne N$ & 1 \\
& & & & -1 & $L\ne N$, $(\breve{g}K)^2=I$ & $\infty$ \\
& & & & -1 & $L\ne N$, $(\breve{g}K)^2=-I$ & $\times$ \\
II-4 & 1 & 0 & $\A=\breve{g}\A^\dagger\breve{g}^{-1}$ & 1 &$\breve{g} \propto \sigma_z$& 2\\
& & & & 1 &$\breve{g} \not\propto \sigma_z$ & 0\\
& & & & -1 &$\breve{g} \propto I$ & 2\\
& & & & -1 &$\breve{g} \not\propto I$& 0\\
III-1 & 0 & 0 & $\A_2=\breve{g}\A_1\breve{g}^{-1}$ & - & - & 3 \\
III-2 & 1 & 1 & $\A_2=\breve{g}\A_1^T\breve{g}^{-1}$ & - & - & 3 & \\
III-3 & 0 & 1 & $\A_2=\breve{g}\A_1^*\breve{g}^{-1}$ & - & - & 3\\
III-4 & 1 & 0 & $\A_2=\breve{g}\A_1^\dagger\breve{g}^{-1}$ & - & - & 3\\
\hlineB{2}
\end{tabular*}
\caption{Codimensions of degeneracy manifold of $\A$ under constraints of different type of symmetries. Most of the symmetry constraints on $\A$ are reproduced from Ref. \onlinecite{100p} except for class II-1 and II-3, where a special gauge is chosen. $\infty$ in the codimension column denotes that eigenvalues of $\A$ cannot be degenerate; $\times$ denotes an entry where we cannot self-consistently impose the symmetry constraint and the conditions on det$\A$ and $\breve{g}$ for a two-by-two unitary $\A$. In the second to last column, $\breve{g}$ is either chosen to be diagonal or off-diagonal.\label{table:fullcodimension}}
\end{table*}

In this appendix, we perform a symmetry analysis of the two-by-two matrix  propagator $\A$  [defined in \qq{eq:prop}{berryconn}], and derive the symmetry-constrained codimensions of the eigenvalue-degeneracies of $\A$.

For a symmetry $g$ which acts on spacetime as $\br \rightarrow \check{g} \br$ and $t\rightarrow (-1)^{s(g)}t$, 
we define $\hat{g}$ as the corresponding operator  on wave functions: $\hat{g} \psi(\br,s)=\sum_{s'}\psi(\check{g}^{-1} \br,s')[D_g]_{s's}$, with $D_g$ the spinor representation of $g$. The sewing matrix $\breve{g}$ is defined as 
\e{
   \breve{g}(\bk)_{nm} :\eq \langle u_{n,g\circ\bk}|\hat{g}(\bk)|u_{m,\bk} \rangle K^{s(g)}, \lin
   \hat{g}(\bk):\eq e^{-i\bk\cdot\hat{\br}}\hat{g}e^{i\bk\cdot\hat{\br}}.
}
Here, $n=1,2$ and $m=1,2$ runs over the degenerate subspace of $\hat{H}_0$; $K^1:=K$ which implements complex conjugation, while $K^0$ is the identity operation. By a basis transformation in the degenerate subspace: 
\e{
|u_{n\bk}\rangle \to \sum_{m=1}^2|u_{m\bk}\rangle V_{mn}(\bk),
}
$\breve{g}$ transforms as
\e{
\breve{g}(\bk) \to V^\dagger(g\circ \bk)\breve{g}(\bk)K^{s(g)}V(\bk)K^{s(g)}.
}
This shall be referred to as a gauge transformation.
Any unitary $\hat{g}$ can be diagonalized by a gauge transformation, such that  $\hat{g}(\bk)$ maps $|u_n\rangle$ to itself, up to a phase factor which we identify as the symmetry eigenvalue. We define order $N$ of $g$ as the smallest positive integer such that $g^N$ is either the identity, or any multiplicative combination of  a $2\pi$ rotation (which acts nontrivially on half-integer-spin representations) and a lattice translation.

The codimension calculation can be  simplified by choosing a convenient gauge; the codimension is of course gauge-invariant. Generally, we have the following rules for two-by-two sewing matrix:
\begin{itemize}
\item If $g\circ\bk\ne\bk$, $\breve{g}(\bk)$ can be transformed into identity matrix by a gauge transformation.
\item If $g\circ\bk=\bk$ and $\hat{g}$ is unitary, $\breve{g}(\bk)$ can be transformed into a diagonal matrix by a gauge transformation.
\item If $g\circ\bk=\bk$ and $\hat{g}$ is antiunitary, $\breve{g}(\bk)$ can be transformed into an off-diagonal matrix (diagonal terms are 0) by a gauge transformation.  For the special case of $\hat{g}$ that is order-2, $\breve{g}$ can be transformed into either $\sigma_x$  or $i\sigma_y$.
\end{itemize}
The first two claims are simple to show, and the last claim is verified  in App. \ref{app:makinggoffdiagonal} below.

The symmetry constraints on $\A$ expressed using sewing matrix is presented in Ref. \onlinecite{100p} and reproduced in Table \ref{table:fullcodimension} for convenience. Generally, these constraints are derived using the following relation
\e{
&\A[g\circ \bk_f\leftarrow g\circ \bk_i] \nonumber\\
&=e^{i\phi}\breve{g}(\bk_f)K^{s(g)}\A[\bk_f\leftarrow \bk_i]K^{s(g)}\breve{g}^{-1}(\bk_i),\label{eq:propsymm}
}
where $\A[\bk_f\leftarrow \bk_i]$ is a segment of the propagator from $\bk_i$ to $\bk_f$ along the cyclotron orbit. $e^{i\phi}$ is a phase factor that appears for $g$ that is a nonsymmorphic element; however this phase always drops out\cite{100p} for closed orbits and will be neglected in the subsequent analysis.

For four of  the ten classes (III 1-4), the symmetry relates the propagators of distinct and disconnected orbits, but each propagator (corresponding to one orbit) is itself unconstrained.  Therefore, the codimensions are still 3. For the other six classes, self constraints of $\A$ are imposed. In particular, in class I-2, II-3, II-4, $\det(\A)$ is quantized at $\pm 1$\cite{topoferm,100p}.

The self constraints of the six classes can be divided into two types: (i) restricting $\A$ to only a submanifold of SU(2) (class I-1, I-2, II-2, II-4); (ii) expressing the propagator $\A$ as a product of propagators over a fraction of the period [e.g., Eq. (\ref{eq:sigmazconstraint})] (class II-1, II-3). For the former case, we define symmetric coordinates by restricting canonical parametrization of $\A$ to that submanifold; in the latter case, we employ the canonical parametrization of the fractional propagators to define symmetric coordinates for the full propagator $\A$. For symmetry constraints of type (i), $\A$ fulfills an equation expressed with sewing matrix at the base point of $\A$ (denoted as $\bk_0$). Therefore, we parametrize the propagator as
\e{
\A = e^{i\theta}\matrixtwo{\alpha}{\beta}{-\beta^*}{\alpha^*}, |\alpha|^2+|\beta|^2=1.
}
and look for how many free parameters are left after the symmetry constraint is imposed, which is codimension of degeneracy manifold. For symmetry constraints of type (ii) we parametrize the first fractional of the propagator $\A_{1/L}$ as
\e{
\A_{1/L} = e^{i\theta}\matrixtwo{\alpha}{\beta}{-\beta^*}{\alpha^*}, |\alpha|^2+|\beta|^2=1.\label{eq:paramA1/N}
}
Then we calculate the full propagator $\A$ and figure out how many parameters is needed to tune $\A$ such that $\A$ is proportional to identity, which is codimension of degeneracy manifold.

\paragraph*{class I-1} According to the general rules of sewing matrix, the sewing matrix $\breve{g}$ can be made diagonal by a gauge transformation
\e{
\breve{g} = \matrixtwo{e^{i\phi_1}}{0}{0}{e^{i\phi_2}}.
}
In this gauge, the symmetry constraints are expressed as 
\e{
\matrixtwo{\alpha}{\beta}{-\beta^*}{\alpha^*}=\matrixtwo{\alpha}{e^{i(\phi_1-\phi_2)}\beta}{-e^{-i(\phi_1-\phi_2)}\beta^*}{\alpha^*}.
}
Therefore, unless $\breve{g}$ is proportional to identity, $\beta=0$ and the only free variable is the phase of $\alpha$.

\paragraph*{class I-2}
By a gauge transformation, the sewing matrix $\breve{g}$ can be made off-diagonal
\e{
\breve{g} = \matrixtwo{0}{1}{\pm 1}{0}.
}
In this gauge, the symmetry constraints are expressed as
\e{
\det({\A})\matrixtwo{\alpha}{\beta}{-\beta^*}{\alpha^*}=\matrixtwo{\alpha}{\mp\beta}{\pm \beta^*}{\alpha^*}.
}
This immediately yeilds corresponding rows in table. \ref{table:fullcodimension}.

\paragraph*{class II-1} In this class (and also class II-3), the full propagator can be constructed from $1/L$ of it, where $L$ is the smallest positive integer fulfilling $g^L\circ \bk_0=\bk_0$. Here, $\bk_0$ is the base point of $\A$. In this class, $L=N$. A good example in this class is $\mathfrak{c}_{Nz}$, $N$-fold rotation in $z$ direction. Generally, by gauge transformation at $g^n\circ \bk_0$, $\breve{g}(g^n\circ \bk_0)$ can be transformed to identity for $0 \le n < N-1$ and the full propagator is
\e{
\A=\breve{g}(g^{N-1}\circ\bk_0)\A_{1/N}^N
}
where 
\e{
\breve{g}(g^{N-1}\circ\bk_0)=\prod_{n=0}^{N-1} \breve{g}(g^n\circ\bk_0)
}
is a constant phase factor determined by the group multiplication rule. For example if $g$ is an $N$-fold rotation,  the phase factor is $+1$ (resp.\ $-1$) for an integer-spin (resp.\ half-integer-spin) representation. For the purpose of determining degeneracies, we may focus on
$\A_{1/N}^N$. Using the parametrization in Eq. (\ref{eq:paramA1/N}), we diagonalize $A_{1/N}$ as 
\e{
A_{1/N} = e^{i\theta}U^{\dagger}\matrixtwo{e^{i\varphi}}{0}{0}{e^{-i\varphi}}U
}
with $e^{i\varphi}=\text{Re}(\alpha)+i\sqrt{\text{Im}(\alpha)^2+|\beta|^2}$. The sufficient and necessary condition for $A_{1/N}^N \propto I$ is $\text{Re}(\alpha)=\cos(n\pi/N)$, with $n\in\mathbb{Z}$. Tuning $\text{Re}(\alpha)=\cos(n\pi/N)$ requires only one parameter, except for $n=0$ and $n=N$, where three parameters are required.

In the Sec. \ref{sec:llquasideg}, we construct $\mathfrak{c}_N$ models by perturbing from $\mathfrak{c}_\infty$ Rashba model and by coupling two otherwise independent orbits by magnetic breakdown. In both constructions, $\varphi$ evolves monotonically with respect to $B$ before perturbation are turned on. $\varphi$ sweeps through $n\pi/N$ where degeneracies of $\A$ are expected. $e^{i\varphi}$ hits $\pm 1$ once in every $N$ degeneracies and therefore the other $N-1$ degenerates are stable, i.e., have codimension 1.

\paragraph*{class II-3} In class II-1, $L=N$ is guaranteed. However, in class II-3, this is generally not true. For symmetry operations still respecting $L=N$, the codimension calculation is similar to what is done in class II-1. The only example of $L\ne N$ is $T\mathfrak{c}_{6z,n\boldsymbol{c}/6}$ ($n\in\{0,1,2,3,4,5\}$), where $L=3$ and $N=6$. We write $g=T\mathfrak{c}_{6z,n\boldsymbol{c}/6}$ and $h=g^3$. By gauge transformation, it is possible to make $\breve{g}(\bk_0)=\breve{g}(g\circ\bk_0)=I$ and $\breve{h}(\bk_0)=\breve{g}(g^2\circ \bk_0)\equiv\breve{h}$. In this gauge, $\breve{h}(g\circ\bk_0)=h^*$. Therefore,
\e{
\A=\breve{h}\A_{1/L}\A_{1/L}^*\A_{1/L}
}
and $\A_{1/L}$ is constrained due to $h$ as 
\e{
\A_{1/L}=\breve{h}^* \A_{1/L} \breve{h}^{-1}.\label{eq:TC6A1/L}
}
Since $h$ is of order two, $\breve{h}$ can either be chosen as $\sigma_x$ or $i\sigma_y$. Then we immediately get $\det \A_{1/L} = \pm 1$. We first study the case $\breve{h}=\sigma_x$. If $\det \A_{1/L} = 1$, we parametrize $\det \A_{1/L}$ as in Eq. (\ref{eq:paramA1/N}) with $\theta=0$. From Eq. (\ref{eq:TC6A1/L}), we get $\beta=0$ and thus 
\e{
\A = \matrixtwo{0}{\alpha^*}{\alpha}{0}
}
and eigenphases of $\A$ are fixed to $0$ and $\pi$. If $\det{\A_{1/L}}=-1$, $\theta=\pi/2$ in Eq.(\ref{eq:paramA1/N}) and we obtain
\e{
\A = \matrixtwo{\beta^{*3}}{0}{0}{\beta^3}
}
and the codimension is 1. If $h=i\sigma_y$, from the fourth entry of class I-2, we deduce $\det \A_{1/L}=1$. Choosing the parametrization in Eq. (\ref{eq:paramA1/N}) with $\theta=0$, we compute the codimension to be 1.

\paragraph*{class II-2 and II-4} Calculation of codimensions for the two classes follow the same pattern as class I-1 and I-2 and is therefore omitted here.

Assuming that the spin-orbit coupling can be turned off continuously without changing the degree ($D=2$) of degeneracy, $\det(A)=1$ for class I-2, II-3, II-4, and thus Tab. \ref{table:fullcodimension} simplified to Tab. \ref{table:codimension-nearlydegen}-\ref{table:codimension-exactlydegen}.

\subsection{Two-by-two sewing matrix of antiunitary symmetry operator can be made off-diagonal\label{app:makinggoffdiagonal}}

Here we derive the third rule for gauge transformations (the third bulleted point in the previous subsection). We will show a gauge exists where $\breve{g}$ (the sewing matrix of an antiunitary symmetry, at a $g$-invariant wavevector) is off-diagonal. We will arrive at this gauge by a sequence of gauge transformations: $\breve{g}\rightarrow \dg{V}\breve{g}V^*$.

\noi{i} The square of an antiunitary operator is unitary, hence $\breve{g}\breve{g}^* \rightarrow V^{-1}\breve{g}\breve{g}^*V$ can be diagonalized. In this diagonalized gauge, it is simple to see that either $\breve{g}$ is off-diagonal or it is a (complex) symmetric unitary matrix. In the former case, our job is done. Let us henceforth deal with the latter.

\noi{ii} Any symmetric unitary matrix can be diagonalized by conjugation with a real orthogonal matrix,\footnote{To appreciate this, consider that any unitary matrix can be expressed as $e^{ih}$ with Hermitian $h$. If this unitary is also symmetric, $h$ must also by symmetric, hence also real. Thus the eigenvectors of a symmetric unitary can always be chosen to be real.} which we identify with $V$ in the next gauge transformation. Henceforth we assume $\breve{g}$ is diagonal. 

\noi{iii} By an appropriately chosen $V$ that is diagonal, we can always simplify $\breve{g}$ to the identity matrix.

\noi{iv} Finally, with $V=e^{-i(\pi/4)\sigma_x}$, we obtain $\breve{g}=i\sigma_x$, with $\sigma_x$ an off-diagonal Pauli matrix. This completes the proof.\\

In the special case that $g$ is order-two, then the associativity condition for $g$ imposes that $\breve{g}\breve{g}^*=\pm I$. In the case of $+I$, then $\breve{g}$ is symmetric, and we may apply (ii-iv) above to simplify $\breve{g}=i\sigma_x$; this can be made real by a final gauge transformation with $V=e^{i\pi/4}\sigma_x$. In the case where $\breve{g}\breve{g}^*=-I$, then $\breve{g}$ must be skew-symmetric (under transpose). Any skew-symmetric unitary matrix is proportional to $i\sigma_y$, up to a phase factor that can be gauged away.

\bibliography{paper}

\begin{thebibliography}{93}%
\makeatletter
\providecommand \@ifxundefined [1]{%
 \@ifx{#1\undefined}
}%
\providecommand \@ifnum [1]{%
 \ifnum #1\expandafter \@firstoftwo
 \else \expandafter \@secondoftwo
 \fi
}%
\providecommand \@ifx [1]{%
 \ifx #1\expandafter \@firstoftwo
 \else \expandafter \@secondoftwo
 \fi
}%
\providecommand \natexlab [1]{#1}%
\providecommand \enquote  [1]{``#1''}%
\providecommand \bibnamefont  [1]{#1}%
\providecommand \bibfnamefont [1]{#1}%
\providecommand \citenamefont [1]{#1}%
\providecommand \href@noop [0]{\@secondoftwo}%
\providecommand \href [0]{\begingroup \@sanitize@url \@href}%
\providecommand \@href[1]{\@@startlink{#1}\@@href}%
\providecommand \@@href[1]{\endgroup#1\@@endlink}%
\providecommand \@sanitize@url [0]{\catcode `\\12\catcode `\$12\catcode
  `\&12\catcode `\#12\catcode `\^12\catcode `\_12\catcode `\%12\relax}%
\providecommand \@@startlink[1]{}%
\providecommand \@@endlink[0]{}%
\providecommand \url  [0]{\begingroup\@sanitize@url \@url }%
\providecommand \@url [1]{\endgroup\@href {#1}{\urlprefix }}%
\providecommand \urlprefix  [0]{URL }%
\providecommand \Eprint [0]{\href }%
\providecommand \doibase [0]{http://dx.doi.org/}%
\providecommand \selectlanguage [0]{\@gobble}%
\providecommand \bibinfo  [0]{\@secondoftwo}%
\providecommand \bibfield  [0]{\@secondoftwo}%
\providecommand \translation [1]{[#1]}%
\providecommand \BibitemOpen [0]{}%
\providecommand \bibitemStop [0]{}%
\providecommand \bibitemNoStop [0]{.\EOS\space}%
\providecommand \EOS [0]{\spacefactor3000\relax}%
\providecommand \BibitemShut  [1]{\csname bibitem#1\endcsname}%
\let\auto@bib@innerbib\@empty
%</preamble>
\bibitem [{\citenamefont {Shubnikov}\ and\ \citenamefont
  {de~Haas}(1930)}]{SdH}%
  \BibitemOpen
  \bibfield  {author} {\bibinfo {author} {\bibfnamefont {L.~W.}\ \bibnamefont
  {Shubnikov}}\ and\ \bibinfo {author} {\bibfnamefont {W.~J.}\ \bibnamefont
  {de~Haas}},\ }\href@noop {} {\bibfield  {journal} {\bibinfo  {journal} {Proc.
  Netherlands Roy. Acad. Sci.}\ }\textbf {\bibinfo {volume} {33}},\ \bibinfo
  {pages} {130} (\bibinfo {year} {1930})}\BibitemShut {NoStop}%
\bibitem [{\citenamefont {de~Haas}\ and\ \citenamefont {van
  Alphen}(1930)}]{dHvA}%
  \BibitemOpen
  \bibfield  {author} {\bibinfo {author} {\bibfnamefont {W.~J.}\ \bibnamefont
  {de~Haas}}\ and\ \bibinfo {author} {\bibfnamefont {P.~M.}\ \bibnamefont {van
  Alphen}},\ }\href@noop {} {\bibfield  {journal} {\bibinfo  {journal} {Proc.
  Netherlands Roy. Acad. Sci.}\ }\textbf {\bibinfo {volume} {33}},\ \bibinfo
  {pages} {1106} (\bibinfo {year} {1930})}\BibitemShut {NoStop}%
\bibitem [{\citenamefont {Onsager}(1952)}]{Onsager}%
  \BibitemOpen
  \bibfield  {author} {\bibinfo {author} {\bibfnamefont {L.}~\bibnamefont
  {Onsager}},\ }\href {\doibase 10.1080/14786440908521019} {\bibfield
  {journal} {\bibinfo  {journal} {The London, Edinburgh, and Dublin
  Philosophical Magazine and Journal of Science}\ }\textbf {\bibinfo {volume}
  {43}},\ \bibinfo {pages} {1006} (\bibinfo {year} {1952})}\BibitemShut
  {NoStop}%
\bibitem [{\citenamefont {Lifshitz}\ and\ \citenamefont
  {Kosevich}(1954)}]{lifshitz_kosevich}%
  \BibitemOpen
  \bibfield  {author} {\bibinfo {author} {\bibfnamefont {L.~M.}\ \bibnamefont
  {Lifshitz}}\ and\ \bibinfo {author} {\bibfnamefont {A.}~\bibnamefont
  {Kosevich}},\ }\href@noop {} {\bibfield  {journal} {\bibinfo  {journal}
  {Dokl. Akad. Nauk SSSR}\ }\textbf {\bibinfo {volume} {96}},\ \bibinfo {pages}
  {963} (\bibinfo {year} {1954})}\BibitemShut {NoStop}%
\bibitem [{\citenamefont {Lifshitz}\ and\ \citenamefont
  {Kosevich}(1956)}]{lifshitz_kosevich_jetp}%
  \BibitemOpen
  \bibfield  {author} {\bibinfo {author} {\bibfnamefont {L.~M.}\ \bibnamefont
  {Lifshitz}}\ and\ \bibinfo {author} {\bibfnamefont {A.}~\bibnamefont
  {Kosevich}},\ }\href@noop {} {\bibfield  {journal} {\bibinfo  {journal} {Sov.
  Phys. JETP}\ }\textbf {\bibinfo {volume} {2}},\ \bibinfo {pages} {636}
  (\bibinfo {year} {1956})}\BibitemShut {NoStop}%
\bibitem [{\citenamefont {Roth}(1962)}]{rotheffham}%
  \BibitemOpen
  \bibfield  {author} {\bibinfo {author} {\bibfnamefont {L.}~\bibnamefont
  {Roth}},\ }\href {\doibase http://dx.doi.org/10.1016/0022-3697(62)90083-5}
  {\bibfield  {journal} {\bibinfo  {journal} {J. Phys. Chem. Solids}\ }\textbf
  {\bibinfo {volume} {23}},\ \bibinfo {pages} {433 } (\bibinfo {year}
  {1962})}\BibitemShut {NoStop}%
\bibitem [{\citenamefont {Roth}(1966)}]{rothmag}%
  \BibitemOpen
  \bibfield  {author} {\bibinfo {author} {\bibfnamefont {L.~M.}\ \bibnamefont
  {Roth}},\ }\href {\doibase 10.1103/PhysRev.145.434} {\bibfield  {journal}
  {\bibinfo  {journal} {Phys. Rev.}\ }\textbf {\bibinfo {volume} {145}},\
  \bibinfo {pages} {434} (\bibinfo {year} {1966})}\BibitemShut {NoStop}%
\bibitem [{\citenamefont {Mikitik}\ and\ \citenamefont
  {Sharlai}(1998)}]{Mikitik_quantizationrule}%
  \BibitemOpen
  \bibfield  {author} {\bibinfo {author} {\bibfnamefont {G.~P.}\ \bibnamefont
  {Mikitik}}\ and\ \bibinfo {author} {\bibfnamefont {Y.~V.}\ \bibnamefont
  {Sharlai}},\ }\href {\doibase 10.1134/1.558717} {\bibfield  {journal}
  {\bibinfo  {journal} {Sov. Phys. JETP}\ }\textbf {\bibinfo {volume} {87}},\
  \bibinfo {pages} {747} (\bibinfo {year} {1998})}\BibitemShut {NoStop}%
\bibitem [{\citenamefont {Alexandradinata}\ \emph {et~al.}(2018)\citenamefont
  {Alexandradinata}, \citenamefont {Wang}, \citenamefont {Duan},\ and\
  \citenamefont {Glazman}}]{topoferm}%
  \BibitemOpen
  \bibfield  {author} {\bibinfo {author} {\bibfnamefont {A.}~\bibnamefont
  {Alexandradinata}}, \bibinfo {author} {\bibfnamefont {C.}~\bibnamefont
  {Wang}}, \bibinfo {author} {\bibfnamefont {W.}~\bibnamefont {Duan}}, \ and\
  \bibinfo {author} {\bibfnamefont {L.}~\bibnamefont {Glazman}},\ }\href
  {\doibase 10.1103/PhysRevX.8.011027} {\bibfield  {journal} {\bibinfo
  {journal} {Phys. Rev. X}\ }\textbf {\bibinfo {volume} {8}},\ \bibinfo {pages}
  {011027} (\bibinfo {year} {2018})}\BibitemShut {NoStop}%
\bibitem [{\citenamefont {Alexandradinata}\ and\ \citenamefont
  {Glazman}(2018)}]{100p}%
  \BibitemOpen
  \bibfield  {author} {\bibinfo {author} {\bibfnamefont {A.}~\bibnamefont
  {Alexandradinata}}\ and\ \bibinfo {author} {\bibfnamefont {L.}~\bibnamefont
  {Glazman}},\ }\href {\doibase 10.1103/PhysRevB.97.144422} {\bibfield
  {journal} {\bibinfo  {journal} {Phys. Rev. B}\ }\textbf {\bibinfo {volume}
  {97}},\ \bibinfo {pages} {144422} (\bibinfo {year} {2018})}\BibitemShut
  {NoStop}%
\bibitem [{Note1()}]{Note1}%
  \BibitemOpen
  \bibinfo {note} {The expressions for $\delta S$ and $S$ in Eq.\ (\ref
  {deltaSvsS}) are asymptotically valid for $E{\gg }m\alpha ^2$.}\BibitemShut
  {Stop}%
\bibitem [{\citenamefont {Kohn}(1959)}]{kohn_effham}%
  \BibitemOpen
  \bibfield  {author} {\bibinfo {author} {\bibfnamefont {W.}~\bibnamefont
  {Kohn}},\ }\href {\doibase 10.1103/PhysRev.115.1460} {\bibfield  {journal}
  {\bibinfo  {journal} {Phys. Rev.}\ }\textbf {\bibinfo {volume} {115}},\
  \bibinfo {pages} {1460} (\bibinfo {year} {1959})}\BibitemShut {NoStop}%
\bibitem [{\citenamefont {Blount}(1962)}]{blount_effham}%
  \BibitemOpen
  \bibfield  {author} {\bibinfo {author} {\bibfnamefont {E.~I.}\ \bibnamefont
  {Blount}},\ }\href {\doibase 10.1103/PhysRev.126.1636} {\bibfield  {journal}
  {\bibinfo  {journal} {Phys. Rev.}\ }\textbf {\bibinfo {volume} {126}},\
  \bibinfo {pages} {1636} (\bibinfo {year} {1962})}\BibitemShut {NoStop}%
\bibitem [{\citenamefont {Wannier}\ and\ \citenamefont
  {Fredkin}(1962)}]{wannier_fredkin}%
  \BibitemOpen
  \bibfield  {author} {\bibinfo {author} {\bibfnamefont {G.~H.}\ \bibnamefont
  {Wannier}}\ and\ \bibinfo {author} {\bibfnamefont {D.~R.}\ \bibnamefont
  {Fredkin}},\ }\href {\doibase 10.1103/PhysRev.125.1910} {\bibfield  {journal}
  {\bibinfo  {journal} {Phys. Rev.}\ }\textbf {\bibinfo {volume} {125}},\
  \bibinfo {pages} {1910} (\bibinfo {year} {1962})}\BibitemShut {NoStop}%
\bibitem [{\citenamefont {Fischbeck}(1970)}]{fischbeck_review}%
  \BibitemOpen
  \bibfield  {author} {\bibinfo {author} {\bibfnamefont {H.~J.}\ \bibnamefont
  {Fischbeck}},\ }\href {\doibase 10.1002/pssb.19700380102} {\bibfield
  {journal} {\bibinfo  {journal} {Physica Status Solidi (b)}\ }\textbf
  {\bibinfo {volume} {38}},\ \bibinfo {pages} {11} (\bibinfo {year}
  {1970})}\BibitemShut {NoStop}%
\bibitem [{\citenamefont {Gao}\ and\ \citenamefont
  {Niu}(2017)}]{gao_zero-field_2017}%
  \BibitemOpen
  \bibfield  {author} {\bibinfo {author} {\bibfnamefont {Y.}~\bibnamefont
  {Gao}}\ and\ \bibinfo {author} {\bibfnamefont {Q.}~\bibnamefont {Niu}},\
  }\href {\doibase 10.1073/pnas.1702595114} {\bibfield  {journal} {\bibinfo
  {journal} {Proc. Natl. Acad. Sci.}\ }\textbf {\bibinfo {volume} {114}},\
  \bibinfo {pages} {7295} (\bibinfo {year} {2017})}\BibitemShut {NoStop}%
\bibitem [{\citenamefont {Soluyanov}\ \emph {et~al.}(2015)\citenamefont
  {Soluyanov}, \citenamefont {Gresch}, \citenamefont {Wang}, \citenamefont
  {Wu}, \citenamefont {Troyer}, \citenamefont {Dai},\ and\ \citenamefont
  {Bernevig}}]{soluyanov_type-ii_2015}%
  \BibitemOpen
  \bibfield  {author} {\bibinfo {author} {\bibfnamefont {A.~A.}\ \bibnamefont
  {Soluyanov}}, \bibinfo {author} {\bibfnamefont {D.}~\bibnamefont {Gresch}},
  \bibinfo {author} {\bibfnamefont {Z.}~\bibnamefont {Wang}}, \bibinfo {author}
  {\bibfnamefont {Q.}~\bibnamefont {Wu}}, \bibinfo {author} {\bibfnamefont
  {M.}~\bibnamefont {Troyer}}, \bibinfo {author} {\bibfnamefont
  {X.}~\bibnamefont {Dai}}, \ and\ \bibinfo {author} {\bibfnamefont {B.~A.}\
  \bibnamefont {Bernevig}},\ }\href {\doibase 10.1038/nature15768} {\bibfield
  {journal} {\bibinfo  {journal} {Nature (London)}\ }\textbf {\bibinfo {volume}
  {527}},\ \bibinfo {pages} {495} (\bibinfo {year} {2015})}\BibitemShut
  {NoStop}%
\bibitem [{\citenamefont {Muechler}\ \emph {et~al.}(2016)\citenamefont
  {Muechler}, \citenamefont {Alexandradinata}, \citenamefont {Neupert},\ and\
  \citenamefont {Car}}]{muechler_tilted_2016}%
  \BibitemOpen
  \bibfield  {author} {\bibinfo {author} {\bibfnamefont {L.}~\bibnamefont
  {Muechler}}, \bibinfo {author} {\bibfnamefont {A.}~\bibnamefont
  {Alexandradinata}}, \bibinfo {author} {\bibfnamefont {T.}~\bibnamefont
  {Neupert}}, \ and\ \bibinfo {author} {\bibfnamefont {R.}~\bibnamefont
  {Car}},\ }\href {\doibase 10.1103/PhysRevX.6.041069} {\bibfield  {journal}
  {\bibinfo  {journal} {Phys. Rev. X}\ }\textbf {\bibinfo {volume} {6}},\
  \bibinfo {pages} {041069} (\bibinfo {year} {2016})}\BibitemShut {NoStop}%
\bibitem [{\citenamefont {Bergholtz}\ \emph {et~al.}(2015)\citenamefont
  {Bergholtz}, \citenamefont {Liu}, \citenamefont {Trescher}, \citenamefont
  {Moessner},\ and\ \citenamefont {Udagawa}}]{bergholtz_topology_2015}%
  \BibitemOpen
  \bibfield  {author} {\bibinfo {author} {\bibfnamefont {E.}~\bibnamefont
  {Bergholtz}}, \bibinfo {author} {\bibfnamefont {Z.}~\bibnamefont {Liu}},
  \bibinfo {author} {\bibfnamefont {M.}~\bibnamefont {Trescher}}, \bibinfo
  {author} {\bibfnamefont {R.}~\bibnamefont {Moessner}}, \ and\ \bibinfo
  {author} {\bibfnamefont {M.}~\bibnamefont {Udagawa}},\ }\href {\doibase
  10.1103/PhysRevLett.114.016806} {\bibfield  {journal} {\bibinfo  {journal}
  {Phys. Rev. Lett.}\ }\textbf {\bibinfo {volume} {114}},\ \bibinfo {pages}
  {016806} (\bibinfo {year} {2015})}\BibitemShut {NoStop}%
\bibitem [{\citenamefont {Kaganov}\ and\ \citenamefont
  {Slutskin}(1983)}]{kaganov_coherent_1983}%
  \BibitemOpen
  \bibfield  {author} {\bibinfo {author} {\bibfnamefont {M.~I.}\ \bibnamefont
  {Kaganov}}\ and\ \bibinfo {author} {\bibfnamefont {A.~A.}\ \bibnamefont
  {Slutskin}},\ }\href {\doibase 10.1016/0370-1573(83)90006-6} {\bibfield
  {journal} {\bibinfo  {journal} {Phys. Rep.}\ }\textbf {\bibinfo {volume}
  {98}},\ \bibinfo {pages} {189} (\bibinfo {year} {1983})}\BibitemShut
  {NoStop}%
\bibitem [{\citenamefont {Slutskin}(1968)}]{slutskin_dynamics_1968}%
  \BibitemOpen
  \bibfield  {author} {\bibinfo {author} {\bibfnamefont {A.}~\bibnamefont
  {Slutskin}},\ }\href@noop {} {\bibfield  {journal} {\bibinfo  {journal}
  {SOVIET PHYSICS JETP}\ }\textbf {\bibinfo {volume} {26}} (\bibinfo {year}
  {1968})}\BibitemShut {NoStop}%
\bibitem [{\citenamefont {Alexandradinata}\ and\ \citenamefont
  {Glazman}(2017)}]{AALG}%
  \BibitemOpen
  \bibfield  {author} {\bibinfo {author} {\bibfnamefont {A.}~\bibnamefont
  {Alexandradinata}}\ and\ \bibinfo {author} {\bibfnamefont {L.}~\bibnamefont
  {Glazman}},\ }\href {\doibase 10.1103/PhysRevLett.119.256601} {\bibfield
  {journal} {\bibinfo  {journal} {Phys. Rev. Lett.}\ }\textbf {\bibinfo
  {volume} {119}},\ \bibinfo {pages} {256601} (\bibinfo {year}
  {2017})}\BibitemShut {NoStop}%
\bibitem [{\citenamefont {O’Brien}\ \emph {et~al.}(2016)\citenamefont
  {O’Brien}, \citenamefont {Diez},\ and\ \citenamefont
  {Beenakker}}]{obrien_magnetic_2016}%
  \BibitemOpen
  \bibfield  {author} {\bibinfo {author} {\bibfnamefont {T.}~\bibnamefont
  {O’Brien}}, \bibinfo {author} {\bibfnamefont {M.}~\bibnamefont {Diez}}, \
  and\ \bibinfo {author} {\bibfnamefont {C.}~\bibnamefont {Beenakker}},\ }\href
  {\doibase 10.1103/PhysRevLett.116.236401} {\bibfield  {journal} {\bibinfo
  {journal} {Phys. Rev. Lett.}\ }\textbf {\bibinfo {volume} {116}},\ \bibinfo
  {pages} {236401} (\bibinfo {year} {2016})}\BibitemShut {NoStop}%
\bibitem [{\citenamefont {Neumann}\ and\ \citenamefont
  {Wigner}(2000)}]{neumann2000behaviour}%
  \BibitemOpen
  \bibfield  {author} {\bibinfo {author} {\bibfnamefont {J.~V.}\ \bibnamefont
  {Neumann}}\ and\ \bibinfo {author} {\bibfnamefont {E.}~\bibnamefont
  {Wigner}},\ }in\ \href@noop {} {\emph {\bibinfo {booktitle} {Quantum
  Chemistry: Classic Scientific Papers}}}\ (\bibinfo  {publisher} {World
  Scientific},\ \bibinfo {year} {2000})\ pp.\ \bibinfo {pages}
  {25--31}\BibitemShut {NoStop}%
\bibitem [{\citenamefont {Peierls}(1933)}]{peierls_substitution}%
  \BibitemOpen
  \bibfield  {author} {\bibinfo {author} {\bibfnamefont {R.}~\bibnamefont
  {Peierls}},\ }\href {\doibase 10.1007/BF01342591} {\bibfield  {journal}
  {\bibinfo  {journal} {Zeitschrift f{\"u}r Physik}\ }\textbf {\bibinfo
  {volume} {80}},\ \bibinfo {pages} {763} (\bibinfo {year} {1933})}\BibitemShut
  {NoStop}%
\bibitem [{\citenamefont {Berry}\ and\ \citenamefont
  {Mount}(1972)}]{berry_mount_review}%
  \BibitemOpen
  \bibfield  {author} {\bibinfo {author} {\bibfnamefont {M.~V.}\ \bibnamefont
  {Berry}}\ and\ \bibinfo {author} {\bibfnamefont {K.~E.}\ \bibnamefont
  {Mount}},\ }\href {http://stacks.iop.org/0034-4885/35/i=1/a=306} {\bibfield
  {journal} {\bibinfo  {journal} {Rep. Prog. Phys.}\ }\textbf {\bibinfo
  {volume} {35}},\ \bibinfo {pages} {315} (\bibinfo {year} {1972})}\BibitemShut
  {NoStop}%
\bibitem [{\citenamefont {Keller}(1958)}]{keller1958}%
  \BibitemOpen
  \bibfield  {author} {\bibinfo {author} {\bibfnamefont {J.~B.}\ \bibnamefont
  {Keller}},\ }\href {\doibase http://dx.doi.org/10.1016/0003-4916(58)90032-0}
  {\bibfield  {journal} {\bibinfo  {journal} {Ann. Phys.}\ }\textbf {\bibinfo
  {volume} {4}},\ \bibinfo {pages} {180 } (\bibinfo {year} {1958})}\BibitemShut
  {NoStop}%
\bibitem [{\citenamefont {Shirley}(1965)}]{shirley_solution_1965}%
  \BibitemOpen
  \bibfield  {author} {\bibinfo {author} {\bibfnamefont {J.~H.}\ \bibnamefont
  {Shirley}},\ }\href {\doibase 10.1103/PhysRev.138.B979} {\bibfield  {journal}
  {\bibinfo  {journal} {Phys. Rev.}\ }\textbf {\bibinfo {volume} {138}},\
  \bibinfo {pages} {B979} (\bibinfo {year} {1965})}\BibitemShut {NoStop}%
\bibitem [{Note2()}]{Note2}%
  \BibitemOpen
  \bibinfo {note} {The Pauli spin-orbit coupling term ${\propto } \protect \bm
  {s}{\cdot } (\protect \bm {E}{\times } \protect \bm {p})$ is traceless in any
  spin-symmetric (pseudo-)spinor basis. Viewing the spin-orbit coupling as a
  small parameter in degenerate perturbation theory, the lowest-order splitting
  of any spin-degenerate energy level is always symmetric about the
  zeroth-order energy. This follows because the zeroth-order, spin-symmetric
  wavefunctions (associated to a spin-degenerate level) are tensor products of
  spin and spatial wavefunctions: ${\setbox \z@ \hbox {\frozen@everymath
  \@emptytoks \mathsurround \z@ $\nulldelimiterspace \z@ \left |\vcenter to\@ne
  \big@size {}\right .$}\box \z@ }\Psi _{\pm }{\setbox \z@ \hbox
  {\frozen@everymath \@emptytoks \mathsurround \z@ $\nulldelimiterspace \z@
  \left \delimiter "526930B \vcenter to\@ne \big@size {}\right .$}\box \z@
  }{:}{=}{\setbox \z@ \hbox {\frozen@everymath \@emptytoks \mathsurround \z@
  $\nulldelimiterspace \z@ \left |\vcenter to\@ne \big@size {}\right .$}\box
  \z@ }{\pm }\protect \cc@accent {"705E}{n}{\setbox \z@ \hbox
  {\frozen@everymath \@emptytoks \mathsurround \z@ $\nulldelimiterspace \z@
  \left \delimiter "526930B \vcenter to\@ne \big@size {}\right .$}\box \z@
  }{\otimes } {\setbox \z@ \hbox {\frozen@everymath \@emptytoks \mathsurround
  \z@ $\nulldelimiterspace \z@ \left |\vcenter to\@ne \big@size {}\right
  .$}\box \z@ }\psi {\setbox \z@ \hbox {\frozen@everymath \@emptytoks
  \mathsurround \z@ $\nulldelimiterspace \z@ \left \delimiter "526930B \vcenter
  to\@ne \big@size {}\right .$}\box \z@ }$ with $\protect \cc@accent
  {"705E}{n}$ a unit vector and $\protect \bm {s}{\cdot } \protect \cc@accent
  {"705E}{n} {\setbox \z@ \hbox {\frozen@everymath \@emptytoks \mathsurround
  \z@ $\nulldelimiterspace \z@ \left |\vcenter to\@ne \big@size {}\right
  .$}\box \z@ }{\pm }\protect \cc@accent {"705E}{n}{\setbox \z@ \hbox
  {\frozen@everymath \@emptytoks \mathsurround \z@ $\nulldelimiterspace \z@
  \left \delimiter "526930B \vcenter to\@ne \big@size {}\right .$}\box \z@
  }{=}{\pm }{\setbox \z@ \hbox {\frozen@everymath \@emptytoks \mathsurround \z@
  $\nulldelimiterspace \z@ \left |\vcenter to\@ne \big@size {}\right .$}\box
  \z@ }\pm \protect \cc@accent {"705E}{n}{\setbox \z@ \hbox {\frozen@everymath
  \@emptytoks \mathsurround \z@ $\nulldelimiterspace \z@ \left \delimiter
  "526930B \vcenter to\@ne \big@size {}\right .$}\box \z@ }$. Hence, ${\setbox
  \z@ \hbox {\frozen@everymath \@emptytoks \mathsurround \z@
  $\nulldelimiterspace \z@ \left \delimiter "426830A \vcenter to\@ne \big@size
  {}\right .$}\box \z@ }\Psi _{\pm } {\setbox \z@ \hbox {\frozen@everymath
  \@emptytoks \mathsurround \z@ $\nulldelimiterspace \z@ \left |\vcenter to\@ne
  \big@size {}\right .$}\box \z@ } \protect \bm {s}{\cdot } (\protect \bm
  {E}{\times } \protect \bm {p}) {\setbox \z@ \hbox {\frozen@everymath
  \@emptytoks \mathsurround \z@ $\nulldelimiterspace \z@ \left |\vcenter to\@ne
  \big@size {}\right .$}\box \z@ } \Psi _{\pm } {\setbox \z@ \hbox
  {\frozen@everymath \@emptytoks \mathsurround \z@ $\nulldelimiterspace \z@
  \left \delimiter "526930B \vcenter to\@ne \big@size {}\right .$}\box \z@
  }{=}{\pm }\protect \cc@accent {"705E}{n}\cdot {\setbox \z@ \hbox
  {\frozen@everymath \@emptytoks \mathsurround \z@ $\nulldelimiterspace \z@
  \left \delimiter "426830A \vcenter to\@ne \big@size {}\right .$}\box \z@
  }\psi {\setbox \z@ \hbox {\frozen@everymath \@emptytoks \mathsurround \z@
  $\nulldelimiterspace \z@ \left |\vcenter to\@ne \big@size {}\right .$}\box
  \z@ } \protect \bm {E}{\times } \protect \bm {p} {\setbox \z@ \hbox
  {\frozen@everymath \@emptytoks \mathsurround \z@ $\nulldelimiterspace \z@
  \left |\vcenter to\@ne \big@size {}\right .$}\box \z@ } \psi {\setbox \z@
  \hbox {\frozen@everymath \@emptytoks \mathsurround \z@ $\nulldelimiterspace
  \z@ \left \delimiter "526930B \vcenter to\@ne \big@size {}\right .$}\box \z@
  }$.}\BibitemShut {Stop}%
\bibitem [{Note3()}]{Note3}%
  \BibitemOpen
  \bibinfo {note} {The neglected term on the right-hand side of Eq.\ (\ref
  {spinsplitwkb}) is smaller (than the kept term) by a factor $\delta v_{\pm
  }/v$, which is the relative change in band speed $v=|\protect \bm {v}|=\hbar
  ^{-1}|\partial \varepsilon /\partial \protect \bm {k}|$ induced by the
  spin-orbit coupling. This factor is assumed to be small. For example in the
  Rashba model, $\delta v_{\pm }/v=m\alpha /\hbar k_E{\sim }\delta S/S{\ll
  }1$.}\BibitemShut {Stop}%
\bibitem [{\citenamefont {Thonhauser}\ \emph {et~al.}(2005)\citenamefont
  {Thonhauser}, \citenamefont {Ceresoli}, \citenamefont {Vanderbilt},\ and\
  \citenamefont {Resta}}]{thonhauser_orbital_2005}%
  \BibitemOpen
  \bibfield  {author} {\bibinfo {author} {\bibfnamefont {T.}~\bibnamefont
  {Thonhauser}}, \bibinfo {author} {\bibfnamefont {D.}~\bibnamefont
  {Ceresoli}}, \bibinfo {author} {\bibfnamefont {D.}~\bibnamefont
  {Vanderbilt}}, \ and\ \bibinfo {author} {\bibfnamefont {R.}~\bibnamefont
  {Resta}},\ }\href {\doibase 10.1103/PhysRevLett.95.137205} {\bibfield
  {journal} {\bibinfo  {journal} {Phys. Rev. Lett.}\ }\textbf {\bibinfo
  {volume} {95}},\ \bibinfo {pages} {137205} (\bibinfo {year}
  {2005})}\BibitemShut {NoStop}%
\bibitem [{\citenamefont {Berry}(1984)}]{berry_quantal_1984}%
  \BibitemOpen
  \bibfield  {author} {\bibinfo {author} {\bibfnamefont {M.~V.}\ \bibnamefont
  {Berry}},\ }\href {http://www.jstor.org/stable/2397741} {\bibfield  {journal}
  {\bibinfo  {journal} {Proc. Royal Soc. Lond. A}\ }\textbf {\bibinfo {volume}
  {392}},\ \bibinfo {pages} {45} (\bibinfo {year} {1984})}\BibitemShut
  {NoStop}%
\bibitem [{\citenamefont {Wilczek}\ and\ \citenamefont
  {Zee}(1984)}]{wilczek_appearance_1984}%
  \BibitemOpen
  \bibfield  {author} {\bibinfo {author} {\bibfnamefont {F.}~\bibnamefont
  {Wilczek}}\ and\ \bibinfo {author} {\bibfnamefont {A.}~\bibnamefont {Zee}},\
  }\href {\doibase 10.1103/PhysRevLett.52.2111} {\bibfield  {journal} {\bibinfo
   {journal} {Phys. Rev. Lett.}\ }\textbf {\bibinfo {volume} {52}},\ \bibinfo
  {pages} {2111} (\bibinfo {year} {1984})}\BibitemShut {NoStop}%
\bibitem [{\citenamefont {Nenciu}(1991)}]{nenciu_review}%
  \BibitemOpen
  \bibfield  {author} {\bibinfo {author} {\bibfnamefont {G.}~\bibnamefont
  {Nenciu}},\ }\href {\doibase 10.1103/RevModPhys.63.91} {\bibfield  {journal}
  {\bibinfo  {journal} {Rev. Mod. Phys.}\ }\textbf {\bibinfo {volume} {63}},\
  \bibinfo {pages} {91} (\bibinfo {year} {1991})}\BibitemShut {NoStop}%
\bibitem [{Note4()}]{Note4}%
  \BibitemOpen
  \bibinfo {note} {Eq.\ (\ref {phaseindependentorbit}) neglects the
  contribution by band $-$ to the single-band orbital moment\cite
  {chang_berry_1996} of band $+$. This contribution is equal to $B\delta
  M_z=il^{-2}\epsilon _{\alpha \beta }(\delta \varepsilon _+-\delta \varepsilon
  _-)\protect \mathfrak {X}^{\alpha }_{+-}\protect \mathfrak {X}^{\beta
  }_{-+}$. $\protect \mathfrak {X}^{\alpha }_{+-}$, being an off-diagonal
  matrix element of the position operator, is generically of order the lattice
  constant $a$. We then estimate $\DOTSI \intop \ilimits@ _0^{T_c} B\delta M_z
  dt/\hbar =O(a^2 \delta S)$, which is negligible compared to $l^2\delta S/2$.
  Note also that the single-band orbital moment simply vanishes in some
  symmetry classes of orbits\cite {100p}.}\BibitemShut {Stop}%
\bibitem [{Note5()}]{Note5}%
  \BibitemOpen
  \bibinfo {note} {An equivalent but less general formulation was proposed
  earlier in Ref. \protect \rev@citealp {rothmag} and Ref. \protect
  \rev@citealp {Mikitik_quantizationrule} for centrosymmetric solids with
  time-reversal symmetry (at zero field). For comparison, it should be
  emphasized that Eqs.\ (\ref {eq:rule}-\ref {berryconn}) applies to solids of
  any symmetry, including magnetically-ordered solids.}\BibitemShut {Stop}%
\bibitem [{\citenamefont {Bychkov}\ and\ \citenamefont
  {Rashba}(1984)}]{bychkov_oscillatory_1984}%
  \BibitemOpen
  \bibfield  {author} {\bibinfo {author} {\bibfnamefont {Y.~A.}\ \bibnamefont
  {Bychkov}}\ and\ \bibinfo {author} {\bibfnamefont {E.~I.}\ \bibnamefont
  {Rashba}},\ }\href {\doibase 10.1088/0022-3719/17/33/015} {\bibfield
  {journal} {\bibinfo  {journal} {J. Phys. C: Solid State Phys.}\ }\textbf
  {\bibinfo {volume} {17}},\ \bibinfo {pages} {6039} (\bibinfo {year}
  {1984})}\BibitemShut {NoStop}%
\bibitem [{Note6()}]{Note6}%
  \BibitemOpen
  \bibinfo {note} {For simplicity, we neglect here off-diagonal elements of
  $M_z$ in the basis of $S_z=\pm 1/2$, with $\protect \cc@accent {"717E}{z}$
  the direction of the field.}\BibitemShut {Stop}%
\bibitem [{Note7()}]{Note7}%
  \BibitemOpen
  \bibinfo {note} {The intra-band Zeeman and orbital-moment couplings vanish
  due spacetime-inversion symmetry\cite {topoferm}.}\BibitemShut {Stop}%
\bibitem [{\citenamefont {Mineev}\ and\ \citenamefont
  {Samokhin}(2005)}]{mineev_haas--van_2005}%
  \BibitemOpen
  \bibfield  {author} {\bibinfo {author} {\bibfnamefont {V.~P.}\ \bibnamefont
  {Mineev}}\ and\ \bibinfo {author} {\bibfnamefont {K.~V.}\ \bibnamefont
  {Samokhin}},\ }\href {\doibase 10.1103/PhysRevB.72.212504} {\bibfield
  {journal} {\bibinfo  {journal} {Phys. Rev. B}\ }\textbf {\bibinfo {volume}
  {72}},\ \bibinfo {pages} {212504} (\bibinfo {year} {2005})}\BibitemShut
  {NoStop}%
\bibitem [{\citenamefont {Kane}\ and\ \citenamefont
  {Blount}(1969)}]{kane_blount}%
  \BibitemOpen
  \bibfield  {author} {\bibinfo {author} {\bibfnamefont {E.~O.}\ \bibnamefont
  {Kane}}\ and\ \bibinfo {author} {\bibfnamefont {E.~I.}\ \bibnamefont
  {Blount}},\ }\href@noop {} {}\bibinfo {howpublished} {in \emph{Tunneling
  phenomenon in solids}, edited by E. Burstein and S. Lundqvist (Plenum Press,
  New York)} (\bibinfo {year} {1969})\BibitemShut {NoStop}%
\bibitem [{Note8()}]{Note8}%
  \BibitemOpen
  \bibinfo {note} {The orbit-averaged energy splitting is of order $\alpha
  k_{\scriptscriptstyle {E}}$ for energies near the Dirac point. $B_\perp
  |{\protect \cal M}_{\scriptscriptstyle {+-}}|$ is simply the Berry
  contribution to the Zeeman interaction, and has the form of the right-most
  term in Eq.\ (\ref {rashbaeffham}); this term is of order $\varepsilon
  _c$.}\BibitemShut {Stop}%
\bibitem [{\citenamefont {Wittig}(2005)}]{wittig_landauzener_2005}%
  \BibitemOpen
  \bibfield  {author} {\bibinfo {author} {\bibfnamefont {C.}~\bibnamefont
  {Wittig}},\ }\href {\doibase 10.1021/jp040627u} {\bibfield  {journal}
  {\bibinfo  {journal} {J. Phys. Chem. B}\ } (\bibinfo {year} {2005}),\
  10.1021/jp040627u}\BibitemShut {NoStop}%
\bibitem [{\citenamefont {{Lifshitz E.M.}}\ and\ \citenamefont {{Landau
  L.D.}}(1991)}]{lifshitz_e.m._quantum_1991}%
  \BibitemOpen
  \bibfield  {author} {\bibinfo {author} {\bibnamefont {{Lifshitz E.M.}}}\ and\
  \bibinfo {author} {\bibnamefont {{Landau L.D.}}},\ }\href
  {http://gen.lib.rus.ec/book/index.php?md5=D81CD60226BEB47ADF5BBFB8790730C8}
  {\emph {\bibinfo {title} {Quantum mechanics: non-relativistic theory}}},\
  \bibinfo {edition} {3rd}\ ed.,\ A-{W} series in advanced physics\ (\bibinfo
  {publisher} {Pergamon},\ \bibinfo {year} {1991})\BibitemShut {NoStop}%
\bibitem [{\citenamefont {Pakmehr}\ \emph {et~al.}(2015)\citenamefont
  {Pakmehr}, \citenamefont {Khaetskii}, \citenamefont {McCombe}, \citenamefont
  {Bhandari}, \citenamefont {Cahay}, \citenamefont {Chiatti}, \citenamefont
  {Fischer}, \citenamefont {Heyn},\ and\ \citenamefont
  {Hansen}}]{pakmehr_g-factor_2015}%
  \BibitemOpen
  \bibfield  {author} {\bibinfo {author} {\bibfnamefont {M.}~\bibnamefont
  {Pakmehr}}, \bibinfo {author} {\bibfnamefont {A.}~\bibnamefont {Khaetskii}},
  \bibinfo {author} {\bibfnamefont {B.~D.}\ \bibnamefont {McCombe}}, \bibinfo
  {author} {\bibfnamefont {N.}~\bibnamefont {Bhandari}}, \bibinfo {author}
  {\bibfnamefont {M.}~\bibnamefont {Cahay}}, \bibinfo {author} {\bibfnamefont
  {O.}~\bibnamefont {Chiatti}}, \bibinfo {author} {\bibfnamefont {S.~F.}\
  \bibnamefont {Fischer}}, \bibinfo {author} {\bibfnamefont {C.}~\bibnamefont
  {Heyn}}, \ and\ \bibinfo {author} {\bibfnamefont {W.}~\bibnamefont
  {Hansen}},\ }\href {\doibase 10.1063/1.4929373} {\bibfield  {journal}
  {\bibinfo  {journal} {Appl. Phys. Lett.}\ }\textbf {\bibinfo {volume}
  {107}},\ \bibinfo {pages} {082107} (\bibinfo {year} {2015})}\BibitemShut
  {NoStop}%
\bibitem [{Note9()}]{Note9}%
  \BibitemOpen
  \bibinfo {note} {For two band models, Berry phases of the two bands sum up to
  0 on arbitrary path}\BibitemShut {NoStop}%
\bibitem [{Note10()}]{Note10}%
  \BibitemOpen
  \bibinfo {note} {The general form of this perturbative calculation can be
  found in Sec. IX-E of Ref. \protect \rev@citealp {100p}.}\BibitemShut {Stop}%
\bibitem [{\citenamefont {Shoenberg}(2009)}]{shoenberg_magnetic_2009}%
  \BibitemOpen
  \bibfield  {author} {\bibinfo {author} {\bibfnamefont {D.}~\bibnamefont
  {Shoenberg}},\ }\href@noop {} {\emph {\bibinfo {title} {Magnetic oscillations
  in metals}}}\ (\bibinfo  {publisher} {Cambridge University Press},\ \bibinfo
  {year} {2009})\BibitemShut {NoStop}%
\bibitem [{\citenamefont {Landau}\ and\ \citenamefont
  {Lifshitz}(2013)}]{landau2013course}%
  \BibitemOpen
  \bibfield  {author} {\bibinfo {author} {\bibfnamefont {L.~D.}\ \bibnamefont
  {Landau}}\ and\ \bibinfo {author} {\bibfnamefont {E.~M.}\ \bibnamefont
  {Lifshitz}},\ }\href@noop {} {\emph {\bibinfo {title} {Statistical Physics,
  Part 1}}}\ (\bibinfo  {publisher} {Elsevier},\ \bibinfo {year}
  {2013})\BibitemShut {NoStop}%
\bibitem [{\citenamefont {Wang}\ \emph {et~al.}(2012)\citenamefont {Wang},
  \citenamefont {Sun}, \citenamefont {Chen}, \citenamefont {Franchini},
  \citenamefont {Xu}, \citenamefont {Weng}, \citenamefont {Dai},\ and\
  \citenamefont {Fang}}]{wang2012dirac}%
  \BibitemOpen
  \bibfield  {author} {\bibinfo {author} {\bibfnamefont {Z.}~\bibnamefont
  {Wang}}, \bibinfo {author} {\bibfnamefont {Y.}~\bibnamefont {Sun}}, \bibinfo
  {author} {\bibfnamefont {X.-Q.}\ \bibnamefont {Chen}}, \bibinfo {author}
  {\bibfnamefont {C.}~\bibnamefont {Franchini}}, \bibinfo {author}
  {\bibfnamefont {G.}~\bibnamefont {Xu}}, \bibinfo {author} {\bibfnamefont
  {H.}~\bibnamefont {Weng}}, \bibinfo {author} {\bibfnamefont {X.}~\bibnamefont
  {Dai}}, \ and\ \bibinfo {author} {\bibfnamefont {Z.}~\bibnamefont {Fang}},\
  }\href@noop {} {\bibfield  {journal} {\bibinfo  {journal} {Phys. Rev. B}\
  }\textbf {\bibinfo {volume} {85}},\ \bibinfo {pages} {195320} (\bibinfo
  {year} {2012})}\BibitemShut {NoStop}%
\bibitem [{\citenamefont {Wan}\ \emph {et~al.}(2011)\citenamefont {Wan},
  \citenamefont {Turner}, \citenamefont {Vishwanath},\ and\ \citenamefont
  {Savrasov}}]{wan2011topological}%
  \BibitemOpen
  \bibfield  {author} {\bibinfo {author} {\bibfnamefont {X.}~\bibnamefont
  {Wan}}, \bibinfo {author} {\bibfnamefont {A.~M.}\ \bibnamefont {Turner}},
  \bibinfo {author} {\bibfnamefont {A.}~\bibnamefont {Vishwanath}}, \ and\
  \bibinfo {author} {\bibfnamefont {S.~Y.}\ \bibnamefont {Savrasov}},\
  }\href@noop {} {\bibfield  {journal} {\bibinfo  {journal} {Phys. Rev. B}\
  }\textbf {\bibinfo {volume} {83}},\ \bibinfo {pages} {205101} (\bibinfo
  {year} {2011})}\BibitemShut {NoStop}%
\bibitem [{\citenamefont {Burkov}\ \emph {et~al.}(2011)\citenamefont {Burkov},
  \citenamefont {Hook},\ and\ \citenamefont {Balents}}]{burkov2011topological}%
  \BibitemOpen
  \bibfield  {author} {\bibinfo {author} {\bibfnamefont {A.}~\bibnamefont
  {Burkov}}, \bibinfo {author} {\bibfnamefont {M.}~\bibnamefont {Hook}}, \ and\
  \bibinfo {author} {\bibfnamefont {L.}~\bibnamefont {Balents}},\ }\href@noop
  {} {\bibfield  {journal} {\bibinfo  {journal} {Phys. Rev. B}\ }\textbf
  {\bibinfo {volume} {84}},\ \bibinfo {pages} {235126} (\bibinfo {year}
  {2011})}\BibitemShut {NoStop}%
\bibitem [{\citenamefont {Neto}\ \emph {et~al.}(2009)\citenamefont {Neto},
  \citenamefont {Guinea}, \citenamefont {Peres}, \citenamefont {Novoselov},\
  and\ \citenamefont {Geim}}]{neto2009electronic}%
  \BibitemOpen
  \bibfield  {author} {\bibinfo {author} {\bibfnamefont {A.~C.}\ \bibnamefont
  {Neto}}, \bibinfo {author} {\bibfnamefont {F.}~\bibnamefont {Guinea}},
  \bibinfo {author} {\bibfnamefont {N.~M.}\ \bibnamefont {Peres}}, \bibinfo
  {author} {\bibfnamefont {K.~S.}\ \bibnamefont {Novoselov}}, \ and\ \bibinfo
  {author} {\bibfnamefont {A.~K.}\ \bibnamefont {Geim}},\ }\href@noop {}
  {\bibfield  {journal} {\bibinfo  {journal} {Rev. Mod. Phys.}\ }\textbf
  {\bibinfo {volume} {81}},\ \bibinfo {pages} {109} (\bibinfo {year}
  {2009})}\BibitemShut {NoStop}%
\bibitem [{\citenamefont {Thouless}\ \emph {et~al.}(1982)\citenamefont
  {Thouless}, \citenamefont {Kohmoto}, \citenamefont {Nightingale},\ and\
  \citenamefont {den Nijs}}]{TKNN}%
  \BibitemOpen
  \bibfield  {author} {\bibinfo {author} {\bibfnamefont {D.~J.}\ \bibnamefont
  {Thouless}}, \bibinfo {author} {\bibfnamefont {M.}~\bibnamefont {Kohmoto}},
  \bibinfo {author} {\bibfnamefont {M.~P.}\ \bibnamefont {Nightingale}}, \ and\
  \bibinfo {author} {\bibfnamefont {M.}~\bibnamefont {den Nijs}},\ }\href@noop
  {} {\bibfield  {journal} {\bibinfo  {journal} {Phys. Rev. Lett.}\ }\textbf
  {\bibinfo {volume} {49}},\ \bibinfo {pages} {405} (\bibinfo {year}
  {1982})}\BibitemShut {NoStop}%
\bibitem [{Note11()}]{Note11}%
  \BibitemOpen
  \bibinfo {note} {The curvature is defined with respect to a line bundle
  formed from the nondegenerate eigenvector of ${\protect \cal
  A}$.}\BibitemShut {Stop}%
\bibitem [{Note12()}]{Note12}%
  \BibitemOpen
  \bibinfo {note} {For $\protect \cc@accent {"7016}{\varphi }(\protect
  \cc@accent {"7016}{\mu })$ to vary $\sim 1$, the Landau-Zener parameter
  $\protect \cc@accent {"7016}{\mu }$ must vary by $\sim 1$ (cf.\ Fig.\ 10 in
  Ref. \protect \rev@citealp {100p}). Moreover, $2\pi \protect \cc@accent
  {"7016}{\mu }{:}{=} (2\pi /8) l^2S_{\scriptscriptstyle {\square }}$\cite
  {AALG}}\BibitemShut {NoStop}%
\bibitem [{\citenamefont {Yakovenko}\ and\ \citenamefont
  {Cooper}(2006)}]{yakovenko_angular_2006}%
  \BibitemOpen
  \bibfield  {author} {\bibinfo {author} {\bibfnamefont {V.~M.}\ \bibnamefont
  {Yakovenko}}\ and\ \bibinfo {author} {\bibfnamefont {B.~K.}\ \bibnamefont
  {Cooper}},\ }\href {\doibase 10.1016/j.physe.2006.03.001} {\bibfield
  {journal} {\bibinfo  {journal} {Physica E: Low-dimensional Systems and
  Nanostructures}\ }\textbf {\bibinfo {volume} {34}},\ \bibinfo {pages} {128}
  (\bibinfo {year} {2006})}\BibitemShut {NoStop}%
\bibitem [{\citenamefont {Pershoguba}\ \emph {et~al.}(2015)\citenamefont
  {Pershoguba}, \citenamefont {Abergel}, \citenamefont {Yakovenko},\ and\
  \citenamefont {Balatsky}}]{pershoguba_effects_2015}%
  \BibitemOpen
  \bibfield  {author} {\bibinfo {author} {\bibfnamefont {S.~S.}\ \bibnamefont
  {Pershoguba}}, \bibinfo {author} {\bibfnamefont {D.~S.~L.}\ \bibnamefont
  {Abergel}}, \bibinfo {author} {\bibfnamefont {V.~M.}\ \bibnamefont
  {Yakovenko}}, \ and\ \bibinfo {author} {\bibfnamefont {A.~V.}\ \bibnamefont
  {Balatsky}},\ }\href {\doibase 10.1103/PhysRevB.91.085418} {\bibfield
  {journal} {\bibinfo  {journal} {Phys. Rev. B}\ }\textbf {\bibinfo {volume}
  {91}},\ \bibinfo {pages} {085418} (\bibinfo {year} {2015})}\BibitemShut
  {NoStop}%
\bibitem [{Note13()}]{Note13}%
  \BibitemOpen
  \bibinfo {note} {These magic angles correspond to zeros of a Bessel function,
  as was first proposed by Yamaji.\cite {yamaji_angle_1989}}\BibitemShut
  {NoStop}%
\bibitem [{\citenamefont {Gusev}\ \emph {et~al.}(2008)\citenamefont {Gusev},
  \citenamefont {Duarte}, \citenamefont {Lamas}, \citenamefont {Bakarov},\ and\
  \citenamefont {Portal}}]{gusev_interlayer_2008}%
  \BibitemOpen
  \bibfield  {author} {\bibinfo {author} {\bibfnamefont {G.~M.}\ \bibnamefont
  {Gusev}}, \bibinfo {author} {\bibfnamefont {C.~A.}\ \bibnamefont {Duarte}},
  \bibinfo {author} {\bibfnamefont {T.~E.}\ \bibnamefont {Lamas}}, \bibinfo
  {author} {\bibfnamefont {A.~K.}\ \bibnamefont {Bakarov}}, \ and\ \bibinfo
  {author} {\bibfnamefont {J.~C.}\ \bibnamefont {Portal}},\ }\href {\doibase
  10.1103/PhysRevB.78.155320} {\bibfield  {journal} {\bibinfo  {journal} {Phys.
  Rev. B}\ }\textbf {\bibinfo {volume} {78}},\ \bibinfo {pages} {155320}
  (\bibinfo {year} {2008})}\BibitemShut {NoStop}%
\bibitem [{\citenamefont {Sakurai}\ and\ \citenamefont
  {Commins}(1995)}]{sakurai1995modern}%
  \BibitemOpen
  \bibfield  {author} {\bibinfo {author} {\bibfnamefont {J.~J.}\ \bibnamefont
  {Sakurai}}\ and\ \bibinfo {author} {\bibfnamefont {E.~D.}\ \bibnamefont
  {Commins}},\ }\href@noop {} {\emph {\bibinfo {title} {Modern quantum
  mechanics, revised edition}}}\ (\bibinfo  {publisher} {AAPT},\ \bibinfo
  {year} {1995})\BibitemShut {NoStop}%
\bibitem [{Note14()}]{Note14}%
  \BibitemOpen
  \bibinfo {note} {For a fixed orbit in class I-1 or II-2, all degeneracy
  manifolds have the same codimension.}\BibitemShut {Stop}%
\bibitem [{\citenamefont {Alexandradinata}\ and\ \citenamefont
  {Bernevig}(2016)}]{alexandradinata_berry-phase_2016}%
  \BibitemOpen
  \bibfield  {author} {\bibinfo {author} {\bibfnamefont {A.}~\bibnamefont
  {Alexandradinata}}\ and\ \bibinfo {author} {\bibfnamefont {B.~A.}\
  \bibnamefont {Bernevig}},\ }\href {\doibase 10.1103/PhysRevB.93.205104}
  {\bibfield  {journal} {\bibinfo  {journal} {Phys. Rev. B}\ }\textbf {\bibinfo
  {volume} {93}},\ \bibinfo {pages} {205104} (\bibinfo {year}
  {2016})}\BibitemShut {NoStop}%
\bibitem [{\citenamefont {B.~Dingle}(1952)}]{Dingle_collisions}%
  \BibitemOpen
  \bibfield  {author} {\bibinfo {author} {\bibfnamefont {R.}~\bibnamefont
  {B.~Dingle}},\ }\href@noop {} {\bibfield  {journal} {\bibinfo  {journal}
  {Proc. Royal Soc. Lond. A}\ }\textbf {\bibinfo {volume} {211}},\ \bibinfo
  {pages} {517} (\bibinfo {year} {1952})}\BibitemShut {NoStop}%
\bibitem [{Note15()}]{Note15}%
  \BibitemOpen
  \bibinfo {note} {For an alternative proof, $\delta \lambda {\equiv }\pi $ is
  equivalent to ${ \protect \cc@accent {"7016}{\protect \cal A}}{=}Ve^{i\sigma
  _z\pi /2}V^{-1}$ for some unitary $V\in U(2)$. The manifold associated to
  $\delta \lambda {\equiv }\pi $ is then the space of unitary matrices that do
  not commute with $e^{i\sigma _z\pi /2}$. The dimension of this manifold is
  the number of linearly-independent Hermitian generators (two) for unitaries
  that do not commute with $e^{i\sigma _z\pi /2}$. The codimension is then
  obtained from deducting two from the total number (three) of
  linearly-independent Hermitian generators for U(2). This approach of
  determining codimension is similar to that in Ref. \protect \rev@citealp
  {holler_non-hermitian_2018}.}\BibitemShut {Stop}%
\bibitem [{\citenamefont {Fuchs}\ \emph {et~al.}(2018)\citenamefont {Fuchs},
  \citenamefont {Piéchon},\ and\ \citenamefont
  {Montambaux}}]{fuchs_landau_2018}%
  \BibitemOpen
  \bibfield  {author} {\bibinfo {author} {\bibfnamefont {J.-N.}\ \bibnamefont
  {Fuchs}}, \bibinfo {author} {\bibfnamefont {F.}~\bibnamefont {Piéchon}}, \
  and\ \bibinfo {author} {\bibfnamefont {G.}~\bibnamefont {Montambaux}},\
  }\href {\doibase 10.21468/SciPostPhys.4.5.024} {\bibfield  {journal}
  {\bibinfo  {journal} {SciPost Phys.}\ }\textbf {\bibinfo {volume} {4}},\
  \bibinfo {pages} {024} (\bibinfo {year} {2018})}\BibitemShut {NoStop}%
\bibitem [{\citenamefont {Das}\ \emph {et~al.}(1989)\citenamefont {Das},
  \citenamefont {Miller}, \citenamefont {Datta}, \citenamefont {Reifenberger},
  \citenamefont {Hong}, \citenamefont {Bhattacharya}, \citenamefont {Singh},\
  and\ \citenamefont {Jaffe}}]{das_evidence_1989}%
  \BibitemOpen
  \bibfield  {author} {\bibinfo {author} {\bibfnamefont {B.}~\bibnamefont
  {Das}}, \bibinfo {author} {\bibfnamefont {D.~C.}\ \bibnamefont {Miller}},
  \bibinfo {author} {\bibfnamefont {S.}~\bibnamefont {Datta}}, \bibinfo
  {author} {\bibfnamefont {R.}~\bibnamefont {Reifenberger}}, \bibinfo {author}
  {\bibfnamefont {W.~P.}\ \bibnamefont {Hong}}, \bibinfo {author}
  {\bibfnamefont {P.~K.}\ \bibnamefont {Bhattacharya}}, \bibinfo {author}
  {\bibfnamefont {J.}~\bibnamefont {Singh}}, \ and\ \bibinfo {author}
  {\bibfnamefont {M.}~\bibnamefont {Jaffe}},\ }\href {\doibase
  10.1103/PhysRevB.39.1411} {\bibfield  {journal} {\bibinfo  {journal} {Phys.
  Rev. B}\ }\textbf {\bibinfo {volume} {39}},\ \bibinfo {pages} {1411}
  (\bibinfo {year} {1989})}\BibitemShut {NoStop}%
\bibitem [{\citenamefont {Hu}\ \emph {et~al.}(1999)\citenamefont {Hu},
  \citenamefont {Nitta}, \citenamefont {Akazaki}, \citenamefont {Takayanagi},
  \citenamefont {Osaka}, \citenamefont {Pfeffer},\ and\ \citenamefont
  {Zawadzki}}]{hu_zero-field_1999}%
  \BibitemOpen
  \bibfield  {author} {\bibinfo {author} {\bibfnamefont {C.-M.}\ \bibnamefont
  {Hu}}, \bibinfo {author} {\bibfnamefont {J.}~\bibnamefont {Nitta}}, \bibinfo
  {author} {\bibfnamefont {T.}~\bibnamefont {Akazaki}}, \bibinfo {author}
  {\bibfnamefont {H.}~\bibnamefont {Takayanagi}}, \bibinfo {author}
  {\bibfnamefont {J.}~\bibnamefont {Osaka}}, \bibinfo {author} {\bibfnamefont
  {P.}~\bibnamefont {Pfeffer}}, \ and\ \bibinfo {author} {\bibfnamefont
  {W.}~\bibnamefont {Zawadzki}},\ }\href {\doibase 10.1103/PhysRevB.60.7736}
  {\bibfield  {journal} {\bibinfo  {journal} {Phys. Rev. B}\ }\textbf {\bibinfo
  {volume} {60}},\ \bibinfo {pages} {7736} (\bibinfo {year}
  {1999})}\BibitemShut {NoStop}%
\bibitem [{\citenamefont {Wilde}\ \emph {et~al.}(2009)\citenamefont {Wilde},
  \citenamefont {Reuter}, \citenamefont {Heyn}, \citenamefont {Wieck},\ and\
  \citenamefont {Grundler}}]{wilde_inversion-asymmetry-induced_2009}%
  \BibitemOpen
  \bibfield  {author} {\bibinfo {author} {\bibfnamefont {M.~A.}\ \bibnamefont
  {Wilde}}, \bibinfo {author} {\bibfnamefont {D.}~\bibnamefont {Reuter}},
  \bibinfo {author} {\bibfnamefont {C.}~\bibnamefont {Heyn}}, \bibinfo {author}
  {\bibfnamefont {A.~D.}\ \bibnamefont {Wieck}}, \ and\ \bibinfo {author}
  {\bibfnamefont {D.}~\bibnamefont {Grundler}},\ }\href {\doibase
  10.1103/PhysRevB.79.125330} {\bibfield  {journal} {\bibinfo  {journal} {Phys.
  Rev. B}\ }\textbf {\bibinfo {volume} {79}},\ \bibinfo {pages} {125330}
  (\bibinfo {year} {2009})}\BibitemShut {NoStop}%
\bibitem [{\citenamefont {Terashima}\ \emph {et~al.}(2008)\citenamefont
  {Terashima}, \citenamefont {Kimata}, \citenamefont {Uji}, \citenamefont
  {Sugawara}, \citenamefont {Kimura}, \citenamefont {Aoki},\ and\ \citenamefont
  {Harima}}]{terashima_fermi_2008}%
  \BibitemOpen
  \bibfield  {author} {\bibinfo {author} {\bibfnamefont {T.}~\bibnamefont
  {Terashima}}, \bibinfo {author} {\bibfnamefont {M.}~\bibnamefont {Kimata}},
  \bibinfo {author} {\bibfnamefont {S.}~\bibnamefont {Uji}}, \bibinfo {author}
  {\bibfnamefont {T.}~\bibnamefont {Sugawara}}, \bibinfo {author}
  {\bibfnamefont {N.}~\bibnamefont {Kimura}}, \bibinfo {author} {\bibfnamefont
  {H.}~\bibnamefont {Aoki}}, \ and\ \bibinfo {author} {\bibfnamefont
  {H.}~\bibnamefont {Harima}},\ }\href {\doibase 10.1103/PhysRevB.78.205107}
  {\bibfield  {journal} {\bibinfo  {journal} {Phys. Rev. B}\ }\textbf {\bibinfo
  {volume} {78}},\ \bibinfo {pages} {205107} (\bibinfo {year}
  {2008})}\BibitemShut {NoStop}%
\bibitem [{\citenamefont {{\={O}}nuki}\ \emph {et~al.}(2014)\citenamefont
  {{\={O}}nuki}, \citenamefont {Nakamura}, \citenamefont {Uejo}, \citenamefont
  {Teruya}, \citenamefont {Hedo}, \citenamefont {Nakama}, \citenamefont
  {Honda},\ and\ \citenamefont {Harima}}]{onuki_chiral-structure-driven_2014}%
  \BibitemOpen
  \bibfield  {author} {\bibinfo {author} {\bibfnamefont {Y.}~\bibnamefont
  {{\={O}}nuki}}, \bibinfo {author} {\bibfnamefont {A.}~\bibnamefont
  {Nakamura}}, \bibinfo {author} {\bibfnamefont {T.}~\bibnamefont {Uejo}},
  \bibinfo {author} {\bibfnamefont {A.}~\bibnamefont {Teruya}}, \bibinfo
  {author} {\bibfnamefont {M.}~\bibnamefont {Hedo}}, \bibinfo {author}
  {\bibfnamefont {T.}~\bibnamefont {Nakama}}, \bibinfo {author} {\bibfnamefont
  {F.}~\bibnamefont {Honda}}, \ and\ \bibinfo {author} {\bibfnamefont
  {H.}~\bibnamefont {Harima}},\ }\href {\doibase 10.7566/JPSJ.83.061018}
  {\bibfield  {journal} {\bibinfo  {journal} {J. Phys. Soc. Jpn.}\ }\textbf
  {\bibinfo {volume} {83}},\ \bibinfo {pages} {061018} (\bibinfo {year}
  {2014})}\BibitemShut {NoStop}%
\bibitem [{\citenamefont {Maurya}\ \emph {et~al.}(2018)\citenamefont {Maurya},
  \citenamefont {Harima}, \citenamefont {Nakamura}, \citenamefont {Shimizu},
  \citenamefont {Homma}, \citenamefont {Li}, \citenamefont {Honda},
  \citenamefont {Sato},\ and\ \citenamefont {Aoki}}]{maurya_splitting_2018}%
  \BibitemOpen
  \bibfield  {author} {\bibinfo {author} {\bibfnamefont {A.}~\bibnamefont
  {Maurya}}, \bibinfo {author} {\bibfnamefont {H.}~\bibnamefont {Harima}},
  \bibinfo {author} {\bibfnamefont {A.}~\bibnamefont {Nakamura}}, \bibinfo
  {author} {\bibfnamefont {Y.}~\bibnamefont {Shimizu}}, \bibinfo {author}
  {\bibfnamefont {Y.}~\bibnamefont {Homma}}, \bibinfo {author} {\bibfnamefont
  {D.}~\bibnamefont {Li}}, \bibinfo {author} {\bibfnamefont {F.}~\bibnamefont
  {Honda}}, \bibinfo {author} {\bibfnamefont {Y.~J.}\ \bibnamefont {Sato}}, \
  and\ \bibinfo {author} {\bibfnamefont {D.}~\bibnamefont {Aoki}},\ }\href
  {\doibase 10.7566/JPSJ.87.044703} {\bibfield  {journal} {\bibinfo  {journal}
  {J. Phys. Soc. Jpn.}\ }\textbf {\bibinfo {volume} {87}},\ \bibinfo {pages}
  {044703} (\bibinfo {year} {2018})}\BibitemShut {NoStop}%
\bibitem [{\citenamefont {Vinter}\ and\ \citenamefont
  {Overhauser}(1980)}]{vinter_resolution_1980}%
  \BibitemOpen
  \bibfield  {author} {\bibinfo {author} {\bibfnamefont {B.}~\bibnamefont
  {Vinter}}\ and\ \bibinfo {author} {\bibfnamefont {A.~W.}\ \bibnamefont
  {Overhauser}},\ }\href {\doibase 10.1103/PhysRevLett.44.47} {\bibfield
  {journal} {\bibinfo  {journal} {Phys. Rev. Lett.}\ }\textbf {\bibinfo
  {volume} {44}},\ \bibinfo {pages} {47} (\bibinfo {year} {1980})}\BibitemShut
  {NoStop}%
\bibitem [{\citenamefont {Jeon}\ \emph {et~al.}(2014)\citenamefont {Jeon},
  \citenamefont {Zhou}, \citenamefont {Gyenis}, \citenamefont {Feldman},
  \citenamefont {Kimchi}, \citenamefont {Potter}, \citenamefont {Gibson},
  \citenamefont {Cava}, \citenamefont {Vishwanath},\ and\ \citenamefont
  {Yazdani}}]{Sangjun_Cd3As2}%
  \BibitemOpen
  \bibfield  {author} {\bibinfo {author} {\bibfnamefont {S.}~\bibnamefont
  {Jeon}}, \bibinfo {author} {\bibfnamefont {B.~B.}\ \bibnamefont {Zhou}},
  \bibinfo {author} {\bibfnamefont {A.}~\bibnamefont {Gyenis}}, \bibinfo
  {author} {\bibfnamefont {B.~E.}\ \bibnamefont {Feldman}}, \bibinfo {author}
  {\bibfnamefont {I.}~\bibnamefont {Kimchi}}, \bibinfo {author} {\bibfnamefont
  {A.~C.}\ \bibnamefont {Potter}}, \bibinfo {author} {\bibfnamefont {Q.~D.}\
  \bibnamefont {Gibson}}, \bibinfo {author} {\bibfnamefont {R.~J.}\
  \bibnamefont {Cava}}, \bibinfo {author} {\bibfnamefont {A.}~\bibnamefont
  {Vishwanath}}, \ and\ \bibinfo {author} {\bibfnamefont {A.}~\bibnamefont
  {Yazdani}},\ }\href {\doibase 10.1038/NMAT4023} {\bibfield  {journal}
  {\bibinfo  {journal} {Nature Materials}\ }\textbf {\bibinfo {volume} {13}},\
  \bibinfo {pages} {851} (\bibinfo {year} {2014})}\BibitemShut {NoStop}%
\bibitem [{\citenamefont {Zeljkovic}\ \emph {et~al.}(2015)\citenamefont
  {Zeljkovic}, \citenamefont {Okada}, \citenamefont {Serbyn}, \citenamefont
  {Sankar}, \citenamefont {Walkup}, \citenamefont {Zhou}, \citenamefont {Liu},
  \citenamefont {Chang}, \citenamefont {Wang}, \citenamefont {Hasan},
  \citenamefont {Chou}, \citenamefont {Lin}, \citenamefont {Bansil},
  \citenamefont {Fu},\ and\ \citenamefont {Madhavan}}]{Ilija_SnTe}%
  \BibitemOpen
  \bibfield  {author} {\bibinfo {author} {\bibfnamefont {I.}~\bibnamefont
  {Zeljkovic}}, \bibinfo {author} {\bibfnamefont {Y.}~\bibnamefont {Okada}},
  \bibinfo {author} {\bibfnamefont {M.}~\bibnamefont {Serbyn}}, \bibinfo
  {author} {\bibfnamefont {R.}~\bibnamefont {Sankar}}, \bibinfo {author}
  {\bibfnamefont {D.}~\bibnamefont {Walkup}}, \bibinfo {author} {\bibfnamefont
  {W.}~\bibnamefont {Zhou}}, \bibinfo {author} {\bibfnamefont {J.}~\bibnamefont
  {Liu}}, \bibinfo {author} {\bibfnamefont {G.}~\bibnamefont {Chang}}, \bibinfo
  {author} {\bibfnamefont {Y.~J.}\ \bibnamefont {Wang}}, \bibinfo {author}
  {\bibfnamefont {M.~Z.}\ \bibnamefont {Hasan}}, \bibinfo {author}
  {\bibfnamefont {F.}~\bibnamefont {Chou}}, \bibinfo {author} {\bibfnamefont
  {H.}~\bibnamefont {Lin}}, \bibinfo {author} {\bibfnamefont {A.}~\bibnamefont
  {Bansil}}, \bibinfo {author} {\bibfnamefont {L.}~\bibnamefont {Fu}}, \ and\
  \bibinfo {author} {\bibfnamefont {V.}~\bibnamefont {Madhavan}},\ }\href
  {\doibase 10.1038/NMAT4215} {\bibfield  {journal} {\bibinfo  {journal}
  {Nature Materials}\ }\textbf {\bibinfo {volume} {14}},\ \bibinfo {pages}
  {318} (\bibinfo {year} {2015})}\BibitemShut {NoStop}%
\bibitem [{\citenamefont {Kushwaha}\ \emph {et~al.}(2015)\citenamefont
  {Kushwaha}, \citenamefont {Krizan}, \citenamefont {Feldman}, \citenamefont
  {Gyenis}, \citenamefont {Randeria}, \citenamefont {Xiong}, \citenamefont
  {Xu}, \citenamefont {Alidoust}, \citenamefont {Belopolski}, \citenamefont
  {Liang}, \citenamefont {Zahid~Hasan}, \citenamefont {Ong}, \citenamefont
  {Yazdani},\ and\ \citenamefont {Cava}}]{kushwaha_Dirac}%
  \BibitemOpen
  \bibfield  {author} {\bibinfo {author} {\bibfnamefont {S.~K.}\ \bibnamefont
  {Kushwaha}}, \bibinfo {author} {\bibfnamefont {J.~W.}\ \bibnamefont
  {Krizan}}, \bibinfo {author} {\bibfnamefont {B.~E.}\ \bibnamefont {Feldman}},
  \bibinfo {author} {\bibfnamefont {A.}~\bibnamefont {Gyenis}}, \bibinfo
  {author} {\bibfnamefont {M.~T.}\ \bibnamefont {Randeria}}, \bibinfo {author}
  {\bibfnamefont {J.}~\bibnamefont {Xiong}}, \bibinfo {author} {\bibfnamefont
  {S.-Y.}\ \bibnamefont {Xu}}, \bibinfo {author} {\bibfnamefont
  {N.}~\bibnamefont {Alidoust}}, \bibinfo {author} {\bibfnamefont
  {I.}~\bibnamefont {Belopolski}}, \bibinfo {author} {\bibfnamefont
  {T.}~\bibnamefont {Liang}}, \bibinfo {author} {\bibfnamefont
  {M.}~\bibnamefont {Zahid~Hasan}}, \bibinfo {author} {\bibfnamefont {N.~P.}\
  \bibnamefont {Ong}}, \bibinfo {author} {\bibfnamefont {A.}~\bibnamefont
  {Yazdani}}, \ and\ \bibinfo {author} {\bibfnamefont {R.~J.}\ \bibnamefont
  {Cava}},\ }\href {\doibase 10.1063/1.4908158} {\bibfield  {journal} {\bibinfo
   {journal} {APL Materials}\ }\textbf {\bibinfo {volume} {3}},\ \bibinfo
  {pages} {041504} (\bibinfo {year} {2015})}\BibitemShut {NoStop}%
\bibitem [{\citenamefont {Fradkin}\ \emph {et~al.}(2010)\citenamefont
  {Fradkin}, \citenamefont {Kivelson}, \citenamefont {Lawler}, \citenamefont
  {Eisenstein},\ and\ \citenamefont {Mackenzie}}]{fradkin_nematic_2010}%
  \BibitemOpen
  \bibfield  {author} {\bibinfo {author} {\bibfnamefont {E.}~\bibnamefont
  {Fradkin}}, \bibinfo {author} {\bibfnamefont {S.~A.}\ \bibnamefont
  {Kivelson}}, \bibinfo {author} {\bibfnamefont {M.~J.}\ \bibnamefont
  {Lawler}}, \bibinfo {author} {\bibfnamefont {J.~P.}\ \bibnamefont
  {Eisenstein}}, \ and\ \bibinfo {author} {\bibfnamefont {A.~P.}\ \bibnamefont
  {Mackenzie}},\ }\href {\doibase 10.1146/annurev-conmatphys-070909-103925}
  {\bibfield  {journal} {\bibinfo  {journal} {Annual Review of Condensed Matter
  Physics}\ }\textbf {\bibinfo {volume} {1}},\ \bibinfo {pages} {153} (\bibinfo
  {year} {2010})}\BibitemShut {NoStop}%
\bibitem [{Note16()}]{Note16}%
  \BibitemOpen
  \bibinfo {note} {This has been observed numerically in \protect \rev@citealp
  {gresch_z2pack:_2017}, and the group-theoretic explanation may be found in
  Sec. VI-D-7 of \protect \rev@citealp {100p} or \protect \rev@citealp
  {li_topological_2017}}\BibitemShut {NoStop}%
\bibitem [{\citenamefont {Bouhon}\ \emph {et~al.}(2018)\citenamefont {Bouhon},
  \citenamefont {Black-Schaffer},\ and\ \citenamefont
  {Slager}}]{bouhon_wilson_2018}%
  \BibitemOpen
  \bibfield  {author} {\bibinfo {author} {\bibfnamefont {A.}~\bibnamefont
  {Bouhon}}, \bibinfo {author} {\bibfnamefont {A.~M.}\ \bibnamefont
  {Black-Schaffer}}, \ and\ \bibinfo {author} {\bibfnamefont {R.-J.}\
  \bibnamefont {Slager}},\ }\href {http://arxiv.org/abs/1804.09719} {\bibfield
  {journal} {\bibinfo  {journal} {arXiv:1804.09719}\ } (\bibinfo {year}
  {2018})}\BibitemShut {NoStop}%
\bibitem [{\citenamefont {Bradlyn}\ \emph {et~al.}(2018)\citenamefont
  {Bradlyn}, \citenamefont {Wang}, \citenamefont {Cano},\ and\ \citenamefont
  {Bernevig}}]{bradlyn_disconnected_2018}%
  \BibitemOpen
  \bibfield  {author} {\bibinfo {author} {\bibfnamefont {B.}~\bibnamefont
  {Bradlyn}}, \bibinfo {author} {\bibfnamefont {Z.}~\bibnamefont {Wang}},
  \bibinfo {author} {\bibfnamefont {J.}~\bibnamefont {Cano}}, \ and\ \bibinfo
  {author} {\bibfnamefont {B.~A.}\ \bibnamefont {Bernevig}},\ }\href
  {http://arxiv.org/abs/1807.09729} {\bibfield  {journal} {\bibinfo  {journal}
  {arXiv:1807.09729 [cond-mat]}\ } (\bibinfo {year} {2018})},\ \bibinfo {note}
  {arXiv: 1807.09729}\BibitemShut {NoStop}%
\bibitem [{\citenamefont {Michel}\ and\ \citenamefont
  {Zak}(2001)}]{michel_elementary_2001}%
  \BibitemOpen
  \bibfield  {author} {\bibinfo {author} {\bibfnamefont {L.}~\bibnamefont
  {Michel}}\ and\ \bibinfo {author} {\bibfnamefont {J.}~\bibnamefont {Zak}},\
  }\href {\doibase 10.1016/S0370-1573(00)00093-4} {\bibfield  {journal}
  {\bibinfo  {journal} {Phys. Rep.}\ }\textbf {\bibinfo {volume} {341}},\
  \bibinfo {pages} {377} (\bibinfo {year} {2001})}\BibitemShut {NoStop}%
\bibitem [{\citenamefont {Wang}\ \emph {et~al.}(2016)\citenamefont {Wang},
  \citenamefont {Alexandradinata}, \citenamefont {Cava},\ and\ \citenamefont
  {Bernevig}}]{wang_hourglass_2016}%
  \BibitemOpen
  \bibfield  {author} {\bibinfo {author} {\bibfnamefont {Z.}~\bibnamefont
  {Wang}}, \bibinfo {author} {\bibfnamefont {A.}~\bibnamefont
  {Alexandradinata}}, \bibinfo {author} {\bibfnamefont {R.~J.}\ \bibnamefont
  {Cava}}, \ and\ \bibinfo {author} {\bibfnamefont {B.~A.}\ \bibnamefont
  {Bernevig}},\ }\href {\doibase 10.1038/nature17410} {\bibfield  {journal}
  {\bibinfo  {journal} {Nature (London)}\ }\textbf {\bibinfo {volume} {532}},\
  \bibinfo {pages} {189} (\bibinfo {year} {2016})}\BibitemShut {NoStop}%
\bibitem [{\citenamefont {Bradlyn}\ \emph {et~al.}(2017)\citenamefont
  {Bradlyn}, \citenamefont {Elcoro}, \citenamefont {Cano}, \citenamefont
  {Vergniory}, \citenamefont {Wang}, \citenamefont {Felser}, \citenamefont
  {Aroyo},\ and\ \citenamefont {Bernevig}}]{bradlyn_topological_2017}%
  \BibitemOpen
  \bibfield  {author} {\bibinfo {author} {\bibfnamefont {B.}~\bibnamefont
  {Bradlyn}}, \bibinfo {author} {\bibfnamefont {L.}~\bibnamefont {Elcoro}},
  \bibinfo {author} {\bibfnamefont {J.}~\bibnamefont {Cano}}, \bibinfo {author}
  {\bibfnamefont {M.~G.}\ \bibnamefont {Vergniory}}, \bibinfo {author}
  {\bibfnamefont {Z.}~\bibnamefont {Wang}}, \bibinfo {author} {\bibfnamefont
  {C.}~\bibnamefont {Felser}}, \bibinfo {author} {\bibfnamefont {M.~I.}\
  \bibnamefont {Aroyo}}, \ and\ \bibinfo {author} {\bibfnamefont {B.~A.}\
  \bibnamefont {Bernevig}},\ }\href {\doibase 10.1038/nature23268} {\bibfield
  {journal} {\bibinfo  {journal} {Nature (London)}\ }\textbf {\bibinfo {volume}
  {547}},\ \bibinfo {pages} {298} (\bibinfo {year} {2017})}\BibitemShut
  {NoStop}%
\bibitem [{\citenamefont {Zil'berman}(1957)}]{zilberman}%
  \BibitemOpen
  \bibfield  {author} {\bibinfo {author} {\bibfnamefont {G.~E.}\ \bibnamefont
  {Zil'berman}},\ }\href@noop {} {\bibfield  {journal} {\bibinfo  {journal}
  {Soviet Physics JETP}\ }\textbf {\bibinfo {volume} {23}},\ \bibinfo {pages}
  {208} (\bibinfo {year} {1957})}\BibitemShut {NoStop}%
\bibitem [{\citenamefont {Culcer}\ \emph {et~al.}(2005)\citenamefont {Culcer},
  \citenamefont {Yao},\ and\ \citenamefont {Niu}}]{culcer_coherent_2005}%
  \BibitemOpen
  \bibfield  {author} {\bibinfo {author} {\bibfnamefont {D.}~\bibnamefont
  {Culcer}}, \bibinfo {author} {\bibfnamefont {Y.}~\bibnamefont {Yao}}, \ and\
  \bibinfo {author} {\bibfnamefont {Q.}~\bibnamefont {Niu}},\ }\href {\doibase
  10.1103/PhysRevB.72.085110} {\bibfield  {journal} {\bibinfo  {journal} {Phys.
  Rev. B}\ }\textbf {\bibinfo {volume} {72}},\ \bibinfo {pages} {085110}
  (\bibinfo {year} {2005})}\BibitemShut {NoStop}%
\bibitem [{\citenamefont {Xiao}\ \emph {et~al.}(2010)\citenamefont {Xiao},
  \citenamefont {Chang},\ and\ \citenamefont {Niu}}]{xiao_berry_2010}%
  \BibitemOpen
  \bibfield  {author} {\bibinfo {author} {\bibfnamefont {D.}~\bibnamefont
  {Xiao}}, \bibinfo {author} {\bibfnamefont {M.-C.}\ \bibnamefont {Chang}}, \
  and\ \bibinfo {author} {\bibfnamefont {Q.}~\bibnamefont {Niu}},\ }\href
  {\doibase 10.1103/RevModPhys.82.1959} {\bibfield  {journal} {\bibinfo
  {journal} {Rev. Mod. Phys.}\ }\textbf {\bibinfo {volume} {82}},\ \bibinfo
  {pages} {1959} (\bibinfo {year} {2010})}\BibitemShut {NoStop}%
\bibitem [{RZf()}]{RZfignote}%
  \BibitemOpen
  \href@noop {} {}\bibinfo {note} {$l^2S\gg 1$ is assured by requiring
  $\epsilon_0>0.4$; the reason for choosing $\epsilon_0<1.0$ is Dirac point
  lying at high energy requires a high cut off of the number of harmonics (cf.
  App. \ref{sec:numerical}) in numerical diagonalization. This requirement is
  equivalent to $\sqrt{0.8m}\alpha/\hbar<Z<\sqrt{2m}\alpha/\hbar$,
  corresponding to the boundaries of the plots in (d).}\BibitemShut {Stop}%
\bibitem [{Note17()}]{Note17}%
  \BibitemOpen
  \bibinfo {note} {To appreciate this, consider that any unitary matrix can be
  expressed as $e^{ih}$ with Hermitian $h$. If this unitary is also symmetric,
  $h$ must also by symmetric, hence also real. Thus the eigenvectors of a
  symmetric unitary can always be chosen to be real.}\BibitemShut {Stop}%
\bibitem [{\citenamefont {Chang}\ and\ \citenamefont
  {Niu}(1996)}]{chang_berry_1996}%
  \BibitemOpen
  \bibfield  {author} {\bibinfo {author} {\bibfnamefont {M.-C.}\ \bibnamefont
  {Chang}}\ and\ \bibinfo {author} {\bibfnamefont {Q.}~\bibnamefont {Niu}},\
  }\href {\doibase 10.1103/PhysRevB.53.7010} {\bibfield  {journal} {\bibinfo
  {journal} {Phys. Rev. B}\ }\textbf {\bibinfo {volume} {53}},\ \bibinfo
  {pages} {7010} (\bibinfo {year} {1996})}\BibitemShut {NoStop}%
\bibitem [{\citenamefont {Yamaji}(1989)}]{yamaji_angle_1989}%
  \BibitemOpen
  \bibfield  {author} {\bibinfo {author} {\bibfnamefont {K.}~\bibnamefont
  {Yamaji}},\ }\href {\doibase 10.1143/JPSJ.58.1520} {\bibfield  {journal}
  {\bibinfo  {journal} {J. Phys. Soc. Jpn.}\ }\textbf {\bibinfo {volume}
  {58}},\ \bibinfo {pages} {1520} (\bibinfo {year} {1989})}\BibitemShut
  {NoStop}%
\bibitem [{\citenamefont {H\"oller}\ \emph {et~al.}(2018)\citenamefont
  {H\"oller}, \citenamefont {Read},\ and\ \citenamefont
  {Harris}}]{holler_non-hermitian_2018}%
  \BibitemOpen
  \bibfield  {author} {\bibinfo {author} {\bibfnamefont {J.}~\bibnamefont
  {H\"oller}}, \bibinfo {author} {\bibfnamefont {N.}~\bibnamefont {Read}}, \
  and\ \bibinfo {author} {\bibfnamefont {J.~G.~E.}\ \bibnamefont {Harris}},\
  }\href {http://arxiv.org/abs/1809.07175} {\bibfield  {journal} {\bibinfo
  {journal} {arXiv:1809.07175}\ } (\bibinfo {year} {2018})}\BibitemShut
  {NoStop}%
\bibitem [{\citenamefont {Gresch}\ \emph {et~al.}(2017)\citenamefont {Gresch},
  \citenamefont {Autès}, \citenamefont {Yazyev}, \citenamefont {Troyer},
  \citenamefont {Vanderbilt}, \citenamefont {Bernevig},\ and\ \citenamefont
  {Soluyanov}}]{gresch_z2pack:_2017}%
  \BibitemOpen
  \bibfield  {author} {\bibinfo {author} {\bibfnamefont {D.}~\bibnamefont
  {Gresch}}, \bibinfo {author} {\bibfnamefont {G.}~\bibnamefont {Autès}},
  \bibinfo {author} {\bibfnamefont {O.~V.}\ \bibnamefont {Yazyev}}, \bibinfo
  {author} {\bibfnamefont {M.}~\bibnamefont {Troyer}}, \bibinfo {author}
  {\bibfnamefont {D.}~\bibnamefont {Vanderbilt}}, \bibinfo {author}
  {\bibfnamefont {B.~A.}\ \bibnamefont {Bernevig}}, \ and\ \bibinfo {author}
  {\bibfnamefont {A.~A.}\ \bibnamefont {Soluyanov}},\ }\href {\doibase
  10.1103/PhysRevB.95.075146} {\bibfield  {journal} {\bibinfo  {journal} {Phys.
  Rev. B}\ }\textbf {\bibinfo {volume} {95}},\ \bibinfo {pages} {075146}
  (\bibinfo {year} {2017})}\BibitemShut {NoStop}%
\bibitem [{\citenamefont {Li}\ \emph {et~al.}(2017)\citenamefont {Li},
  \citenamefont {Yan}, \citenamefont {Wang},\ and\ \citenamefont
  {Held}}]{li_topological_2017}%
  \BibitemOpen
  \bibfield  {author} {\bibinfo {author} {\bibfnamefont {G.}~\bibnamefont
  {Li}}, \bibinfo {author} {\bibfnamefont {B.}~\bibnamefont {Yan}}, \bibinfo
  {author} {\bibfnamefont {Z.}~\bibnamefont {Wang}}, \ and\ \bibinfo {author}
  {\bibfnamefont {K.}~\bibnamefont {Held}},\ }\href {\doibase
  10.1103/PhysRevB.95.035102} {\bibfield  {journal} {\bibinfo  {journal} {Phys.
  Rev. B}\ }\textbf {\bibinfo {volume} {95}},\ \bibinfo {pages} {035102}
  (\bibinfo {year} {2017})}\BibitemShut {NoStop}%
\end{thebibliography}%

\end{document}